\documentclass[iop]{emulateapj-rtx4} 
\shortauthors{Sekanina}
\shorttitle{Kreutz Sungrazers in AD 363?}
\slugcomment{Version \today }
\newcommand{\Rsun}{$R_{\mbox{\scriptsize \boldmath $\odot$}}\!$}

\newcommand{\rasun}{$\alpha_{\mbox{\tiny \boldmath $\odot$}}$}
\newcommand{\decsun}{$\delta_{\mbox{\tiny \boldmath $\odot$}}$}
\newcommand{\azsun}{${\cal A}_{\mbox{\tiny \boldmath $\odot$}}$}
\newcommand{\elsun}{${\cal E}_{\mbox{\tiny \boldmath $\odot$}}$}
\newcommand{\elng}{$\langle \epsilon \rangle$}
\newcommand{\vis}{$\langle \Im \rangle$}
\begin{document}
\title{UNPRECEDENTED DAYLIGHT DISPLAY OF KREUTZ SUNGRAZERS IN AD 363?}
\author{Zdenek Sekanina}
\affil{Jet Propulsion Laboratory, California Institute of Technology,
  4800 Oak Grove Drive, Pasadena, CA 91109, U.S.A.}
\email{Zdenek.Sekanina@jpl.nasa.gov.}

\begin{abstract} 
In the context of the recently proposed contact-binary model (Sekanina 2021),       
I investigate the circumstances of the first perihelion passage of the Kreutz       
sungrazers in orbits with barycentric periods near 735~yr, following the initial    
near-aphelion splitting of the presumed progenitor, Aristotle's comet of 372~BC.    
Given favorable conditions at this breakup and at episodes of secondary             
fragmentation in its aftermath, the fragments should have arrived at their first    
perihelion nearly simultaneously, reminiscent of the anticipated outcome for the    
two-superfragment model's perihelion return of AD~356 (Sekanina \& Chodas           
2004).  The relevant case of a swarm of Kreutz sungrazers is examined to appraise   
possible scientific ramifications of the brief remark by Ammianus Marcellinus,      
a Roman historian, that {\it ``in broad daylight comets were seen"\/} in late       
AD~363, only seven years later.  The tested scenario, which does not contradict     
Ammianus' narrative and is consistent with the contact-binary model, involves a     
set of ten sungrazers visible in the daytime, all reaching perihelion over a        
period of 4.6~days.  As part of this work, I comment on the role of the rapidly     
developing, brilliant post-perihelion tail; revise the apparent magnitude typical   
for the first and last naked-eye sightings; compare the visibility conditions       
in full daylight, in twilight, and at night; and, for the first time, present       
circumstantial evidence that favors comet X/1106~C1 as the parent to C/1843~D1      
rather than to C/1882~R1 and C/1965~S1.                                             
\end{abstract} 

%
\section{Introduction}  
For a short period of time after the discovery of the sungrazer C/1882~R1, the
Great September Comet of 1882, its possible identity with C/1843~D1, the Great
March Comet of 1843, was a point of contention.  The winning argument against
the identity relied entirely on their long orbital periods (e.g., Tebbutt 1882,
Plummer 1889, Lynn 1903).  I find it puzzling that the major, nearly 20$^\circ$
difference in the longitude of the ascending node (and equivalent differences in
the other angular elements) as well as the discrepancy of 0.5~{\Rsun}\,, or nearly
50~percent, in the perihelion distance were never employed in the arguments
against the identity.  An answer to the question of why not was later offered by
Lynn (1903), when he said in his remarks on the two spectacular comets that ``what
events may have been occasioned in their past history by their close approaches
to the Sun when in perihelion can only be matter of speculation.''  Thus, as late
as the beginning of the 20th century, major changes in the orbit orientation and
perihelion arc (but apparently not the period) were believed possible as a result
of the sungrazers' interaction with the Sun's corona, even though Kreutz (1888,
1891) and others were able to successfully link preperihelion and post-perihelion
astrometric observations of C/1882~R1 by a gravitational orbit in compliance with
the Newtonian theory.

The nature of the sources causing the significant orbital differences between
C/1843~D1 and C/1882~R1 (and other Kreutz sungrazers), especially in the nodal
longitude and perihelion distance, was first seriously confronted by \"{O}pik
(1966) and, in particular, Marsden (1967).  In these independent efforts the observed
discrepancies in the orbital elements were interpreted as cumulative effects of the
indirect planetary perturbations over a large number of revolutions about the Sun.
Marsden did not appear to be very happy about the explanation (his sophisticated
approach to the problem notwithstanding), but he realized that there was no
alternative, insofar as all Kreutz comets were products of the process of tidal
breakup at perihelion, in close proximity of the Sun.  In the 1960s, this
still was an overwhelmingly prevailing view, reinforced eight decades earlier by
the extensive fragmentation of C/1882~R1, thoroughly documented by the observed
post-perihelion multiplicity of the comet's nucleus (Kreutz 1888, 1891), and
further strengthened by evidence of fragmentation of C/1965~S1, which was
investigated, among others, by Marsden (1967) himself. 

\section{Cascading Nature of Fragmentation} 
Two decades ago I demonstrated that a fragment separating from its parent in a Kreutz
sungrazer orbit at aphelion, presumed to be at a heliocentric distance of $\sim$170~AU
(equivalent to{\vspace{-0.04cm}} an orbital period of 784~yr), with a velocity of
5~m~s$^{-1}$ in the direction normal to the orbital plane is perturbed by
27$^\circ$ in its nodal longitude, while a separation velocity of the same
magnitude in the transverse direction exerts a perturbation of 0.7\,\,{\Rsun}\,
in the fragment's perihelion distance (Sekanina 2002).  In that study I also
argued that the process of {\it nontidal\/} fragmentation at large heliocentric
distance --- first proposed for dwarf sungrazers (Sekanina 2000) imaged by the
coronagraphs on board the Solar and Heliospheric Observatory (SOHO) --- has obviously
been affecting the entire Kreutz sungrazer system, including the progenitor.

It is well-known that no perceptible perturbations of any fragment's angular
elements and perihelion distance can be achieved as a result of the parent's
breakup close to the Sun.  A separation velocity of 5~m~s$^{-1}$ at perihelion
does, however, change the fragment's orbital period by $\sim$400 years or
more.  In general, the orbital distribution of Kreutz sungrazers shows
evidence of significant perturbations of all elements, thereby suggesting that
nuclear fragmentation has cascading nature, occurring throughout the orbit.
This conclusion is important, because it implies a {\it much higher number
of fragmentation events per unit time\/} than if they were confined to close
proximity of perihelion.  Higher fragmentation rates lead in turn to a {\it
shorter lifespan\/} of the Kreutz system, especially restricting survival
of objects as intrinsically bright and massive as C/1843~D1 and
C/1882~R1.\footnote{Predicated on my own study of their light curves, I
referred to C/1882~R1 as the brightest Kreutz sungrazer and to C/1843~D1
as the second brightest (Sekanina 2002).  The fact is that opinions differ
widely on which of the two comets was more spectacular (and, presumably,
more massive), with no strong preference either way (Section~5.5).}

\section{Two-Superfragment Model and\\Simultaneous Arrivals of Fragments\\at
 First Perihelion}  
Marsden's (1967) description of the Kreutz group (which at the time consisted of
no more than a dozen known sungrazers, including six questionable ones) in terms
of two subgroups had major influence on the further evolution of ideas on the
subject.  Equally impressive was Marsden's result of integration of the orbits of
comets C/1965~S1 and C/1882~R1 back in time to the early 12th century, prompting
his suggestion that the spectacular comet X/1106~C1 was the most promising
candidate for their common parent.

A version of the two-superfragment model for the Kreutz system, proposed in
Sekanina (2002), was implemented in a follow-up study (Sekanina \& Chodas 2004),
which adopted Marsden's suggested relationship among X/1106~C1, C/1882~R1, and
C/1965~S1.  The sizable gaps between the longitudes of the ascending node and
perihelion distances of C/1843~D1 and C/1882~R1 were consistent with their birth
as two superfragments in the progenitor's initial breakup at 50~AU from the Sun
early in the 4th century AD.  The observed evolution of the naked-eye members
of the Kreutz system was then modeled as a result of progressive fragmentation
of the two superfragments over a period of time a little longer than two
revolutions about the Sun, with C/1843~D1 and C/1882~R1 as their respective
largest surviving masses.

The two-superfragment model has been an extremely useful tool in pursuit of
getting insight into the evolution of the Kreutz system.  However, because of new
developments since 2004, the model has become in some ways obsolete, as I recently
argued (Sekanina 2021; referred to hereafter as Paper~1).  Yet, two of the model's
features continue to be valid:\ (1)~the {\small \bf progenitor fragmented at a very
large heliocentric distance}; and (2)~the {\small \bf fragments} --- whether two
or more --- {\small \bf arrived at their first perihelion} (in the 4th century AD)
{\small \bf nearly simultaneously}.  In fact, the {\it modeled\/} superfragments
arrived at their first perihelion, in AD~356 or 30~years after the progenitor's
breakup at 50~AU from the Sun, merely seven days apart.  Accordingly, a promising
historical account of the Kreutz system's first appearance should include a reference
to {\it multiple comets arriving at nearly the~same~time\/}.

\section{Weaknesses of Two-Superfragment Model\\and Paradigm of a Contact
 Binary} 
Evidence on C/1843 D1 and C/1882 R1 suggests that the two superfragments may
have been of approximately equal mass.  If the progenitor were a solid body of
quasi-spherical shape, its fracture through the center into two halves appears
a priori to be an unlikely scenario.  On the other hand, if, in line with
Paper~1, the progenitor was a {\small \bf contact binary}, consisting of two
about equally large lobes that eons ago coalesced into a single body at a very
low relative velocity, its splitting along the bridge linking the two lobes at
a fairly recent time sounds as a plausible event. 

Developing the two-superfragment{\vspace{-0.02cm}} model, Sekanina \& Chodas
(2004) had to tolerate separation velocities in a general range of 6--10~m~s$^{-1}$.
Compared to the separation velocities of fragments of most known split comets (e.g.,
Sekanina 1982), these numbers are too high by a factor of three or more.  To avoid
such high separation velocities, one should accept that the progenitor fragmented
in the general proximity of aphelion rather than at 50~AU.

The focus of another controversy linked to the two-superfragment model was the
already mentioned role of the brilliant comet (and a very probable sungrazer)
X/1106~C1.  Its relationship to C/1882~R1 and C/1965~S1 is an open question;
in Section~5.6 I present the first circumstantial evidence that X/1106~C1 was
the parent to C/1843~D1 rather than to C/1965~S1 and C/1882~R1.

Most difficulties that the two-superfragment model encounters have been
removed by the introduction of a contact-binary model.  Only two problems remain.
One is the missing second spectacular sungrazer in the early 12th century that
served as the parent to C/1882~R1, if X/1106~C1 was the parent to C/1843~D1.
Even though there is a chance that the second sungrazer was bright enough to have
been visible in broad daylight, it apparently was missed, if it arrived between
May and early August.  Short of its future accidental discovery in a historical
record, this problem will remain unsolved.

The other problem is the absence of evidence for two (or more) bright sungrazers,
arriving nearly simultaneously in AD~356, one revolution about the Sun before
X/1106~C1.  In this context it is most encouraging to note the brief remark by
{\small \bf Ammianus Marcellinus}, a reputable Roman historian, conveying that
towards the end of AD~363 {\small \bf ``in broad daylight comets were seen''}
from Antioch on the Orontes (Rolfe 1940), his residence at the time.  His note
refers to an event of the right kind, which occurred only seven years after
the time that satisfies the two-supefragment model!  Besides, Seargent (2009)
speculated that the daylight comets could have been {\it ``several sungrazing
fragments close together''\/}, given that members of the Kreutz system {\it
``very late in the year would have had a strong southerly declination and might
have been seen \ldots only in the daytime close to perihelion''\/} from much
of the northern hemisphere.  Even though the account by Ammianus is brutally
short, I deem it worthwhile to attempt reconstructing a scenario of daytime
multiple-fragment spectacle, orchestrated by Kreutz sun\-grazers and consistent
with the contact-binary model.

\begin{table*}
\vspace{-4.2cm}
\hspace{0.66cm}
\centerline{
\scalebox{1}{
\includegraphics{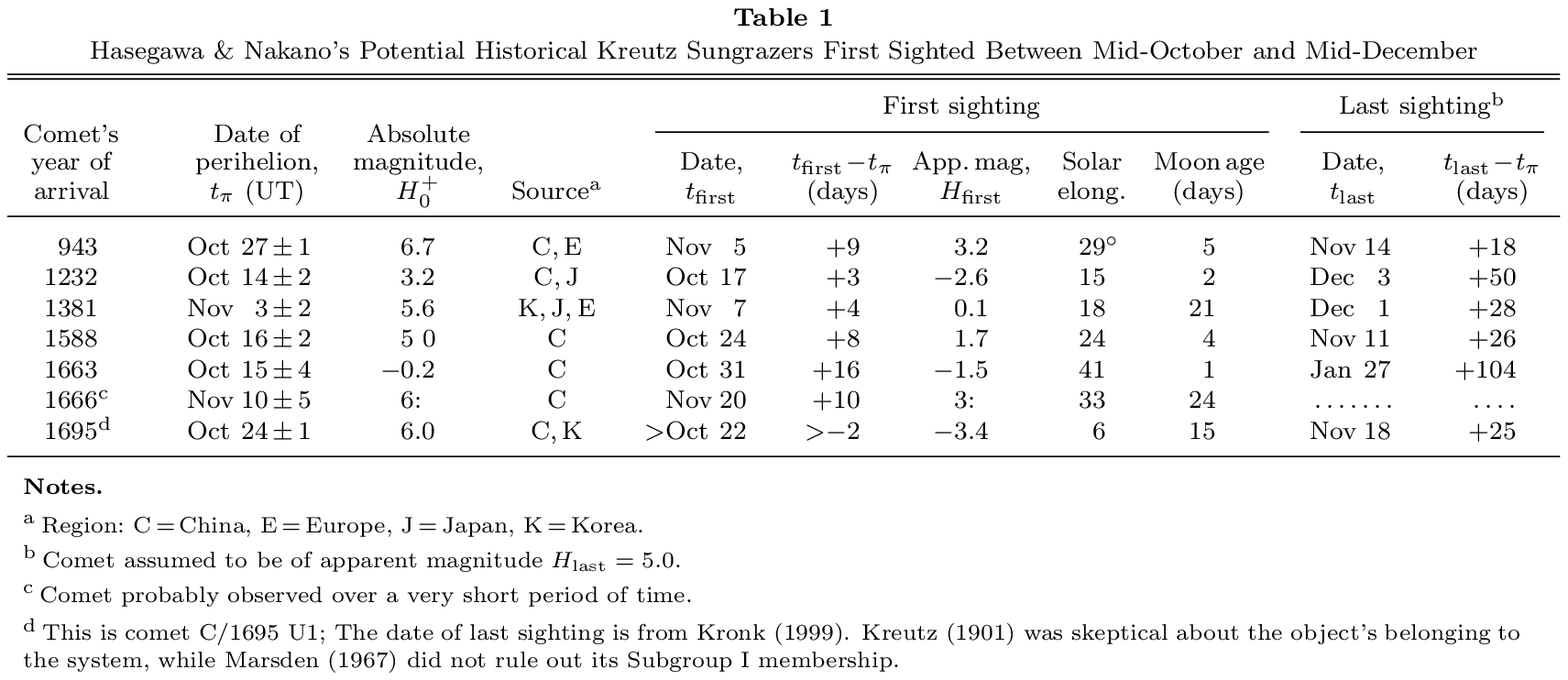}}}
\vspace{-17.6cm}
\end{table*}

\section{The Case for Daytime Kreutz Sungrazers\\in AD~363:\ Preliminaries}
%
Details on the spectacle of daylight comets seen by Ammianus will never be
known.  A scenario described in Section~6 simulates a display that may or may
not resemble his experience.  Dictated by the documented time of Ammianus'
arrival at Antioch, the comets could not have been sighted before mid-October
(e.g., Drijvers 2011, 2022).  On the other hand, they were probably seen before
Emperor Jovian's departure from Antioch ``on a day in the dead of winter''
(Rolfe 1940).  While he may have left as early as November (e.g., Drijvers
2011, 2022), I conservatively adopt that the comets could have been observed
as late as mid-December.  The two-months period defines the time window of
interest to this study.  The objective is being pursued in stages.

\subsection{Potential Historical Kreutz Sungrazers Discovered\\Between Mid-October and
 Mid-December} 
To learn more about such members of the Kreutz system, I first examine Hasegawa
\& Nakano's (2001) list of 24~potential historical Kreutz sungrazers up to
AD~1702 to single out the ones discovered between mid-October and mid-December.
Table~1 shows that there were seven such objects.  The {\it post-perihelion\/}
absolute magnitudes $H_0$ (i.e., reduced{\vspace{-0.04cm}} to unit distances
from the Earth, $\Delta$,~and the Sun, $r$; later called $H_0^+$) are modified
values (extended to one decimal) introduced by Hasegawa \& Nakano on the
assumptions that the comets moved in an average Kreutz sungrazing orbit
(similar to that of Population~Ia in Paper~1) and{\vspace{-0.04cm}} at the
time of last sighting were of apparent visual magnitude~5, obeying a
\mbox{$\Delta^{-2} r^{-4}$} law.
%

The comets in Table 1 exhibit two striking peculiarities:\ (i)~a strongly uneven
temporal distribution, with four of the seven potential sungrazers discovered in
the first quarter of the two-month period, six in the first half, but none in the
last three weeks, apparently an effect of gradually increasing deep southerly
declinations; and (ii)~at least six and quite possibly all seven were not discovered
until after perihelion.  The lack of preperihelion sightings may seem strange
given that five of the seven comets in Table~1 passed perihelion within a week
of October~21, the perihelion date of C/1965~S1 (Ikeya-Seki), whose apparition
has been deemed favorable.

In order to understand the features of Table~1, I examine (i)~Hasegawa \&
Nakano's (2001) assumption regarding the apparent brightness at the last naked-eye
sighting; (ii)~the problem of the apparent brightness of a comet at naked-eye
discovery; and (iii)~the expected shape of Kreutz sungrazers' preperihelion
and post-perihelion light curves.  In the following sections I separately
address each of these issues.

\subsection{Post-Perihelion Tails of Kreutz Sungrazers and\\Their Last
 Naked-Eye Sightings} 
Bright Kreutz sungrazers are unique in that they display conspicuous {\it
post-perihelion\/} tails,\footnote{This excessive brightness is by no means
a property of bright sungrazers' {\it preperihelion\/} tails, except possibly
near perihelion.} which carry enormous amounts of microscopic dust grains
ejected from the nucleus at very small heliocentric distances just after
perihelion but no longer subjected to instant sublimation.  The supply
of this dust is so substantial that the {\small \bf tail becomes much
brighter than the head}.

Compelling arguments are consistently provided by observations of at least
three sungrazers.  Roemer (1966) reported that comet Ikeya-Seki's {\it ``tail
nearly 10\,$^\circ$ in length could still be detected with the naked eye
[even though the] brightness of the head \ldots was about magnitude 7.4''\/}
37~days after perihelion.\footnote{From the comet's light curve (Sekanina
2002) I find for this time a total visual magnitude of the head equaling
6.9 (see Table~3 below).}  Similarly, the tail of the more recent sungrazer
C/2011~W3 (Lovejoy) was detected with the naked eye by several observers
(I.\,Cooper, R.\,Kaufman, M.\,Mattiazzo, R.\,McNaught)\footnote{See {\tt
https://groups.io/g/comets-ml}, messages 19113, 19120, 19121, 19134,
19146, and 19147.} as late as 20 or more days after perihelion, when the
brightness of the disintegrating head was estimated at magnitude 6.6--8.1
(Green 2012).  And back in the late 19th century, Gould (1883b) reported
that the Great September Comet of 1882 {\it ``was last seen by the
unaided eye on [1883]~March~7, at which time it was already very faint in
the telescope.''}  The formula for the post-perihelion light curve of
C/1882~R1 (Sekanina 2002) suggests that the comet's head could not be
brighter than magnitude 6.6 at the time, more than 170~days (!) after
perihelion; it again was the tail that Gould was referring to.

The prodigious brightness of the post-perihelion tails of major Kreutz
sungrazers can be documented even more convincingly.  One line of argument
is based on Weinberg \& Beeson's (1976) photometric investigation of
comet Ikeya-Seki on four days after perihelion, examining primarily the
polarization of the comet's light.  From scans of a tail image obtained
at Mt.\,Haleakala, Maui, Hawaii, on 1965 October 29.60~UT, the authors
determined the surface brightness along the tail's axis as a function
of the angular distance from the nucleus at 1$^\circ$ intervals.  The
comet's head was 5$^\circ\!$.3 below the horizon, so that the data refer
to only the outer two thirds of the tail.  Weinberg \& Beeson did not
compute the tail's total magnitude, but its estimate is offered in the
following.

The surface-brightness distribution, $B(b,a)$, across the tail at a
given angular distance from the nucleus, $b$, is approximated by the
Gaussian function,
\begin{equation}
B(b,a) = B_0(b) \exp \! \left[ -\frac{a^2}{2 \sigma_b^2} \right], 
\end{equation}
where $a$ is an angular distance from the tail's axis, $B_0(b)$ is the
surface brightness at the point on the axis at distance $b$ from the
nucleus, and $\sigma_b$ is the distance from the axis at a point of the
steepest surface-brightness gradient, measuring the optically perceived
half-width of the tail at distance $b$.  The values of $B_0(b)$ were
published by Weinberg \& Beeson, while the values of $\sigma_b$ over
the range of distances $b$ were measured by the present author from the
tail's width on an image in their paper.  This image was taken with
the same instrument one day earlier and shows the entire comet.  The
measurements of $\sigma_b$ were made on the assumption that the ratio
$\sigma_b/b$ as a function of $b$ did not change significantly in
24~hours.  The total apparent brightness of the tail between any two
distances $b_1$ and $b_2$ from the nucleus is given by
\begin{eqnarray}
{\cal B}(b_1,b_2) & = & \! \int_{b_1}^{b_2} \!\! B_0(b) \, db \!
 \int_{-\infty}^{+\infty} \!\! \exp \! \left[ -\frac{a^2}{2 \sigma_b^2}
 \right] da \nonumber \\
  & = & \sqrt{2 \pi} \! \int_{b_1}^{b_2} \!\! B_0(b) \, \sigma_b \, db. 
\end{eqnarray}
Weinberg \& Beeson tabulate $B_0(b)$ in units of S$_{10}$,~the number of visual
magnitude~10 stars per square degree.  The product of $B_0(b) \, \sigma_b$ is
integrated~numeri\-cally over the available range of distances to derive the
visual brightness of the visible part of the tail (at distances of more than
6$^\circ\!$.8 from the nucleus) equaling \mbox{${\cal B}(6^\circ\!.8,\infty) =
42,400$ magnitude~10 stars}, or a visual magnitude of $-$1.6.  Furthermore,
extrapolating the Weinberg \& Beeson's curve of surface brightness back to the
nucleus, the tail's total brightness is crudely estimated at about \mbox{${\cal
B}(0,\infty) = 94,000$ magnitude~10 stars}, equivalent to a total visual
magnitude of
\begin{equation}
H = -2.4 \: {\rm mag}. 
\end{equation}
The comet's head at the time had a total magnitude of +2.5, so its brightness
was just about 1~percent of the tail's brightness!

An independent estimate is based on a generic model of comet activity and
a sizable sungrazer's production rate of dust, in line with{\vspace{-0.05cm}}
the results obtained by Weinberg \& Beeson (1976) and by others.  Let
$\dot{Z}_0$ be the average sublimation rate from a unit surface area of a comet
at a heliocentric distance of 1~AU, $\mu_{\rm mol}$ the average mass of a molecule
(assuming a number of sublimating species), and $R$ the radius of the comet's
nucleus.  If $f$ is the fraction of the nucleus surface that is active and
$\kappa$ is the dust-to-gas mass production rate ratio, the comet's mass production
rate of dust at 1~AU equals
\begin{equation}
\dot{\cal M}_0 = 4 \pi \kappa f R^2 \mu_{\rm mol} \, \dot{Z}_0. 
\end{equation}
Let this mass production rate vary inversely as the square of heliocentric
distance, $r(t)$,
\begin{equation}
\dot{\cal M}(t) = \dot{\cal M}_0 \left( \! \frac{r_0}{r} \! \right)^{\!2}, 
\end{equation}
where \mbox{$r_0 = 1$ AU}.  The post-perihelion dust tail of a sungrazer
is made up of only post-perihelion ejecta, starting at a time $t_{\rm
beg}$, as all preperihelion particles were sublimated at perihelion.  Thus, at
time $t_{\rm obs}$ the mass of dust in the tail amounts to
\begin{eqnarray}
{\cal M}(t) & = & \! \int_{t_{\rm beg}}^{t_{\rm obs}} \!\!\! \dot{\cal M}(t)
 \,dt = \dot{\cal M}_0 r_0^2 \! \int_{t_{\rm beg}}^{t_{\rm obs}}
 \frac{dt}{r^2} \nonumber \\[0.2cm]
   & = & \frac{\dot{\cal M}_0 r_0^2}{k_0} \sqrt{\frac{2}{q}} \!\left( \!
  \arccos \!\sqrt{\frac{q}{r_{\rm obs}}} \!-\! \arccos \!
  \sqrt{\frac{q}{r_{\rm beg}}} \right) \:\!\!\!, 
\end{eqnarray}
where $k_0$ is the Gaussian gravitational constant, $q$ is the perihelion
distance of the sungrazer's orbit (approximated by a parabola), \mbox{$r_{\rm
beg} \!=\! r(t_{\rm beg})$} and \mbox{$r_{\rm obs} \!=\! r(t_{\rm obs})$}.
The apparent visual magnitude, $H$, that this mass of dust has at a given
time depends on its cross-sectional area $X$, geometric albedo $p$, the
function $\Phi(\psi)$ of the phase angle $\psi$, as well as on the distances
from Earth, $\Delta$, and the Sun, $r$:
\begin{equation}
H = 15.38 -2.5 \log X -2.5 \log \left[p\,\Phi(\psi)\,\Delta^{-2} r^{-2}
 \right], 
\end{equation}
where $X$ is in km$^2$, $\Delta$ and $r$ in AU, and \mbox{$\Phi(0^\circ) = 1$}.
The total cross-sectional area of the dust in the tail is related to its mass
${\cal M}$ via the size and mass distributions and the bulk density.  The tail's
total cross-sectional area can be written as \mbox{$X = \frac{1}{4} \pi \langle
s^2 \rangle\,{\cal N}$} and{\vspace{-0.06cm}} its total mass as \mbox{${\cal M}
= \frac{1}{6} \pi \rho \langle s^3 \rangle \, {\cal N}$}, where ${\cal N}$ is
the{\vspace{-0.09cm}} number of dust particles in the tail, \mbox{$\langle s^2
\rangle^{\frac{1}{2}}$} their mean diameter{\vspace{-0.05cm}} determined by the
cross-sectional distribution, \mbox{$\langle s^3 \rangle^{\frac{1}{3}}$} their
mean diameter determined by the mass distribution, and $\rho$ their bulk density.
By eliminating ${\cal N}$ from the expressions for $X$ and ${\cal M}$ one obtains
\begin{equation}
X = \frac{3{\cal M}}{2 \rho} \frac{\langle s^2 \rangle}{\langle s^3 \rangle}.
\end{equation}
Inserting first from Equation (6) into (8) and then from (8) into (7), one
obtains the tail's apparent magnitude. 

Assuming next that the sublimation{\vspace{-0.04cm}} of water ice dominates,
I take \mbox{$\dot{Z}_0 = 3 \times \! 10^{17}$\,molecules cm$^{-2}$ s$^{-1}$}
and \mbox{$\mu_{\rm mol} = 3 \times \! 10^{-23}$\,g}.  Adopting further for
the sungrazer Ikeya-Seki \mbox{$R = 6$ km}, \mbox{$f = 0.2$}, and \mbox{$\kappa
= 3$}, I get for the mass production rate of dust at 1~AU from the Sun 
\begin{equation}
\dot{\cal M}_0 = 2.4 \times \! 10^7\,{\rm g}\:{\rm s}^{-1} = 2.1 \times \!
 10^{12}\,{\rm g}\:{\rm day}^{-1}. 
\end{equation}
On October 29.60 UT, the time of Weinberg \& Beeson's (1976) observation, the
comet was 0.447~AU from the Sun and 1.048~AU from Earth.  The mass of dust in
the tail then equaled (in grams)
\begin{equation}
{\cal M} = 2.0 \times \! 10^{15} (1.44 - U), 
\end{equation}
where $U$ depends on the earliest post-perihelion dust ejecta surviving
sublimation in the tail.  The value of $U$ varies from zero at perihelion,
\mbox{$r_{\rm beg} = q = 1.67$ {\Rsun}\,}, to 0.31 for 1.84~{\Rsun}\,, to 0.42
for 2.0~{\Rsun}\,, and to 0.61 for 2.5~{\Rsun}\,, so that the expression in the
parenthesis of Equation~(10) is near unity.  Adopting now in Equation~(8)
a grain density of 2.5~g~cm$^{-3}$ and a ratio \mbox{$\langle s^3 \rangle /
\langle s^2 \rangle = 4 \times \! 10^{-4}$\,cm}, describing an $s^{-2.8}$
cumulative distribution with the limits of 0.1~$\mu$m and 1~cm, I obtain
\begin{equation}
X = 3.0 \times \! 10^{8}\,{\rm km}^2. 
\end{equation}
Moving from the cross section to the apparent magnitude, strictly one should
consider the implications of the tail as an extended feature.  In this very
approximate exercise, I note that the linear length of the nearly 20$^\circ$
long tail was about 0.54~AU, extending from the comet's head at 0.45~AU from
the Sun to nearly 1~AU at the outer end.  The tail's median point was near
0.62~AU from the Sun and 1.16~AU from Earth and the phase angle was 54$^\circ$.
Assuming a geometric albedo of 0.04 and using the phase law for dust-rich
comets by Marcus (2007), this generic model gives for the total apparent
visual magnitude of the imaged tail of Ikeya-Seki
\begin{equation}
H = -2.1 \: {\rm mag}, 
\end{equation}
a value that agrees with the estimate based on the surface-brightness
measurements to within a few tenths of a magnitude, suggesting the plausibility
of the procedure and chosen parametric values.  Most importantly, both
approaches show that at the time the tail was brighter than the head by a
factor of nearly 100.   

In the light of these results, Hasegawa \& Nakano's (2001) assumption that
after perihelion the sungrazers were observed with the naked eye until the
brightness of the head declined to apparent magnitude~5 is greatly off the
mark.  Instead, evidence shows that a sungrazer's {\small \bf tail is last
sighted with the unaided eye when the object's head has faded to about
magnitude~7}, implying that both the absolute magnitudes and the magnitudes
at the time of first sighting in Table~1 should be fainter by 2~mag or so.
This means that with the exception of the comet of 1663 and, to a lesser
degree, 1232, the tabulated comets were intrinsically rather faint, most
of them fainter than comet Ikeya-Seki, although not as faint as comet
Lovejoy.  The brightness of the comets from Table~1 is further addressed
in Section~5.7.

\subsection{Apparent Magnitude of Historical Comets\\at First Sighting with the
 Naked Eye} 
How bright on the average were historical comets when first detected with the
naked eye?  The most appropriate object to use in addressing this issue is
surely comet 1P, because before E.\,Halley's 1705 discovery of its periodicity,
the comet had been detected with the naked eye as a new object at every return
to perihelion back to 12~BC (Ho 1962, Kiang 1972) and, including the probable
Babylonian accounts in 87~BC and 164~BC (Stephenson et al.\ 1985), apparently
all the way back to 240~BC.  The solution to the problem of naked-eye discovery
was in the past sought on the assumptions that (i)~Halley's light curve does
not vary from one apparition to the next, as suggested first in his investigation
by Holetschek (1896); and (ii)~the brightness variations with heliocentric
distance are satisfactorily represented by the comet's light curve at the
1910 apparition.

The first assumption is probably valid, at least approximately, even though
Ferrin \& Gil (1988) argued that Halley's comet had been fading at a rate of
0.055 magnitude per revolution.  Over two millennia, this rate would accumulate
to a total of nearly 1.5~magnitude, an effect that is not obvious from the
available data.  Neither is this rate of decreasing activity supported by
the invariability of the nongravitational perturbations of the comet's orbital
motion established by Yeomans \& Kiang (1981), even though the radial component,
the driver of such secular variations, was rather poorly determined.  In any case,
I do not see compelling evidence for a material long-term fading or a rationale
for its search in the limited data set available.  Unlike the first assumption,
the second one is demonstrably incorrect and leads to grossly inaccurate
results, which are revised below.

Let a comet's apparent brightness vary with geocentric distance $\Delta$
and heliocentric distance $r$ according to a $\Delta^{-2} r^{-n}$ law, so
that its apparent magnitude $H(\Delta,r)$ is given by
\begin{equation}
H(\Delta, r) = H_0 + 5 \log \Delta + 2.5 n \log r,  
\end{equation}
where $H_0$ is the absolute magnitude (at \mbox{$\Delta = r = 1$ AU}).  If
the light curve is asymmetric relative to perihelion, one should distinguish
between the preperihelion parameters, $H_0^-$, $n^-$, and post-perihelion
parameters, $H_0^+$, $n^+$, because they may carry different physical
meanings.  Once known, they allow the calculation of the comet's apparent
magnitude at first sighting, $H_{\rm first}$, from the respective distances
$\Delta_{\rm first}$ and $r_{\rm first}$.

Broughton (1979) employed this approach by plotting the heliocentric distance
against the geocentric distance at first sighting for a select subset of 19
returns of Halley's comet and determined that the best fit from seven entries
referring to preperihelion discoveries was reached for a power \mbox{$n^- = 4.6$},
which yielded \mbox{$H_{\rm first} \!-\! H_0^- = -2.2$}.  He argued that when
he assumed \mbox{$H_{\rm first} = 3.5$}, the absolute magnitude came out to be
\mbox{$H_0^- = 5.7$}, very close to the value obtained by Ernst (1911) in his
1910 preperihelion light-curve formula
\begin{equation}
H(\Delta,r) = 5.8 + 5 \log \Delta + 13.5 \log r,  
\end{equation}
even though the parameter $n^-$ by Ernst was higher than Broughton's value by 0.8.

\begin{table*}
\vspace{-4.2cm}
\hspace{0.5cm}
\centerline{
\scalebox{1}{
\includegraphics{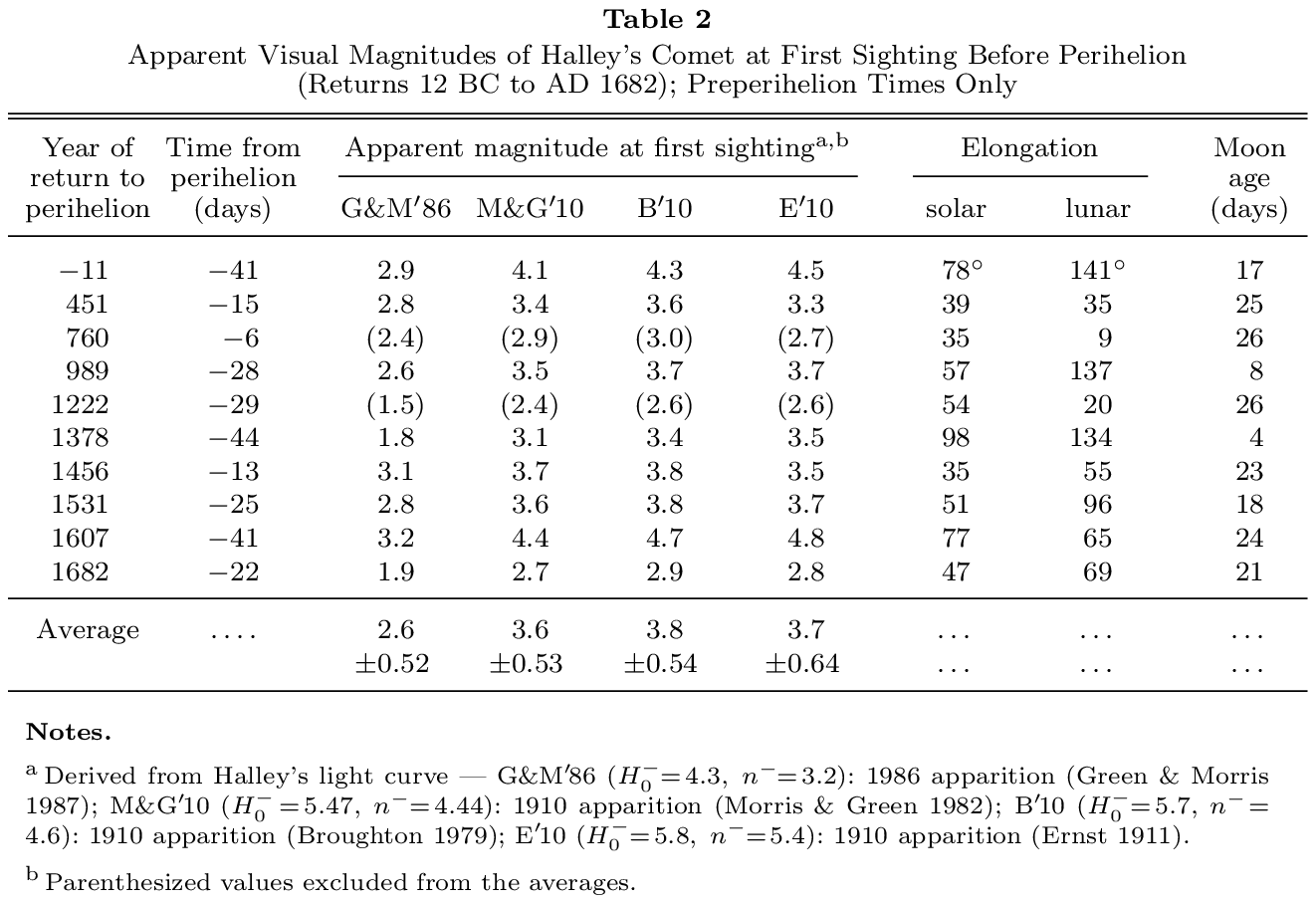}}}
\vspace{-15.6cm}
\end{table*}

The present study is limited to examining the apparent magnitudes of Halley's
comet at the first naked-eye sighting before perihelion and is terminated by
the 1682 apparition, the last one before the predicted returns.  This condition
leaves a total of 10~entries of the 19 listed by Broughton (1979).  From the
$\Delta$ and $r$ data that he provided, I calculated the apparent magnitudes
at first sighting based on four different laws.  Besides the light curve by
Broughton (\mbox{$H_0^- \!= 5.7$}, \mbox{$n^- \!= 4.6$}) and the preperihelion
light curve by Ernst [Equation~(14)], I also include another 1910 light curve,
derived by Morris \& Green (1982) ({\mbox{$H_0^- \!= 5.47$}, \mbox{$n^- \!=
4.44$}), used by Stephenson et al.\ (1985), and, most importantly,{\vspace{-0.02cm}}
the {\it 1986\/} light curve by Green \& Morris (1987) (\mbox{$H_0^- \!= 4.3$},
\mbox{$n^- \!= 3.2$}).

The results, summarized in Table 2, are striking in two respects.  One, even
after removing the two entries when the Moon was less than 30$^\circ$ from
the comet (suggesting that the comet was brighter than at routine discovery;
parenthesized in the table), the average naked-eye brightness at first sighting
found from the superior 1986 light curve, is apparent magnitude of 2.6, at
least 1~mag brighter than the previously used numbers based on the 1910 light
curves.  And two, the choice among the 1910 light-curve laws had almost no
effect on the resulting magnitude.  In view of what follows I also may add
that in none of the listed apparitions was the comet discovered at a solar
elongation of less than 35$^\circ$.

The disparity between the 1986- and 1910-based results is caused by the comet's
systematically underestimated total brightness in 1910.  Also, the higher values
of $n$ indicate that this bias was increasing with heliocentric distance.  The
new value of the apparent magnitude at the first naked-eye sighting means of
course that detection with the naked eye requires a comet to be brighter than
previously thought.

Yau et al.\ (1994) examined historical records on 109P/Swift-Tuttle, another
bright periodic comet, in the manner as Broughton did in the case of 1P/Halley.
Unfortunately, only a very few apparitions were found and the magnitudes at first
sighting varied by nearly 3~mag, thus offering no statistically meaningful
information to complement the findings from the Halley data.

Hasegawa \& Nakano's (2001) brightness overestimate at the last sighting and
the underestimated average brightness needed to detect a comet with the naked eye
conspired to make {\it preperihelion\/} discovery of the seven Kreutz candidates
in Table~1 extremely difficult, thereby explaining the second of the two striking
effects commented on in Section~5.1.  The first one is linked, as already noted,
to the orbital orientation of the Kreutz sungrazers, which gradually become deep
southern-hemisphere objects as time advances from mid-October to mid-December.
This is clearly apparent from Marsden's (1967) perennial ephemeris.

\subsection{Kreutz Sungrazer C/1965 S1 (Ikeya-Seki)}  
I already noted that comet Ikeya-Seki was a great spectacle thanks in part to the
favorable timing of its arrival, with the perihelion point reached in the second
half of October.  Why then all potential historical Kreutz sungrazers coming to
perihelion at about the same time of the year (Table~1) were not observed before
perihelion?  This issue is addressed by examining at what point in its orbit would
comet Ikeya-Seki have been discovered had it arrived before the era of telescopic
comet hunters.

\begin{table}[ht]
\vspace{-4.12cm}
\hspace{5.15cm}
\centerline{
\scalebox{0.99}{
\includegraphics{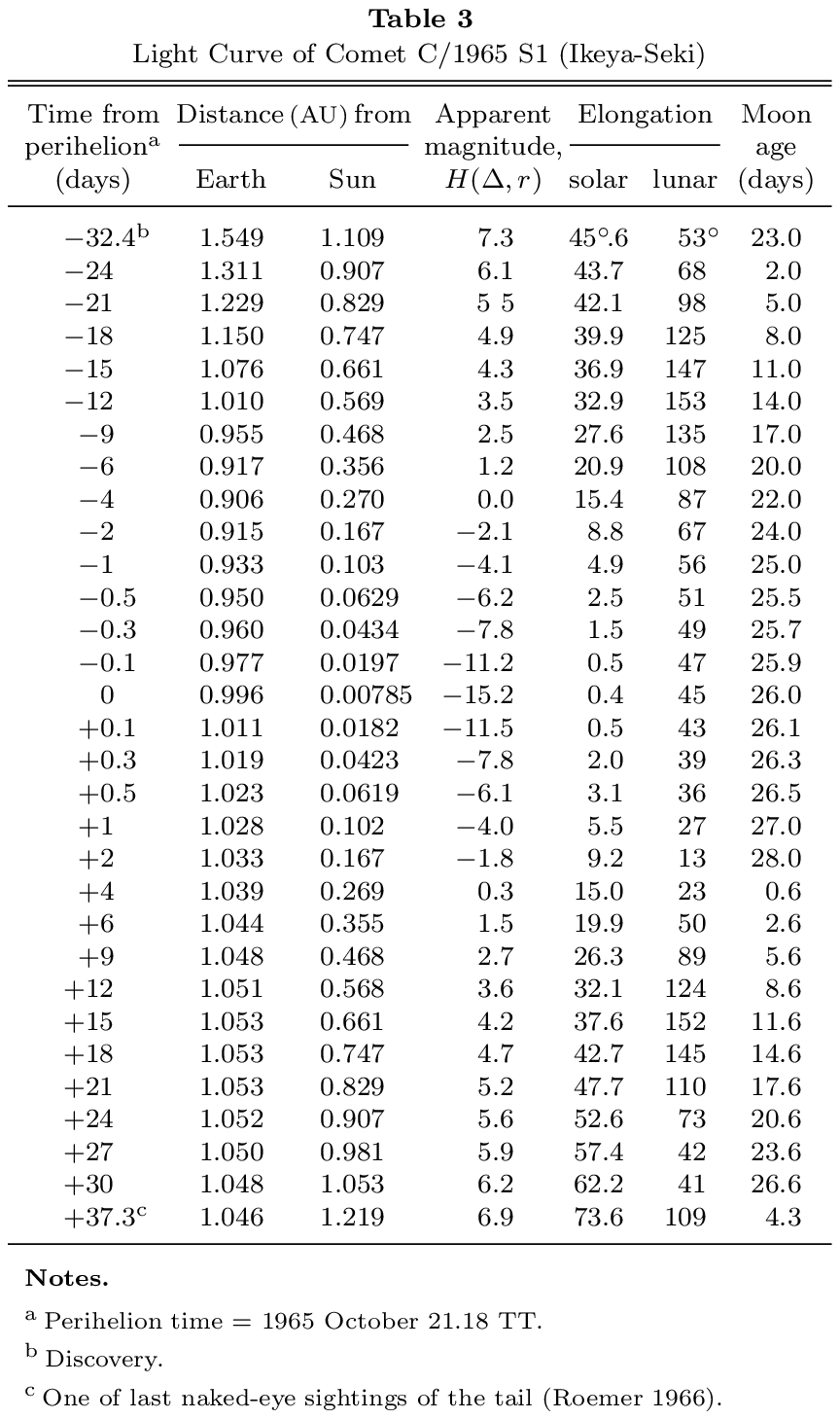}}}
\vspace{-10.3cm}
\end{table}

Interestingly, comet Ikeya-Seki was discovered (almost simultaneously by the two
observers) only about 50~hours after it had reached a maximum solar elongation since
mid-May 1965.  Its light curve was symmetric relative to perihelion{\vspace{-0.02cm}}
(Sekanina 2002), closely following a law given by Equation\,(13) with
\mbox{$H_0^- \!=\! H_0^+ \!= 5.9$} and \mbox{$n^- \!=\! n^+ \!= 4.0$}.  As a
function of time from perihelion, \mbox{$t \!-\! t_\pi$}, the comet's apparent
magnitude and solar and lunar elongations are presented in Table~3.  One finds
somewhat unexpectedly that, upon its approach, the comet brightened to the
average naked-eye first-sighting magnitude of 2.6 (Section~5.3) only nine days
before perihelion at a solar elongation of 28$^\circ$!  In order to be discovered
with the naked eye as a morning object before or around the beginning of the
astronomical twilight (with the Sun 18$^\circ$ below the horizon), it would have
to have been detected at an elevation of less than 10$^\circ$ above the horizon
under conditions of severe atmospheric extinction.

The comprehensive algorithm by Schaefer (1993, 1998) for the limiting magnitude
of a star to be seen with the naked eye was employed to address the issue of
discovering Ikeya-Seki in the pre-telescopic era.  I considered the most favorable
circumstances, namely, the comet's minute, nearly-stellar head (to approximately
fit the conditions to which the algorithm applies), its solar elongation equaling
the difference between the elevations of the comet and the Sun (i.e., the comet and
the Sun located at the same azimuth).  Obviously, at the given solar elongation of
28$^\circ$ the observing conditions, as a function of the comet's elevation above
the horizon, deteriorated from an optimum position in either direction, because of
either excessive atmospheric extinction or advancing twilight.  Application of
Schaefer's theory suggests that the optimum conditions occurred when the comet
reached an elevation of 9$^\circ$; at that time the naked-eye detection limit
at its location was magnitude~4.1, or 1.5~magnitude fainter than the discovery
magnitude~2.6.  The naked-eye detection limit equaled magnitude~2.6 when the
comet's elevation was 4$^\circ\!$.6 (because of the extinction) or 13$^\circ\!$.6
(because of advancing twilight); thus,{\vspace{-0.04cm}} the time available to the
potential discoverer was at best about $\frac{1}{2}$ hour, more probably shorter.
As for the tail at this time, it was relatively short and not bright enough to
significantly improve the likelihood of detection.  I conclude that the chance
of naked-eye discovery of comet Ikeya-Seki in this narrow window of time about
nine days preperihelion, while not nil, appears to have been relatively poor.

\begin{figure}[t]
\vspace{-7.15cm}
\hspace{2.52cm}
\centerline{
\scalebox{0.70}{
\includegraphics{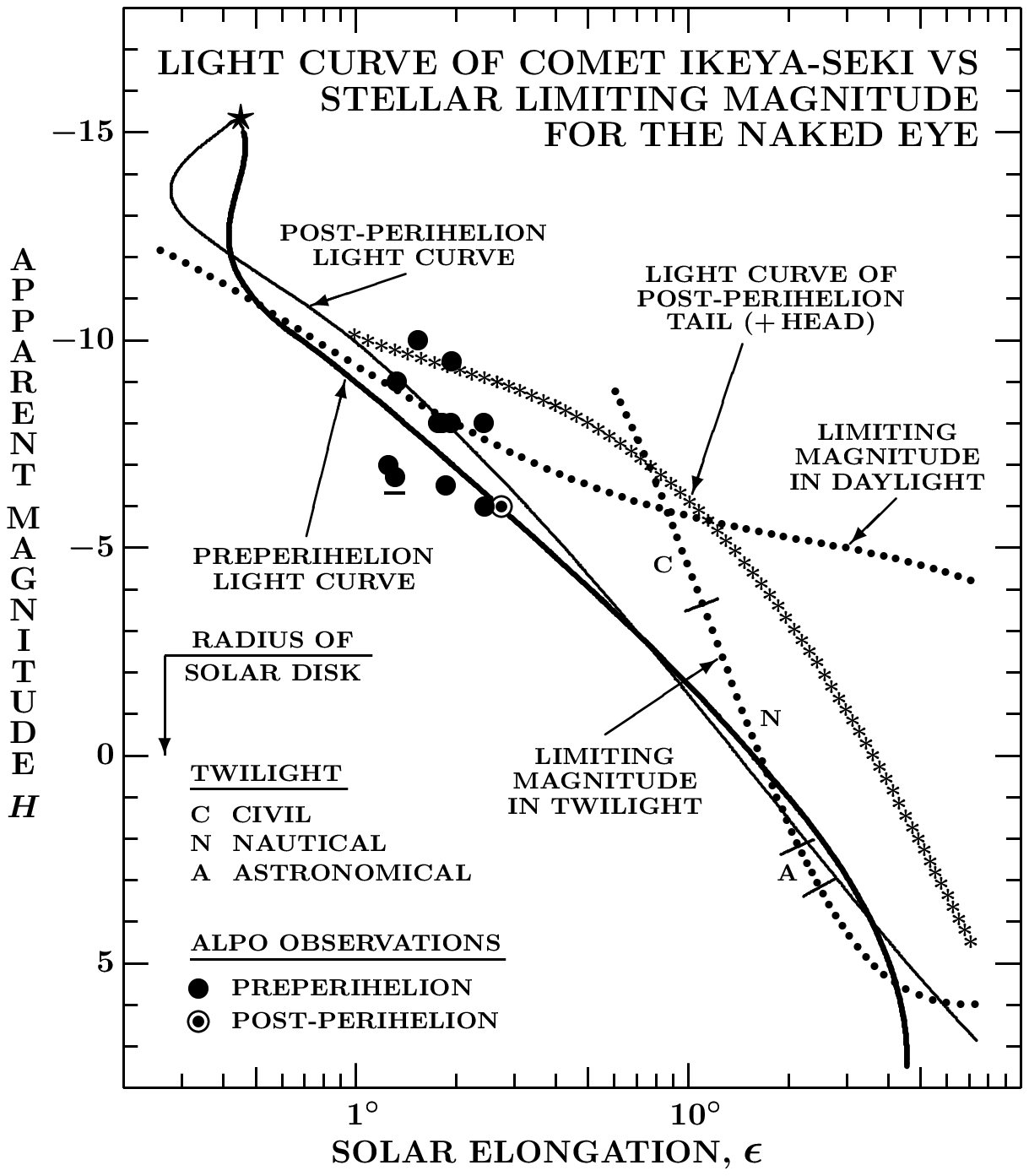}}}
\vspace{-3.95cm}
\caption{Comparison of the visual light curve of Comet Ikeya-Seki with Schaefer's
naked-eye limiting magnitude for stellar objects in a plot of the apparent
visual magnitude $H$ as a function of the solar elongation of the comet's head.
Next to its preperihelion and post-perihelion branches of the light curve (the
perihelion point being marked by the star), the figure also presents an estimated
light curve of the comet's post-perihelion tail (+\,head) and magnitude estimates
by several ALPO members; the observation by Meisel, who used a bulb photometer,
is underscored.  The limiting magnitude is displayed for daylight conditions (both
the Sun and comet assumed at an elevation of 45$^\circ$), as well as in twilight
(astronomical. nautical, and civil) and at nighttime.  The twilight curve fits an
optimum case of the Sun and comet located at the same azimuth, as described in the
text.{\vspace{0.55cm}}}
\end{figure}

Next I examine the possibility of the comet's naked-eye discovery in broad daylight,
whether before or after perihelion, or later after perihelion.  In Figure~1 I plot
the apparent visual magnitude of the head of comet Ikeya-Seki calculated from the
light curve (Sekanina 2002), both preperihelion and post-perihelion, as a function
of solar elongation of the comet's head.  The figure also shows (i)~the naked-eye
magnitude estimates by members of the Association of Lunar and Planetary Observers
(ALPO), summarized by Milon (1966); and (ii)~an estimated apparent magnitude of
the comet's dominant post-perihelion tail, based on the approximate results of
Section~5.2.  These brightness data are compared with the elongation dependent
limiting magnitude for stellar objects detected with the naked eye, as derived
from Schaefer's theory for daytime observation in the case of the Sun's and comet's
elevations of 45$^\circ$, as well as for twilight and nighttime observation.  The
daytime limiting brightness declines with increasing elevation at a rate between
0.01 and 0.03~magnitude per degree.

Figure 1 shows several interesting features.  An important point to make is that the
comet's preperihelion apparent brightness increased with decreasing solar elongation
less steeply than did the limiting brightness for the naked eye in twilight, but
more steeply than it did in broad daylight, so that while the comet's head was too
faint to be detected at 5--15$^\circ$ from the Sun, it was bright enough for
detection at smaller elongations.  This seemingly peculiar correlation supports
the existence of the ``sun-comets'' reported by Strom (2002) from his inspection
of historical Chinese sources and explains fairly common reports of daylight comets
(not only Kreutz sungrazers) in {\it close\/} proximity to the Sun.

Another notable feature of the figure is the distribution in the plot of daytime
naked-eye magnitude estimates of the comet by the ALPO observers.  The scatter
of some 4~mag shows the high degree of uncertainty involved in the estimates,
which were not used in constructing the light curve.  Yet, two effects are
worth noting:\ (i)~most estimates are decidedly not below the standard
light curve, thereby showing no evidence of the $r^{-4}$ law's flattening
out near perihelion (down to at least 0.04~AU from the Sun), unlike for
numerous other comets; and (ii)~some of the estimates are below Schaefer's
limiting magnitude curve, including the only one among the plotted data points
that was determined{\vspace{-0.04cm}} with some precision (it is underscored
in Figure~1).\footnote{This brightness determination, by D.\,Meisel, was
made with help of a simple bulb photometer, by comparing the brightness of
the comet's head with that of the Sun; this measurement is in good agreement
with the standard light curve.}  It is possible that the brighter estimates
included a modest contribution from the tail, which was widely reported to be up
to 3$^\circ$ in length between October~20.7 and 20.9~UT, that is \mbox{7--12}~hr
before perihelion.\footnote{Interestingly, Milon (1966) reported only a single
daytime detection of the comet and its tail with the naked eye the following
day, about 10~hours after perihelion.}

\begin{figure}[t]
\vspace{-7.85cm}
\hspace{2.95cm}
\centerline{
\scalebox{0.755}{
\includegraphics{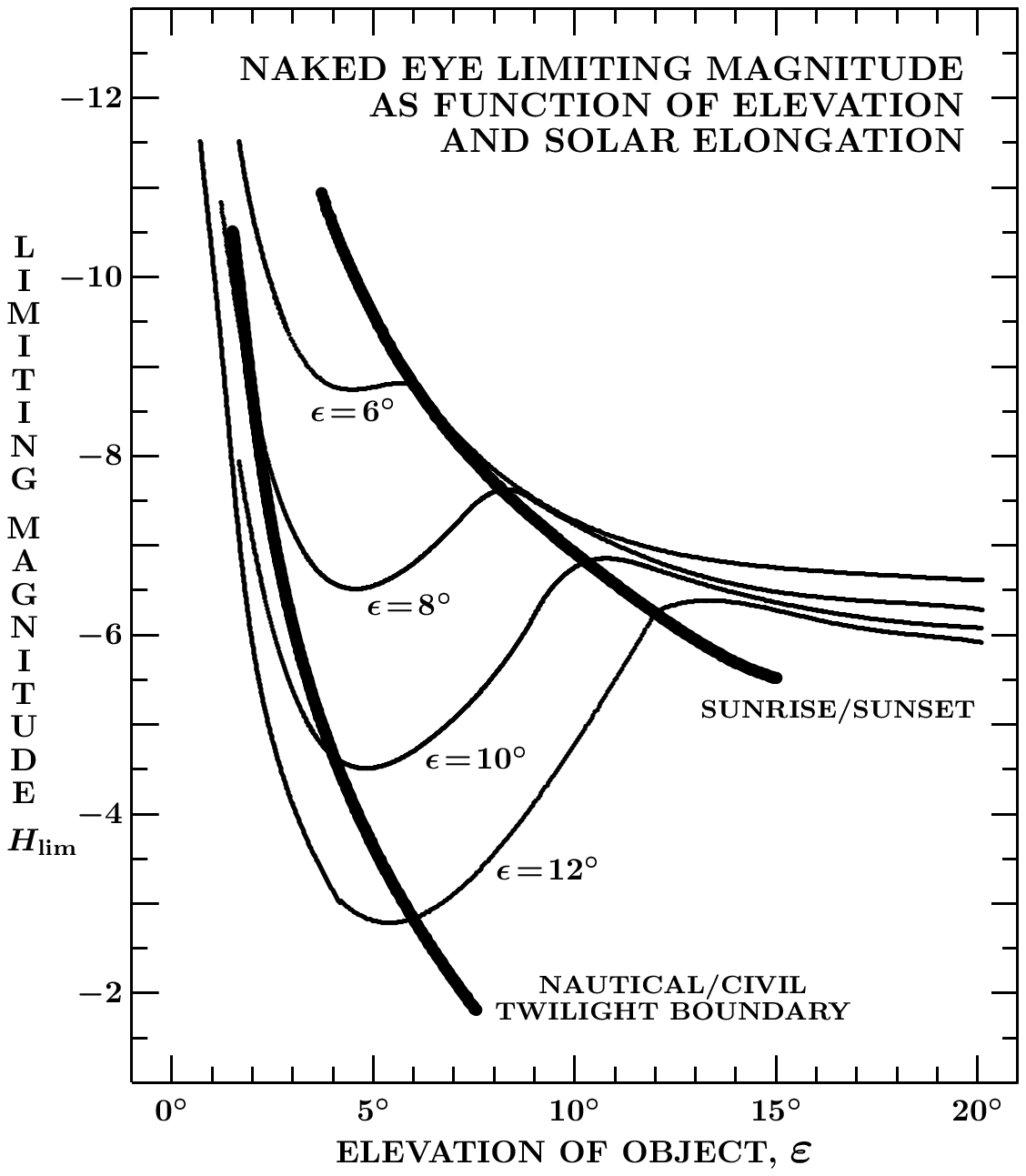}}}
\vspace{-5cm}
\caption{Schaefer's (1998) limiting magnitude of a stellar object for
the naked eye, $H_{\rm lim}$, as a function of the object's elevation,
$\cal E$, at the solar elongations, $\epsilon$, of 6$^\circ\!$,
8$^\circ\!$, 10$^\circ\!$, and 12$^\circ\!$.  The object and the Sun
are assumed to have the same azimuth, so that the elongation equals
the difference between their elevations.  The two heavy dotted curves
show sunrise/sunset and the boundary between the civil and nautical
twilight, respectively.  Note the considerable depth of the minimum
near elevation 5$^\circ$ for elongation 12$^\circ$, which contrasts
with the extremely shallow minimum at elevation 4$^\circ\!$.5 for
elongation 6$^\circ$.  Also note that the twilight minimum for
elongation 12$^\circ$ is substantially fainter than the daylight
limiting magnitude, while the opposite is true for elongation
6$^\circ$.{\vspace{0.6cm}}}
\end{figure}

An unusual feature in Figure 1 is the upward trend of the twilight branch
of the limiting-magnitude curve after crossing the daylight branch, indicating
that daylight conditions for discovering an object of a given brightness with
the naked eye may be superior to the twilight conditions.  To further examine
this finding, I plot in Figure~2 a limiting magnitude for the naked eye at
four solar elongations between 6$^\circ$ and 12$^\circ$ as a function of the
object's elevation in the case when the object and the Sun are located
at the same azimuth, the difference between their elevations then
equaling the elongation.  Figure~2 shows enormous changes of the
limiting magnitude in the critical range of elongations:\ while much fainter
objects could be detected in twilight than daylight at a solar elongation of
12$^\circ$ (displaying a deep minimum at an elevation of 5$^\circ\!$.5), the
opposite is true at 6$^\circ$ (where a very shallow minimum is exhibited at
4$^\circ\!$.5), with a transition region in between.  This is the
effect that causes the twilight curve crossing the daylight curve near the
elongation of 8--9$^\circ$ in Figure~1.

If comet Ikeya-Seki were to be discovered with the naked eye in broad daylight
{\it prior to perihelion\/}, this would most probably occur in immediate proximity
of the Sun's disk, as only then did its brightness exceed, {\it very briefly\/},
the detection limit by a sizable margin.  However, if the comet even then remained
undetected, a {\it rapidly developing, spectacular dust tail was about to blaze
into view within days after perihelion\/} (Section~5.2; the 
asterisked curve in Figure~1).\footnote{One has to keep in mind that thanks to
its large dimensions, the tail extended to much greater solar elongations where
viewing conditions were more favorable.}  The reader will notice that this
``foolproof'' scenario of naked-eye discovery shortly after perihelion places comet
Ikeya-Seki squarely in the category of the historical Kreutz sungrazer candidates
listed in Table~1 (cf.\ also Table~5).  
%

\subsection{Kreutz Sungrazers C/1843 D1 and C/1882 R1} 
The observing conditions for the two giant members of the Kreutz system, C/1843~D1
and C/1882~R1, were relatively favorable.  A major physical difference between them
was that the nucleus of the latter became multiple after perihelion as a result of
extensive near-perihelion fragmentation, whereas the nucleus of the former remained
single.  This disparity was reflected dramatically in the post-perihelion visibility
and light curves of the two objects, as discussed below.

According to Encke (1843), anonymous reports of C/1843~D1 published in New
York newspapers mentioned that the comet was first detected on February~5 or
22~days before perihelion.  On the other hand, in his updated ``Directory of
Elements of Cometary Orbits'' Galle (1894) remarked that the first observations
of C/1882~R1 were made on September~1, 16~days before perihelion, by unknown
individuals from the Gulf of Guinea and from Cape of Good Hope; independently,
Kreutz (1888) noted that the Gulf sighting was made by the crew of an Italian
ship.  Both comets were discovered fortuitously, with no telescopic aid; the
findings from the historical discoveries of Halley's comet therefore apply.

Until very recently, there was very little information available on the
preperihelion light curve of either of the two comets.  The situation changed
for C/1882~R1 thanks to a comprehensive paper by Orchiston et al.\ (2020)
on the comet's observations in New Zealand.  The authors mention that the
nucleus of the comet was of magnitude~2 when observed from Wellington on the
morning of September~9 and from Dunedin before sunrise on September~11, and
almost equal to a star of magnitude~1 when sighted from Gisborne around 5~AM on
September~11; the solar elongation of the comet's head was then only 21$^\circ$.
Orchiston et al.\ add that an article in an Auckland newspaper stated that on
September~12 the comet (not its nucleus) was as bright as Jupiter, which might
be an exaggeration.  The New Zealand preperihelion brightness data are
complemented by W.~H.~Finlay's estimate of magnitude~3 (allegedly the nucleus)
on September~7.2~UT (Gill 1882) and Elkin's (1882) magnitude estimate of~3--4
on September~8.2~UT.  In addition, Eddie (1883) stated that the comet's
nucleus equaled Jupiter in brilliancy on September~13.1~UT and again on
September~15.1~UT,\footnote{This could not both be correct, because it is
inconceivable that the comet did not brighten over a period of 48~hours,
during which its heliocentric distance dropped from 0.3~AU to 0.2~AU on its way
to perihelion.} and that it was brighter than Venus around September~17.3~UT,
in full daylight and shortly before reaching perihelion.  A second-hand report
published by Gould (1883a) that described the comet ``as being as bright as
Venus'' before sunrise on September~5, nearly two weeks before perihelion, is
entirely outside the realm of possibility.

Most of the assembled preperihelion magnitude estimates suggest that the head
of the Great September Comet of 1882 brightened as the inverse 4th power of
heliocentric distance, just as did Ikeya-Seki.  The $r^{-4}$ law indicates
that at the time of first sighting on September~1.2~UT the difference
between the apparent and absolute brightness was \mbox{$H_{\rm first}
\!-\! H_0^- = -0.8$ mag}.  The comet was then 30$^\circ$ from the Sun, an
elongation only slightly greater than in the hypothetical case considered
above for Ikeya-Seki.  Yet, since the two independent and essentially
simultaneous (although not first-hand) reports exist from widely separated
locations, there are no grounds to dispute the veracity of the sightings.
Besides, additional independent accounts exist from the subsequent days.
The comet apparently did display a tail.  What could be questioned is the
apparent magnitude, but the Halley-based value of 2.6~mag (Section~5.3)
provides a conservative estimate, resulting in the absolute magnitude of
\mbox{$H_0^- = 3.4$ mag}, or 2.5~magnitudes brighter than Ikeya-Seki.  It is
2~magnitudes brighter than the absolute magnitude derived from the reported
brightness estimates of the nuclear condensation mentioned above.

To address the controversy about whether the Great March Comet of 1843 was
intrinsically brighter or fainter than the Great September Comet of 1882, it is
appropriate to compare both objects in terms of the first naked-eye sighting.  For
the 1843 comet the application of the $r^{-4}$ law gives for the discovery date
of February~5.9~UT a difference of \mbox{$H_{\rm first} \!-\! H_0^- = -0.9$ mag}.
The comet was 54$^\circ$ from the Sun, but there was some interference from the
Moon, which was 6~days old and less than 50$^\circ$ from the comet.  The adopted
Halley-based apparent magnitude of 2.6 implies the absolute magnitude of
\mbox{$H_0^- = 3.5$ mag}.  Given the uncertainties involved, I tentatively
conclude that the two comets were intrinsically of equal brightness along the
preperihelion leg of the orbit.  If the inverse fourth-power law should apply
all the way to perihelion (as Figure~1 suggests for Ikeya-Seki), the 1882
sungrazer would have reached a peak apparent magnitude of $-$17.7, the 1843
sungrazer $-$19.1 on account of its slightly smaller perihelion distance.

The absolute brightness and the light curve inferred for the 1843 sungrazer
can crudely be tested on three probable sightings reported from the period of
time shortly before perihelion (Herrick 1843b, Peirce 1844).  The first of
the three allegedly occurred on February~19.9~UT from Bermuda; the comet is
calculated to have been of apparent magnitude $-$0.3 at a solar elongation of
26$^\circ$.  For the second sighting, on February~24.0~UT from Philadelphia,
Penn., the apparent magnitude is derived as $-$2.3 at an elongation of
16$^\circ$.  And for the third, on February~26.9~UT from Puerto Rico, the
predicted apparent magnitude is $-$6.4 at an elongation of 6$^\circ$.  The
comet was an evening object and the first two sightings appear to have probably
taken place during twilight.  For the solar elongation of 26$^\circ$ the Schaefer
theory predicts in the best possible case the naked eye limiting magnitudes of
+2.3 at elevation 15$^\circ$ (nautical twilight), +2.9 at elevation 10$^\circ$
(astronomical twilight), and +2.7 at elevation 5$^\circ$.  For the solar
elongation of 16$^\circ$ the limiting magnitudes in the best case are $-$2.1
at elevation 12$^\circ$ (civil twilight), $-$0.2 at elevation 8$^\circ$
(nautical twilight), and $-$1.4 at elevation 4$^\circ$ (beginning of astronomical
twilight).  The Puerto Rico sighting appears to have been made most probably around
midday; the predicted limiting magnitude is $-$6.1 at an elevation 60$^\circ$
and $-$6.6 at 30$^\circ$.  The comparisons suggest that the Bermuda sighting
should have been very easy, the Philadelphia sighting fairly easy, while the Puerto
Rican sighting was difficult but still doable.  One of course cannot rule out
that the comet was in fact intrinsically brighter than adopted.

As already hinted, the 1843 and 1882 sungrazers differed dramatically in their
post-perihelion behavior.  I determined a light curve of the 1882 comet from a number
of visual estimates between 0.6~AU and 4.4~AU and found \mbox{$H_0^+ \!=\! -0.2$ mag}
and \mbox{$n^+ \!= 3.3$} (Sekanina 2002), much flatter than the comet's preperihelion
light curve.  The sungrazer was observed for the last time on 1883 June~1, or 257~days
after perihelion, with a 28-cm equatorial of the C\'ordoba Observatory (Gould 1883b).
It was then less than 40$^\circ$ from the Sun and its total apparent magnitude
calculated from the above photometric parameters was 8.7; the nuclear condensation
measured for astrometry was 2--3~mag fainter.

By contrast, the 1843 sungrazer was fading after perihelion much more rapidly.  It
was observed for the last time on 1843 April~19, 51~days after perihelion, with a
15-cm refractor of the Cape Observatory (Maclear 1851).  A tentative post-perihelion
light curve, based primarily on Warner's (1980) published extracts from C.\,Piazzi
Smyth's journal of his Cape observations and supplemented by isolated brightness
estimates reported by Herrick (1843a), Kendall (1843), and Simms (1845), suggests
that the brightness of the comet's condensation was declining according to an
$r^{-4}$ or a slightly steeper law.  If one accepts that the power $n^+$ correlates
with the extent to which the nucleus fragmented at perihelion, attaining a value the
lower the greater the extent, the 1882 comet (six major fragments and \mbox{$n^+ \!=
3.3$}) and Ikeya-Seki (two major fragments and \mbox{$n^+ \!= 4.0$}) imply that
the apparently nonfragmenting sungrazer of 1843 should indeed have had \mbox{$n^+ \!>
4$}.  This is consistent with the well-known fact that sungrazers normally fade quite
rapidly.

In Section 5.2 I noted that the apparent magnitude of three sungrazers --- the Great
Comet of 1882, Ikeya-Seki, and Lovejoy --- at the time of last naked-eye sighting was
always close to \mbox{$H_{\rm last} \!= 7$ mag}.  According to Kronk (2004), the
Great Comet of 1843 was last time seen with the naked eye by F.\,W.\,L.\,Leichhardt
on 1843 April~11, 43~days after perihelion.  With $n^+$ in the range \mbox{$4.0 \leq
n^+ \! \leq 4.2$} one{\vspace{-0.05cm}} finds \mbox{$H_{\rm last} \!-\! H_0^+ = +2.4$
mag} and therefore \mbox{$H_0^+ = 4.6$ mag}; the post-perihelion absolute magnitude
is nominally 1.1~mag fainter than the preperihelion absolute magnitude.  In order
to connect with the preperihelion light curve at perihelion, it is required that
\mbox{$n^+ \!= 4.2$}, in line with the condition based on the last naked-eye detection.

\subsection{Probable Kreutz Sungrazer X/1106 C1} 
Based on the constraints to the apparent magnitudes at the first and last naked-eye
sightings, a consistent picture emerges on the relationship between the two giant
sungrazers, the Great Comets of 1843 and 1882.  In terms of light-curve variations,
the two objects appear to have behaved similarly before perihelion, but very
differently after perihelion.  The most profound disparity was noted in the time
of last sighting of their tails with the naked eye (171~days after perihelion for
1882 vs 43~days for 1843) and their heads telescopically (257~days for 1882 vs
51~days for 1843).  As noted, these enormous differences were associated with
the fact that the 1882 sungrazer fragmented profusely at perihelion, whereas the
nucleus of the 1843 sungrazer remained single.

There are obvious evolutionary ramifications for the Kreutz system, as the 1843
and 1882 sungrazers are considered the largest surviving masses of Lobe~I and
Lobe~II, respectively.  The propensity for {\it near-perihelion\/} fragmentation
is distinctly higher for products of Lobe~II than Lobe~I.  Although obviously
implying different morphologies of the two lobes in the framework of the
contact-binary model of Paper~1, the discrepancy is of course model independent.

As a fundamental property, the susceptibility to near-perihelion fragmentation is
expected to be hereditary, and this is supported (i)~by the multiple nuclei of the
1882 sungrazer and Ikeya-Seki, both products of Lobe~II; and (ii)~by Marsden's
(1967) compelling demonstration of their common origin, that is, a breakup of
their parent.  Given the correlation between near-perihelion fragmentation
and the duration of post-perihelion visibility,\footnote{Because the duration
of visibility depends on the intrinsic brightness of the object, the correlation
with the extent of fragmentation can meaningfully be examined only among objects
of comparable pre-breakup intrinsic brightness.}
a parent sungrazer and its most massive fragment should be visible with the
naked eye over comparably long post-perihelion periods of time.  The continuing
controversy about the population membership of comet X/1106~C1 --- granted it was
a Kreutz sungrazer --- should be settled by the time of its last sighting:\ if
historical sources should indicate that it was visible over 40--60~days after
perihelion, it probably was a fragment of Lobe~I (a member of Population~I) and
the parent to C/1843~D1; if it was seen for longer than $\sim$150~days, it should
have been a fragment of Lobe~II (a member of Population~II) and the parent to
C/1882~R1 and C/1965~S1.

Historical records, summarized by Kronk (1999), suggest that comet X/1106~C1 was
discovered shortly after perihelion.  It was extensively observed in Europe, the
Middle East, as well as the Far East.  The comet was first seen from Belgium in
broad daylight on February~2, probably just a few days after perihelion.  Within
the following two weeks, the comet was detected in the evening sky in a number
of European countries (including England, Scotland, France, Holland, Italy, and
Germany), as well as in Palestine, Armenia, Japan, Korea, and China.  The tail
apparently grew rapidly in length, reaching 100$^\circ$ on February~9.  The comet
was under observation throughout February and into March.  Kronk (1999) writes that
European chroniclers differed greatly on how long was the comet seen, the shortest
period being about 15~days and the longest about 70~days, disappearing between
late February and the second half of April.  The Far Eastern sources indicated that
the comet remained visible for more than 30 days from the time of first sighting,
the estimate that was also adopted by Hasegawa \& Nakano (2001).  This suggests
that the comet was still seen after March~10.  Seargent (2009) mentions an Armenian
text, which implies that the comet disappeared on April~3.  And a Chinese source,
quoted by Hasegawa \& Nakano, noted that on April~9 a comet ``changed and
disappeared,'' a somewhat cryptic comment on an object that may or may not have
been identical with X/1106~C1.

\begin{table}
\vspace{-4.15cm}
\hspace{5.3cm}
\centerline{
\scalebox{1}{
\includegraphics{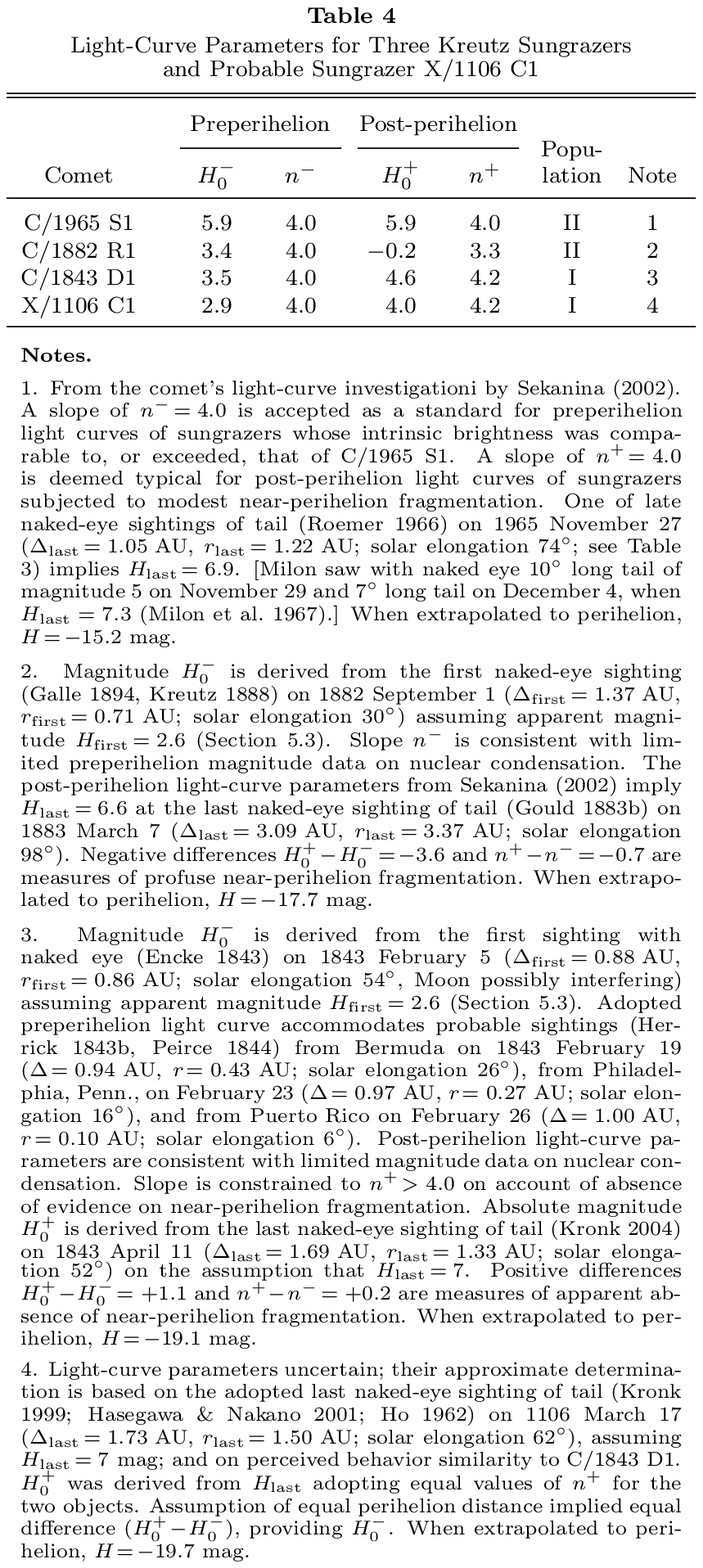}}}
\vspace{-5.7cm}
\end{table}

Adopting, rather conservatively, March 17 as the date of the final sighting (according
to {\it Historia Hierosolymitana\/}) and January~26 as the nominal perihelion date
(Hasegawa \& Nakano 2001), the comet was seen until 50~days after perihelion.  Even
in the extreme case it should have been seen for less than three months from perihelion.
This duration is comparable to, or moderately exceeding, the 43~days from perihelion
found above for C/1843~D1, but it is considerably shorter than the nearly six
months, the length of time established for C/1882~R1.  One can therefore rather
safely conclude that the post-perihelion light curve of X/1106~C1 resembled that of
C/1843~D1 and that the former comet was in all probability the parent of the latter
comet, both failing apparently to fragment profusely near perihelion.

Accepting that on 1106 March 17 the comet's head was of apparent magnitude 7 and
the brightness variations followed an $r^{-4}$ or $r^{-4.2}$ law, the post-perihelion
absolute magnitude of X/1106~C1 is computed to have amounted to \mbox{$H_0^+ = 4.0$}.
Scaling up the case of C/1843~D1, the predicted preperihelion value{\vspace{-0.05cm}}
for X/1106~C1 becomes \mbox{$H_0^- = 2.9$}.  The adopted light-curve parameters for
C/1965~S1, C/1882~R1, C/1843~D1, and X/1106~C1, discussed in Sections~5.4--5.6, are
summarized in Table~4, accompanied by extensive narrative.

In spite of uncertainties in the tabulated data, especially for X/1106~C1, the
differences{\vspace{-0.04cm}} in the physical meaning of the light-curve parameters,
$H_0^-$ and $H_0^+$ in particular, are very apparent.  The preperihelion brightening
of these major sungrazers tends to follow a power law with \mbox{$n^- \!= 4$} and
their degree of conspicuousness is described by $H_0^-$.\footnote{This characterization
does not apply to dwarf Kreutz sungrazers, which exhibit a very different behavior
(see Ye et al.\ 2014).}  This parameter allows one to conclude that C/1843~D1 and
C/1882~R1 were both of about equal dimensions when arriving at perihelion and X/1106~C1
may have been a little larger, while C/1965~S1 was a distinctly smaller fragment in
comparison.

On the other hand, $H_0^+$ and especially{\vspace{-0.05cm}} the differences
\mbox{$H_0^+ \!-\! H_0^-$} and \mbox{$n^+ \!-\! n^-$} describe the severity
of~the near-perihelion fragmentation effects on the comet's nucleus and its
activity:\ the more vigorous the effects are, the brighter the comet becomes
after perihelion and the more negative are the differences
\mbox{$H_0^+ \!-\! H_0^-$} and \mbox{$n^+ \!-\! n^-$}.{\vspace{-0.3cm}}
%

\subsection{Potential Kreutz Sungrazers from Table 1 Revisited} 
I already pointed out in Sections 5.2 and 5.3 that most of the Kreutz candidates in
Table~1 are objects intrinsically fainter than the sungrazer Ikeya-Seki.  Three of
the seven comets do however display, at least nominally, an anomaly of a kind that
deserves a comment.

In the reversed chronological order, the first of these is C/1695~U1.  It is the
only one of the seven for which the data nominally do not rule out the possibility
of preperihelion discovery.  However, closer inspection leads to a somewhat different
conclusion.  The Chinese source that Hasegawa \& Nakano (2001) refer to says that {\it
after\/} October~22, {\it ``at the fifth watch, a comet appeared.''\/}  Unfortunately,
the report does not offer any information either on how long after the date, or how
many watches were there per day.  If, for example, just a single one, the message
would effectively say that the comet was discovered on October~26 or 27, or a few days
{\it after\/} perihelion.  This guess can be compared with Kronk's (1999) account,
according to which the comet was discovered by P.\,Jacob, a French Jesuit living in
Brazil, on October~28.  Kronk writes that the comet was last seen by J.\,Bovet in
India on November~18, which is consistent with a Korean source, quoted by Hasegawa
\& Nakano, that the comet {\it ``began to disappear \ldots a few days''\/} after
November~7.  The comet almost certainly was not seen before perihelion and its
intrinsic brightness is estimated at \mbox{$H_0^+ \simeq 7$}, near an average among
the seven comets.

The time of last sighting of the comet of 1663 in Table~1 immediately raises a red
flag.  This is the only object among the seven whose post-perihelion behavior exhibits
striking similarity to that of C/1882~R1, to the extent that one could predict the
former comet's extensive near-perihelion fragmentation and Population~II membership.
With an{\vspace{-0.06cm}} estimated \mbox{$n^+ \!\simeq 3.5$} one finds
\mbox{$H_0^+ \!\simeq 2.5$} and \mbox{$H_0^- \!\simeq 5$}, much fainter than
Hasegawa \& Nakano{\nopagebreak} {\nopagebreak}(2001) suggest, but still above
average.{\pagebreak} 

\begin{table}[t]
\vspace{-4.17cm}
\hspace{5.2cm}
\centerline{
\scalebox{1}{
\includegraphics{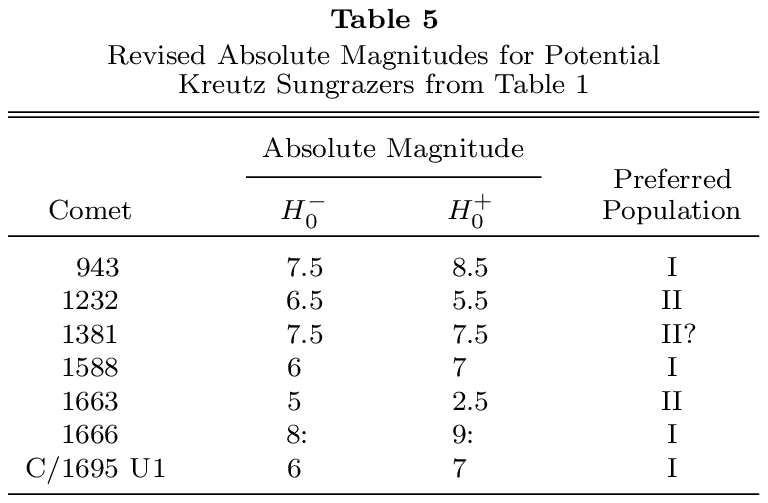}}}
\vspace{-20.1cm}
\end{table}

The period of post-perihelion visibility of the comet of 1232 also appears to have
been a little longer than average.  It probably was another member of Population~II
subjected to some fragmentation near perihelion, perhaps more extensive than Ikeya-Seki.

Table 5 summarizes the revised absolute magnitudes for the seven Kreutz candidates
from Table~1.  While the original magnitude estimates implied that four of the objects
were intrinsically brighter, two about equal, and only one fainter than Ikeya-Seki,
the revised numbers reverse that trend:\ four were fainter, two equal, and only one
brighter.  Also worth noting is that all seven were substantially fainter than
X/1106~C1, C/1843~D1, and C/1882~R1 before perihelion.

\section{Simulation of Ammianus' Daytime Comets by\\Contact-Binary Model
 of Kreutz System} 
The appearance of a comet has historically been one of the most popular portents of
an impending ominous event, such as a natural disaster or the death of an influential
man.  One certainly could find instances of fabricating, by an unscrupulous or sloppy
chronicler, the sudden appearance of a comet when none was available at the right time.

This possibility can safely be dismissed when it comes to Ammianus' remark on {\it
comets seen in broad daylight\/}, which does not fit a run-of-the-mill account of a
comet foretelling the death of an emperor:\ (i)~use of the plural is unprecedented and for a good reason, as
the appearance of two or more bright comets in the sky simultaneously (or nearly so) is an unrivaled
event that has to be seen to believe it; (ii)~the reference to broad daylight
%
%
further enhances the exceptional nature of the incident, making it highly
interesting astronomically but offering no obvious value to a portent;
(iii)~Ammianus' extensive elaboration on the nature of comets that follows the brief remark
has no place in a fabricated story; and (iv)~the timing is odd, several months
{\it after\/} the death of Ammianus' beloved Emperor Julian; a concocted omen would
have been moved by the narrator to an earlier time.

The first two points are so extraordinary that they clearly show the historian's
motivation by a real event.  Together with the third point, they strongly suggest
that Ammianus acted like anyone else would when greatly impressed by a singular
personal experience of spotting brilliant comets in full daylight, having witnessed
no such spectacle ever before.  The last point implies that the idea of making the
story up never entered his mind.

Although it is impossible to estimate how many comets Ammianus observed and over
how many days, the historian's remark does not rule out a chance of their nearly
simultaneous arrival at perihelion and his sighting of more than one comet near
the Sun at a time.  Considering the arguments presented in Section~5.4 in relation
to the limiting magnitude, it is unlikely that the comets were far from the Sun
in the sky.  In the following, I address the general features of the distribution
of daytime comets over the sky that Ammianus should have seen, if they were products
of a fragmented Kreutz progenitor and their motions and evolution were described by
the contact-binary model introduced in Paper~1.  Next to the primary breakup, the
procedure focuses on a sequence of secondary, cascading-fragmentation events that
followed over a limited period of time, all but one having taken place in the
aphelion region of the orbit, at heliocentric distances of about 160~AU.

The orbital distribution of the fragments, including the times of their arrival at
perihelion in AD~363, is determined by the separation velocity vectors.  The ranges of
angular elements, the longitudes of the ascending node in particular, are effects of the
out-of-plane (normal) component $V_{\rm N}$ of the separation velocity, while the range
of perihelion distances is an effect of its transverse component $V_{\rm T}$.  This
component also slightly influences the times of the fragments' first perihelion passage,
an effect that is declining with the fragmentation event advancing past aphelion.  The
radial component $V_{\rm R}$, which is indeterminate and assumed to be nil, could
potentially exert greater effects, at a rate of about 5~days per 0.1~m~s$^{-1}$.

If the separation velocity is of rotational nature, as it appears to be the case,
the direction of its vector becomes an issue of the spin-axis orientation of the
parent --- the progenitor sungrazer --- at the time of the initial fragmentation
event.  A near-zero radial separation velocity implies the progenitor spinning
along an axis pointing essentially at the Sun.

%
\subsection{Contact-Binary Model and Perihelion Times of\\Kreutz Sungrazers
 in AD 363}  

The case of a zero radial component of the separation velocity leads effectively
to the scenario of maximum orbital concentration of bright Kreutz sungrazers, the
progenitor's major fragments, in a swarm that arrived at the first perihelion in
late AD~363.  Evidence based on observations from both the ground and aboard the
Solar and Heliospheric Observatory (SOHO), as presented in Paper~1, shows the Kreutz
system made up of eight populations, two having branches.  The eight populations
are I, Ia, II, IIa, III, IIIa, IV, and Pre-I.  In addition, Population~Pe is a
side branch to Population~I and Population~IIa$^\ast$ a low-perihelion branch to
Population~IIa.  As described, the swarm under consideration consists of 10~primary
objects, for which the terms {\it sungrazer\/} and {\it fragment\/} are used
interchangeably in this scenario's in-depth examination that follows.

\begin{figure*}
\vspace{-12.02cm}
\hspace{-0.2cm}
\centerline{
\scalebox{0.875}{
\includegraphics{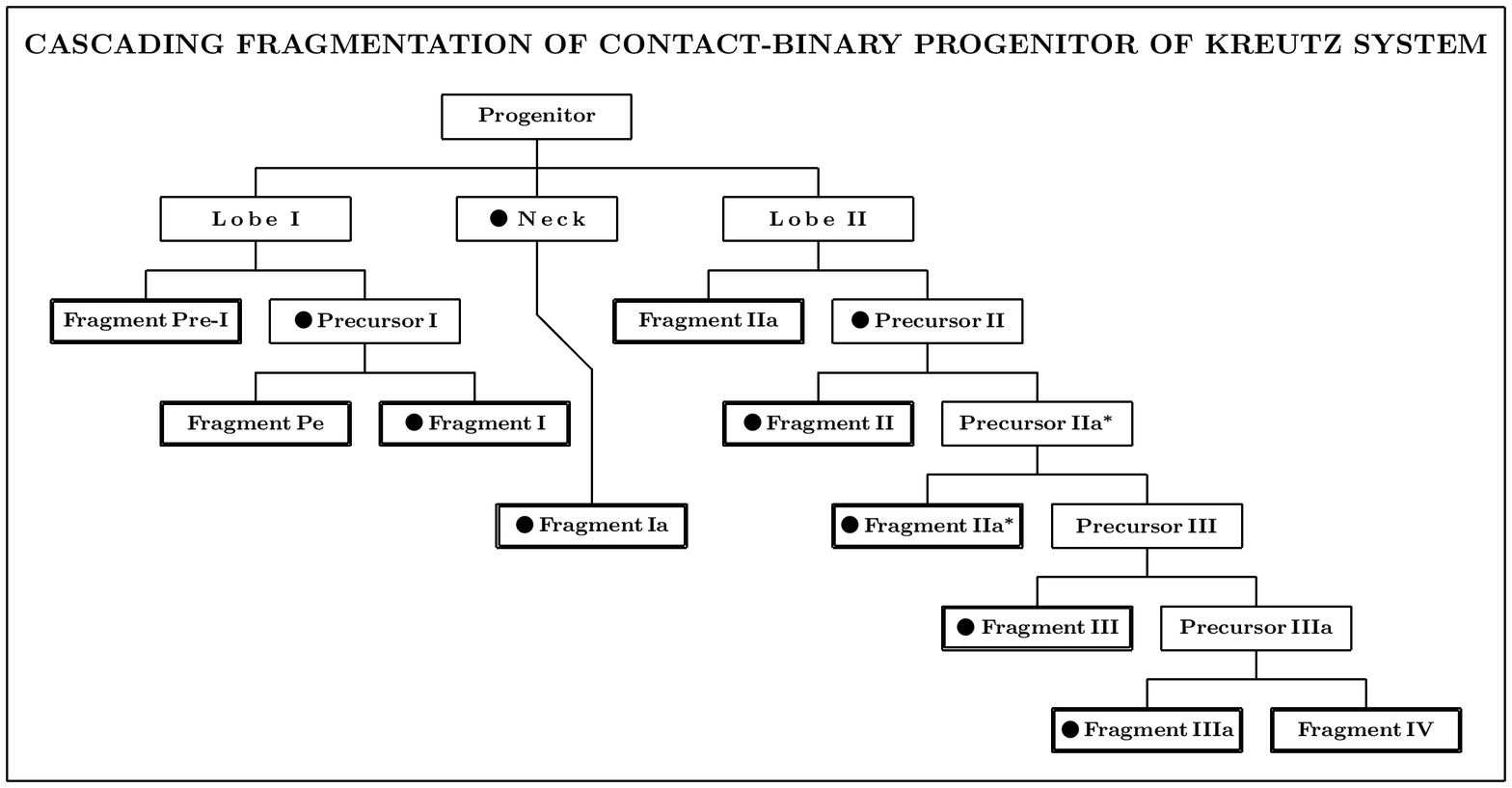}}}
\vspace{-4.45cm}
\caption{Chart of cascading fragmentation of the modeled contact-binary
progenitor of the Kreutz system.  The bullets indicate the sungrazers
that move in the same orbit as their immediate parent.  The sungrazers
in heavily framed boxes arrived at the first perihelion in AD~363 and
are used here to generate the swarm of daylight comets.  Their orbits
are approximated as follows: Fragment~I by C/1843~D1; Fragment~II by
C/1882~R1; Fragment~Pe by C/1963~R1; Fragment~IIa by C/1970~K1; and
Fragment~III by C/2011~W3.  The orbits for Fragments~Ia, Pre-I,
IIa{\boldmath $^\ast$}, IIIa, and IV were derived as described in
Paper~1.{\vspace{0.56cm}}}
\end{figure*}

The orbits of the populations were in Paper~1 approximated by the osculating
orbits of the bright Kreutz sungrazers when available; for the rest (such as IIIa
or IV) I used averages of the orbits of the population members among the SOHO dwarf
sungrazers imaged exclusively with the C2 coronagraph, their angular elements
corrected for major effects of a nongravitational acceleration.  Paper~1 also
described a sequence of breakup events from which the primary fragments were born
and in which they acquired momenta making them move in an array of fairly similar
orbits and arrive at perihelion in AD~363 at only slightly differing times.

The initial fragmentation of the contact-binary progenitor is assumed to have
resulted in three fragments, two lobes --- the earliest precursors of Populations~I
and II --- and the neck, the precursor of Population~Ia.  The model assumptions
were described in detail in Paper~1.  Here I only recall the requirement of perfect
symmetry to simplify the computations, which dictated that the lobes be released
in exactly opposite directions with separation velocities of the same magnitude.
The neck was let to continue moving in the progenitor's unperturbed orbit, arriving
at perihelion in AD~363 at the time the progenitor would have, had it not broken up.
To get into an orbit with a smaller perihelion distance, Lobe~I had to separate from
the rest of the progenitor with a transverse velocity $V_{\rm T}$ in the general
direction opposite the orbital-velocity vector.\footnote{It is noted that this and
other separation velocities differ slightly from their values in Paper~1; the reason
for these differences is a minor change of less than 10~AU in the aphelion distance of
the progenitor; in this paper it equals just under 163~AU, fitting the adopted orbital
period of 734.9~yr.}  This transverse separation velocity would have caused Lobe~I to
arrive at perihelion more than 1~day earlier than would the progenitor (or the neck).
On the other hand, Lobe~II had to separate with a transverse velocity in the direction
of the orbital-velocity vector, which would have caused it to arrive at perihelion about
1$\frac{1}{2}$~days later than would the neck (or nearly 3~days later than Lobe~I).
As the lobes did not survive intact, their major fragments are assumed to have arrived
at those times.  This rule applies to all other fragments. 

The cascading fragmentation sequence is shown in Figure~3.  The events are assumed to
have occurred --- in the depicted order (the time increasing from the top down) ---
in the aphelion region of the progenitor's orbit, with the exception of Fragment~Pe,
a precursor of C/1963~R1, which is suggested to have separated from Lobe~I later, on the
way from aphelion to perihelion, at about 70~AU.\footnote{This fragmentation time is
only crudely estimated.}  The fragmentation events are favored to have occurred at
aphelion because the same effect in the critical orbital elements was then achieved
with a minimum separation velocity, even though this minimum was flat:\ the change
is confined to less than 10~percent of the minimum value over an orbital arc around the
aphelion point that the comet travels in nearly 300~yr, or 40~percent of the orbital
period!  This tolerance explains extremely wide choice for the fragmentation time.
%

\begin{table*}
\vspace{-4.2cm}
\hspace{0.6cm}
\centerline{
\scalebox{1}{
\includegraphics{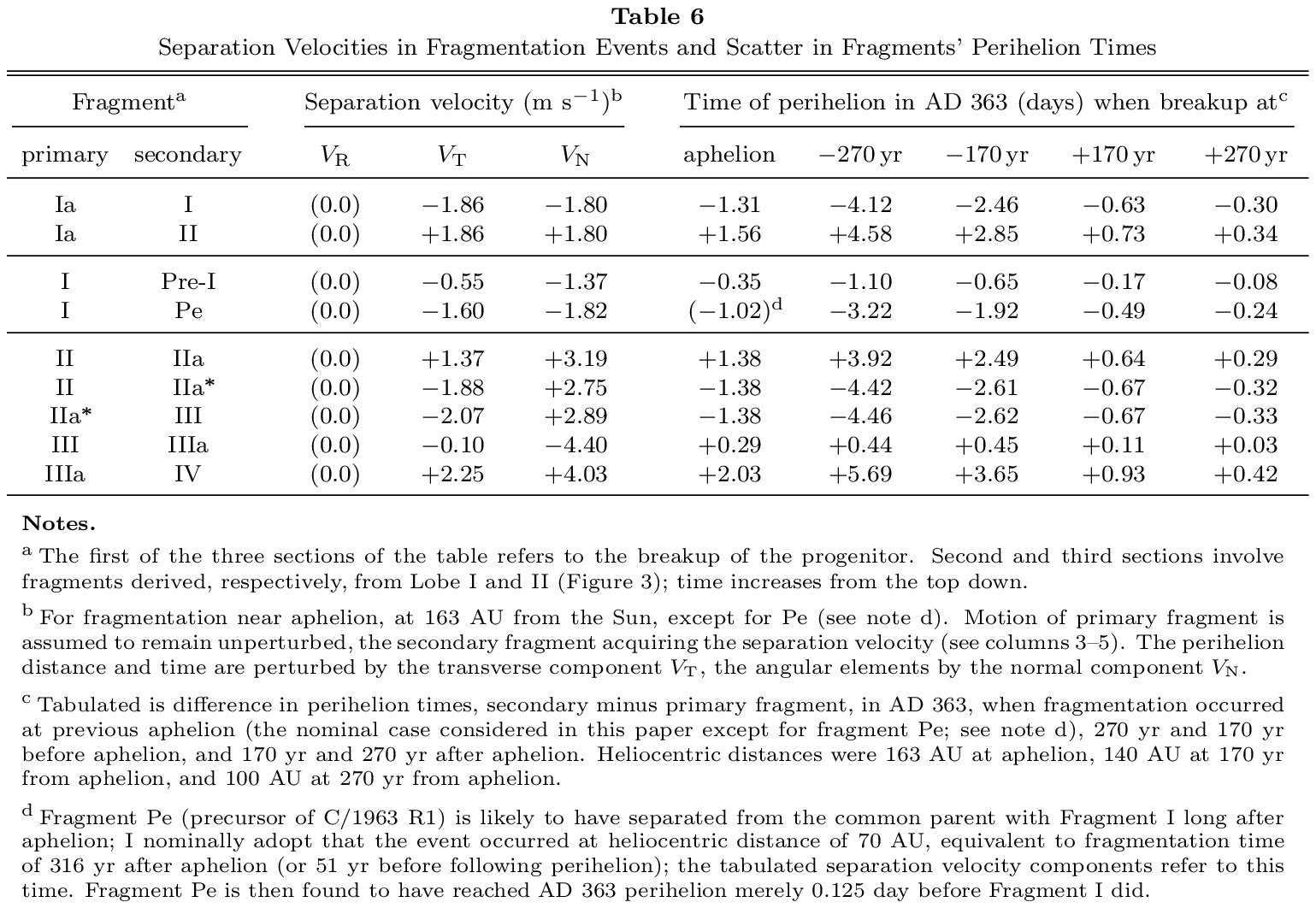}}}
\vspace{-14.4cm}
\end{table*} 

Only the separation of Fragment Pe occurred at a location clearly different from
aphelion; the time was dictated by requiring that the magnitude of the separation
velocity be comparable to those of Lobes~I and II from the progenitor.  Table~6
demonstrates that for most fragments the total separation velocity was on the
order of \mbox{2--3 m~s$^{-1}$} and always lower than 5~m~s$^{-1}$.  A possible
slight increase in the separation velocity with time that the table might show,
could --- if genuine --- present evidence on the fragments' spin-up.  The
relationships between the components of the separation velocity and the
perturbations of the orbital elements are governed by the equations presented
in Appendix~A.

The differences in the perihelion times of the various pairs among the 10~primary
fragments, listed in Table~6, were streamlined by referring each time to that of
one particular sungrazer, the obvious choice being Fragment~Ia.  This streamlining
was readily achieved by properly adding up the tabulated differences.  Designating
the perihelion time of Fragment~Y$_0$ by $t_\pi({\rm Y}_0)$, the difference between
this time and the perihelion time of Fragment~Ia, $t_\pi$(Ia), is given by the
expression
%
%
\begin{eqnarray}
t_\pi({\rm Y}_0) \!-\! t_\pi({\rm Ia}) & = &
  \left[\,t_\pi({\rm Y}_n) \!-\! t_\pi({\rm Ia})\right] \nonumber \\[-0.05cm]
  & & + \sum_{k=0}^{n-1} \left[\, t_\pi({\rm Y}_k) \!-\!
  t_\pi({\rm Y}_{k+1}) \right], 
\end{eqnarray}
where Y$_n$ is either Fragment I or Fragment II, while Y$_1$, \ldots, Y$_{n-1}$
are the other fragments in Table~6, all related to either Lobe~I or II.  For
the chosen scenario, the parameters are, together with all other elements
(taken from Paper~1), listed in Table~7.  The middle of the time window of
interest (see the beginning of Section~5), has been adopted for the reference
time, equal to the perihelion time of the fragmented progenitor's center of
mass.  This time has been equated with the perihelion time of Fragment~Ia, thus
\mbox{$t_\pi({\rm Ia}) = {\rm Nov}$ 15.00}.  The overall range of perihelion
times of the 10~progenitor fragments is 4.6~days.  Since the perturbations
of both the perihelion distance and perihelion time were exerted by the
transverse component of the separation velocity, one would expect a
correlation between the two elements.  Figure~4 confirms that this is
indeed so.  The correlation is perfect for the nine sungrazers that were
products of fragmentation events in the aphelion region, while Fragment~Pe,
separating from its parent at 70~AU from the Sun, shows a minor deviation from
the fit.  

In order to be able to judge the visibility of the ten sungrazers, I adopted
a simple photometric model, based on the findings in Section~5, and employed it
with the Schaefer (1998) algorithm for the limiting magnitude by the unaided eye.
Following the conclusions from Section~5.6 I assumed that (i)~all fragments on
their way to the 363 perihelion were brightening in accordance with the $r^{-4}$
law; (ii)~the difference{\vspace{-0.06cm}} in $H_0^-$ between two consecutive
generations was 0.5~mag (as it was, approximately, between X/1106~C1 and C/1843~D1);
and (iii)~the absolute magnitude of the fragment of the last generation in
Figure~3, Fragment~IV, was comparable with those of C/1843~D1 and C/1882~R1.  As
a result,{\vspace{-0.06cm} the progenitor was assigned \mbox{$H_0^- \!= 1$},
Fragments~I and{\vspace{-0.06cm}} II \mbox{$H_0^- \!= 1.5$} each, etc., except
that \mbox{$H_0^- \!= 3$} for Fragment~Ia.  Furthermore, to define the
post-perihelion curves of the sungrazers, I assigned a number of subfragments
(or ``perihelion'' fragments), $\nu$, into which each arriving sungrazer is expected
to have broken at the 363 perihelion passage, and subtracted 0.2 per subfragment
from the value of \mbox{$n^+ \!= 4.2$} for a sungrazer that did not break up at
perihelion at all (\mbox{$\nu \!=\! 1$}).  The post-perihelion absolute magnitude
$H_0^+$ was then determined from
\begin{equation}
H_0^+ = H_0^- + 2.5 (4 \!-\!n^+) \log q, 
\end{equation}

\begin{table*}
\vspace{-4.12cm}
\hspace{0.5cm}
\centerline{
\scalebox{1}{
\includegraphics{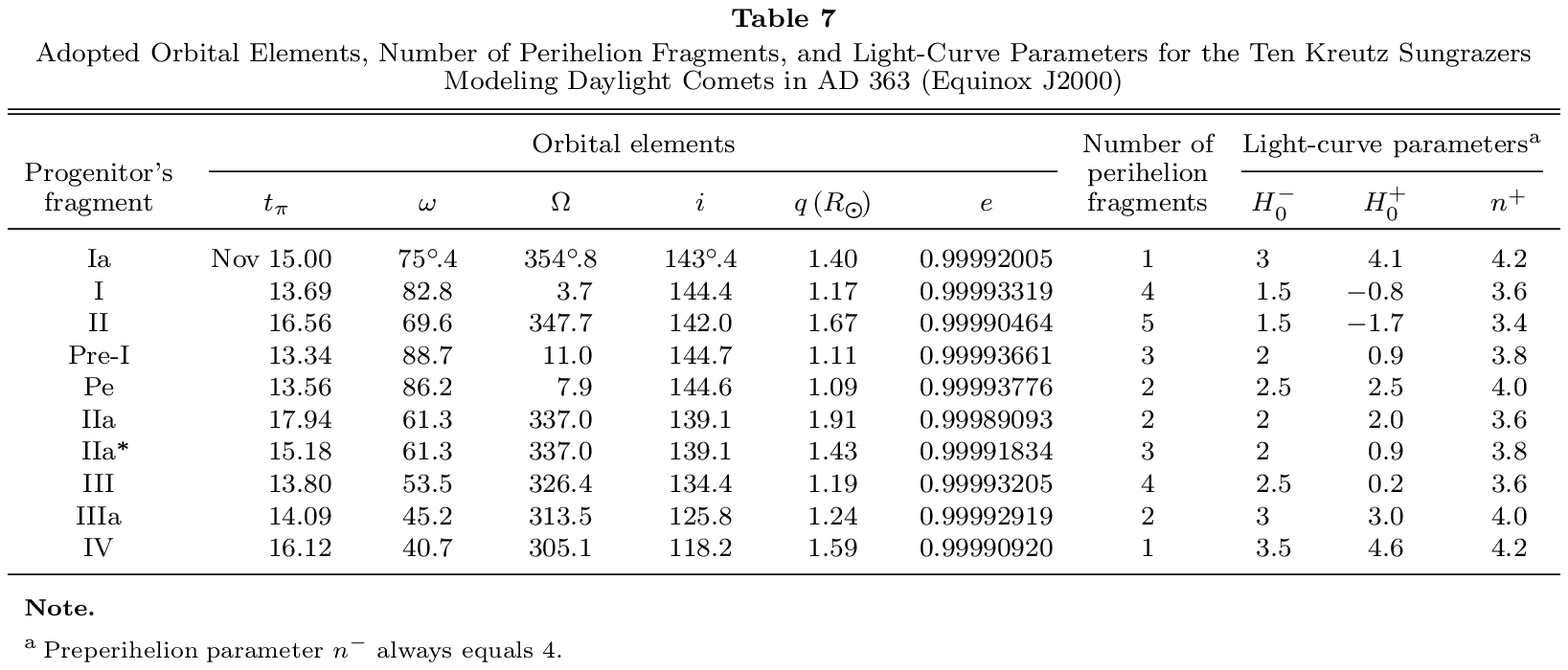}}}
\vspace{-18.1cm}
\end{table*}

\noindent
where $q$ is the perihelion distance in AU.  The adopted values of the light-curve
parameters are listed on the right side of Table~7.  Because the phase angle is
subjected to rapid and sizable variations in close proximity of perihelion and
strong effects of forward scattering are expected to occur, the phase term was
added to the light-curve formula, accounting for the phase effect by employing
the method introduced by Marcus (2007).

To use the photometric model in determining under what conditions is a sungrazer
visible to the naked eye, I define a simple {\it visibility index\/} $\Im$ as a
difference between Shaefer's limiting magnitude for the naked eye, $H_{\rm
lim}$, and the sungrazer's apparent magnitude, $H$,
\begin{equation}
\Im = H_{\rm lim} - H. 
\end{equation}
The sungrazer is said to be {\it clearly visible\/}~with~the~naked
eye~when~$\Im > +1.5$~mag;~it~is~{\it probably visible\/}~when
$+0.5 < \Im \leq +1.5$ mag, and {\it potentially visible\/}~when
$-0.5 \leq \Im \leq +0.5$ mag. It is {\it probably invisible\/}~when
\mbox{$-1.5 \leq \Im <\! -0.5$\,mag;~and {\it invisible\/} when $\Im \!<\!
-1.5$\,mag}.

\begin{figure}[b]
\vspace{-10.3cm}
\hspace{3cm}
\centerline{
\scalebox{0.76}{
\includegraphics{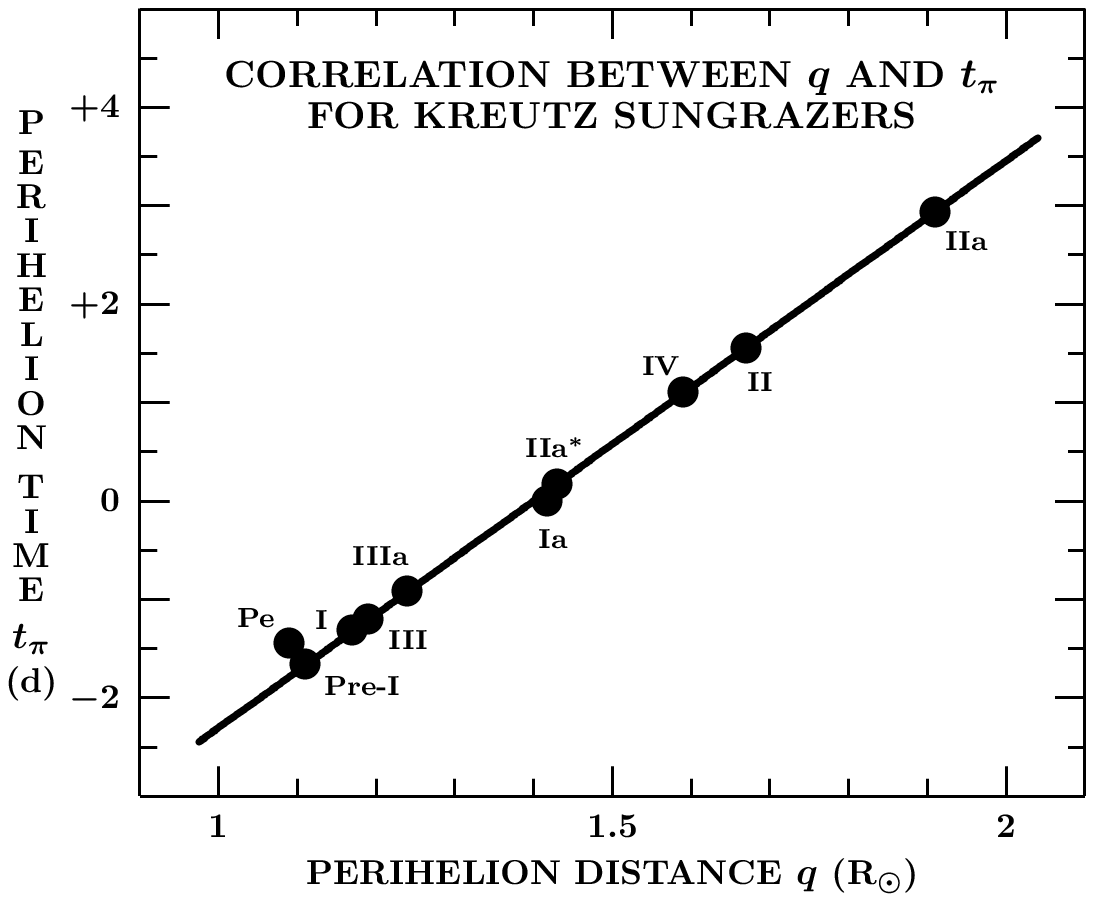}}}
\vspace{-5.1cm}
\caption{Correlation between the perihelion distance, $q$, and the
perihelion time, $t_\pi$, of the Kreutz sungrazers, derived from the
perturbations of their parents' orbits acquired at the breakup of
the progenitor and the sequence of subsequent secondary breakups,
as illustrated in Figure~3.  The position of Fragment~Pe is off the
correlation line, because it separated from its parent at a heliocentric
distance much smaller than the aphelion distance.{\vspace{-0.05cm}}}
\end{figure}

The final data set for the computations concerns Ammianus' observing site, Antioch
on the Orontes.  Overall information is in Appendix~B; specifically, the geographic
latitude of the location, \mbox{$\phi = +36^\circ\!.2$}.\footnote{There is no need
for the geographic longitude for reasons explained in the text.}  For the
computation of the limiting magnitude, I assume an altitude of 70~meters above
sea level, an average November temperature of 15$^\circ$C (an average of a maximum
20$^\circ$C and a minimum 10$^\circ$C), and 55~percent humidity.{\vspace{-0.08cm}}

\subsection{The Computations} 
The objective is to model a swarm of Kreutz sungrazers, fragments of a
contact-binary progenitor, arriving at the first perihelion passage to find out whether
the result fits Ammianus' {\it comets seen in broad daylight\/}.  One needs to examine
a scenario in the correct time of the year and in an approximately correct sky
projection, but there is no need to actually perform the computer runs specifically
in the year 363.  This fits the tolerance documented by use of the osculating orbital
elements, with the indirect planetary perturbations over nearly two millennia neglected,
and by ignoring atmospheric refraction and the effects of the equation of time.
Expressing the apparent motions of the Kreutz sungrazers relative to the Sun in
the sky as a function of local solar time, the Sun culminates exactly at noon,
when the local sidereal time equals the Sun's right ascension, with no need to
introduce the geographic longitude.  The approximations result
in positional errors that are deemed acceptable.

For each of the ten sungrazers and each selected time of appearance the data file
provides an ephemeris entry that contains the right ascension, $\alpha$, and
declination, $\delta$, phase angle, $\psi$, solar elongation, $\epsilon$,
position angle of the prolonged radius vector, $P_{\rm RV}$, and distances from
the Earth, $\Delta$, and the Sun, $r$.  The Sun's right ascension, {\rasun}, and
declination, {\decsun}, are calculated from
\begin{eqnarray}
\sin \mbox{\decsun} & = & \sin \delta \cos \epsilon
  - \cos \delta \sin \epsilon \cos P_{\rm RV}, \nonumber \\[-0.05cm]
\sin (\alpha \!-\! \mbox{\rasun}) & = & \sec \mbox{\decsun} \sin \epsilon
   \sin P_{\rm RV}, 
\end{eqnarray}
where \mbox{$|\alpha \!-\! \mbox{\rasun}| < 90^\circ$} as long as \mbox{$\cos \epsilon
> \sin \delta \sin \mbox{\decsun}$}.

In order to display the positions of the Kreutz sungrazers relative to the Sun in the
Antioch sky, the equatorial coordinates need to be converted to the local horizontal
coordinates.  The position of each fragment is then given by its azimuth, $\cal A$,
reckoned positive from the south through the west, and by its elevation, ${\cal E}$,
reckoned positive from the ideal local horizon to the zenith.  The relationship between
the horizontal and equatorial coordinates is given by the transformation formulas
\begin{eqnarray}
\cos {\cal E} \sin {\cal A} & = & \cos \delta \sin (\Theta \!-\! \alpha), \nonumber \\
\cos {\cal E} \cos {\cal A} & = & -\cos \phi \cos \delta + \sin \phi \cos \delta
 \cos (\Theta \!-\! \alpha), \nonumber \\
\sin {\cal E} & = & \sin \phi \sin \delta + \cos \phi \cos \delta \cos (\Theta \!-\!
 \alpha), 
\end{eqnarray}
where $\phi$ is the geographic latitude of Antioch, $\Theta$ is the local sidereal
time, which is related to the local solar time $t$ (in hours) by
\begin{equation}
\Theta(t) = \mbox{\rasun}^{\!\!\!\!\ast} + \zeta \, (t \!-\! 12),  
\end{equation}
{\rasun}$^{\!\!\!\!\ast}$ is the right ascension of the Sun at local noon, and
\mbox{$\zeta = 1.00273791$}.  Equations~(19) of course equally apply to the Sun,
thereby providing its azimuth {\azsun} and elevation {\elsun}.  The positions of
the sungrazers relative to the Sun in the sky are then given by the solar elongation
$\epsilon$ (from the ephemeris) and by a zenithal position angle $P_{\rm Z}$, which
is in the local coordinate system reckoned positive from the Sun's zenithal direction
clockwise through the west,
\begin{equation}
\sin P_{\rm Z} = \csc \epsilon \cos {\cal E} \sin ({\cal A} \!-\! \mbox{\azsun}), 
\end{equation}
with the sign of $\cos P_{\rm Z}$ equaling the sign of the expression \mbox{$\cos
\epsilon \!-\! \sin {\cal E} \sin \mbox{\elsun}$}.

The visibility index compares a quantity that is a function of the sungrazer's
photometric and orbital properties [i.e., its apparent magnitude \mbox{$H =
H(\Delta,r,\psi$;\,{\sl phot\/})}, where {\sl phot\/} stands for the light-curve
law (preperihelion vs post-perihelion)] with a quantity that is a function of the
observing conditions [i.e., the limiting magnitude \mbox{$H_{\rm lim} = H_{\rm
lim}(\epsilon,{\cal E},\,$\mbox{\elsun};\,{\sl loc\/})}, where {\sl loc\/} defines the
observing site].  The distinction between the two quantities is that, in general,
the latter varies substantially on much shorter time scales (often a small fraction
of an hour) than the former.  Since all scenarios involve close proximity to the
Sun, the effects of the Moon are deemed insignificant and are ignored.

\subsection{The Results} 
As noted in Table 7, a swarm of ten major sungrazers of the Kreutz progenitor is
investigated for an assumed perihelion time of November~15.0, the midpoint of the
two-month long period of interest, as established at the beginning of Section~5.
The time frame used is the local solar time, which is adequate for the purposes of
this study.  The progenitor's orbit is assigned to Fragment~Ia (Figure~3); the
perihelion times of the other nine sungrazers range over a period of 4.6~days, from
November~13.34 (Fragment~Pre-I) to November~17.94 (Fragment~IIa), being strongly
correlated with the perihelion distance, as shown in Figure~4.

A computer code has been written that calculates, for the geographic latitude of
Antioch, (a)~the horizontal coordinates of the ten sungrazers and the Sun, using the
objects' ephemerides; and (b)~the apparent magnitudes of the sungrazers, employing
the prescribed light-curve parameters, the ephemerides, and the Marcus (2007)
dust scattering law.  The period of time covered by the daily ephemerides extends
from November~1 to November 25, each day from 4:48 to 19:12 local solar time at
a 36~minute step, thereby totaling 625~individual entries for each sungrazer.

\begin{figure*}
\vspace{-10.16cm}
\hspace{-0.05cm}
\centerline{
\scalebox{0.87}{
\includegraphics{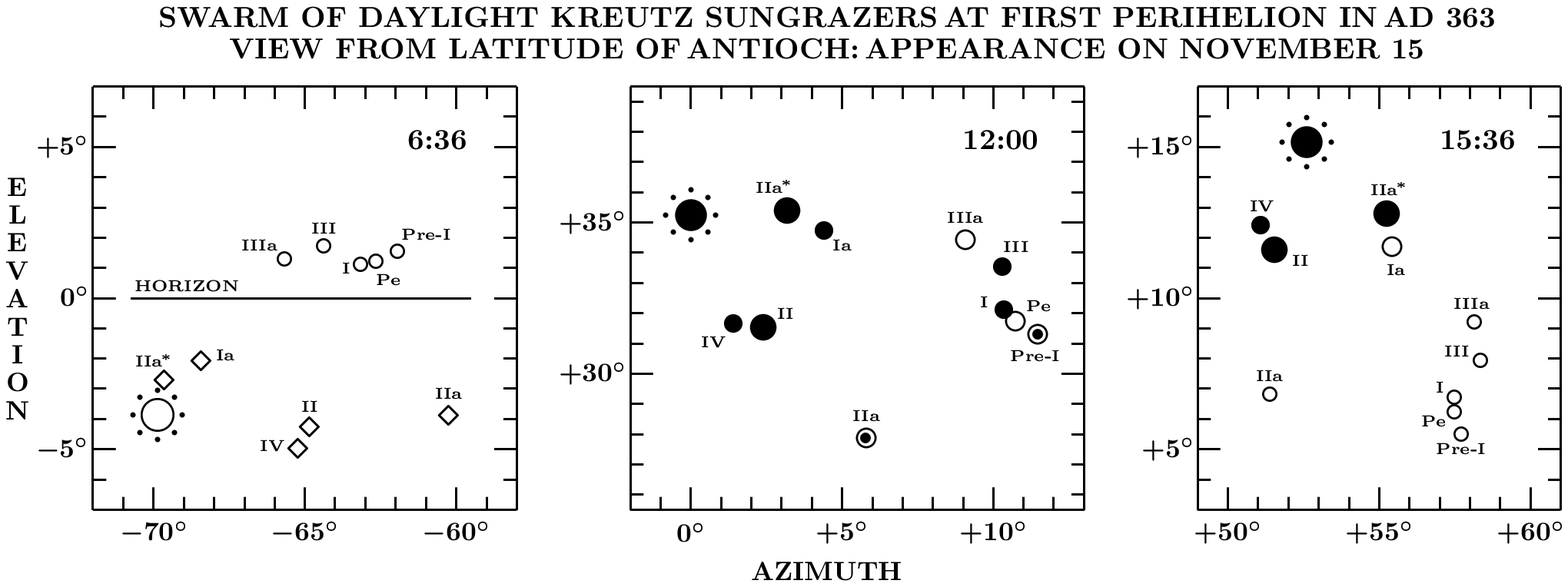}}}
\vspace{-9.2cm}
\caption{Simulation of the swarm of ten Kreutz sungrazers in projection
onto the plane of the sky, viewed, near the Sun, from the latitude of
Antioch at, respectively, 6:36, noon, and 15:36 local solar time on
363 November 15.  The horizontal coordinate system has
the azimuth reckoned positive from the south through the
west, the elevation reckoned positive from the horizon to the zenith.
The sungrazers' orbital motions and photometric properties are
described by the parameters listed in Table~7.  The symbols depict
each sungrazer's visibility to the naked eye, defined by the index
$\Im$ (in magnitudes):\ large solid circles stand for \mbox{$\Im >
+1.5$} (clearly visible); medium solid circles for \mbox{$+0.5 <
\Im \leq \! +1.5$} (probably visible); circled dots for \mbox{$-0.5
\leq \Im \leq \! +0.5$} (potentially visible); medium open circles
for \mbox{$-1.5 \leq \Im < \! -0.5$} (probably invisible); small open
circles for \mbox{$\Im < \! -1.5$} (invisible); and small diamonds
for fragments that are below the horizon.{\vspace{0.7cm}}}
\end{figure*}

The trajectories of the sungrazers in the sky follow the expected pattern given by
the projected spatial orientation of their orbits.  Relative to the Sun the comets
approach from the southwest and recede to the west-southwest.  They pass between
the Sun and Earth on their way to perihelion; 1--2~hours before reaching it, their
phase angle exceeds briefly 150$^\circ$, their brightness then being strongly
enhanced by forward-scattering effects.  They recede from perihelion on the far
side of the Sun, reckoned from the Earth.  The limited, yet clearly apparent, orbital
scatter is prompted by the almost 70$^\circ$ wide range of nodal longitudes and the
associated range of nearly 30$^\circ$ in the inclination.

In Section 5.4 I showed for comet Ikeya-Seki that the preperihelion light curve still
followed the $r^{-4}$ law as late as 6~hours before perihelion and possibly closer.
However, there is no evidence either way whether the brightness could reach magnitudes
near, or in excess of, $-$20~mag over a period of, say, a fraction of an hour around
perihelion for the most brilliant sungrazers.  To a degree, this question is academic,
because even if attained, these magnitudes should be highly transient.  It suffices
to mention that the apparent brightness of a Kreutz sungrazer that follows an inverse
fourth power law varies by fully 3~magnitudes\footnote{Not accounting for the effect
of forward scattering.} over an arc of 180$^\circ$ in true anomaly centered on the
perihelion point and traveled in 1.9 to 4.4~hours (depending on the perihelion
distance).  While in the tables that follow I strictly adhere to the power law, the
magnitudes brighter than about $-$15 are not to be taken literally, meaning only that
the sungrazer was then surely bright enough to be seen with the naked eye even at close
proximity of the Sun.  Indeed, Schaefer's algorithm indicates that a stellar object
in contact with the Sun's disk in the sky is visible to the naked eye when its
brightness exceeds magnitude $-$12 to $-$13.

The results are presented in three figures and a series of comprehensive tabulations
in Appendix~C plus a summary table.  The figures are to offer views of the diverse
appearances of the sungrazers' sky display relative to the Sun on particular days
and to illustrate the overall evolution of the swarm.  By contrast, the tables
provide the in-depth positional and visibility data.   

Figure 5 and Table C-1 show the development of the swarm of sungrazers on November~15.
At 6:36 local solar time the Sun was still nearly 4$^\circ$ below the horizon (civil
twilight).  Of the ten sungrazers, only five were above the horizon, although all by
less than 2$^\circ$.  None of them was visible to the naked eye because of twilight
and atmospheric extinction.  By 7:12, both the Sun and all ten comets moved above
the horizon and Fragment~IIa{\boldmath $^\ast$} became visible to the naked eye.
The conditions further improved by 7:48, when three additional sungrazers joined
IIa{\boldmath $^\ast$}, now more than 10$^\circ$ above the horizon, in becoming
visible and from this time on they could be referred to as a swarm.

By noon, the number of visible comets grew to the day's peak of six.  They were
located from 2$^\circ\!$.6 to 10$^\circ\!$.4 from the Sun, to the west through
south-southwest of it, and as bright as magnitude $-$11.  It is not surprising that
the brightest sungrazer was IIa{\boldmath $^\ast$}, as it passed perihelion only
8~hours earlier.  The second brightest, II, still had more than 1~day to go to
perihelion, but its brightness was aided a little by the effect of forward scattering.
Additional two pieces were at the limit of visibility.  Only IIIa and Pe remained
hidden from sight, either being fainter than magnitude $-$6 at the time.

While in the morning of November 15 the sungrazers rose essentially with the Sun,
they all set before sunset in the afternoon.  Figure~5 and Table C-1 show that
by 15:36 only three remained visible; by 16:48, when the Sun was still almost
3$^\circ$ above the horizon, all but one comet had set and none was visible.
Overall, no sungrazer was seen on this date before sunrise or after sunset.

\begin{figure*}[t]
\vspace{-2.25cm}
\hspace{-0.08cm}
\centerline{
\scalebox{0.865}{
\includegraphics{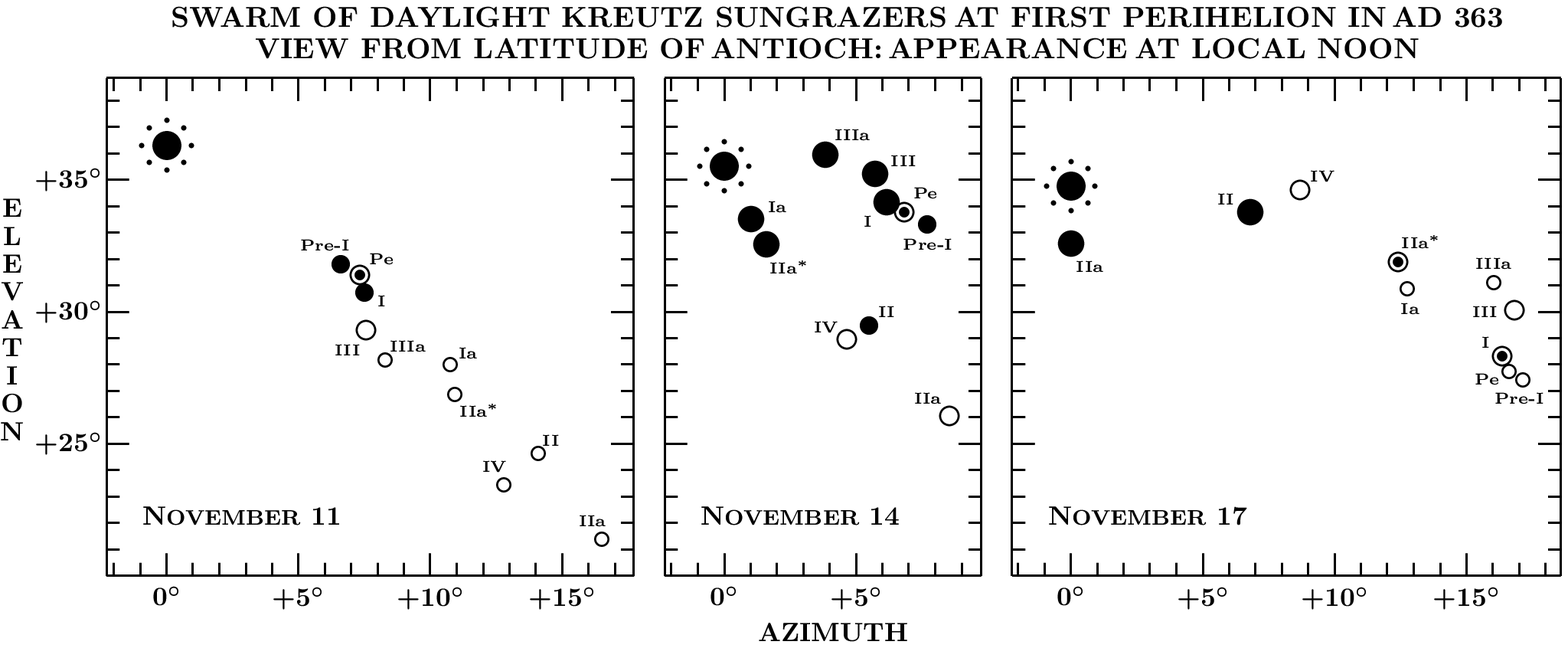}}}
\vspace{-16.35cm}
\caption{Simulation of the swarm of ten Kreutz sungrazers in projection onto the
plane of the sky, viewed, near the Sun, from the latitude of Antioch at local
noon on 363 November 11, 14, and 17.  See the caption to Figure 5 for more
explanation.{\vspace{0.6cm}}}
\end{figure*}

Figure 6 exhibits the noon appearance of the swarm on three days around the time
of peak performance, on November~11, 14, and 17.  Together with Figure~5 these
data constrain the duration of the swarm to at least two days but fewer than
seven days, as on November~11 and 17 only two sungrazers were visible.

Figure 7 allows one to inspect the visibility of sungrazers at times long before
and/or after the peak in mid-November.  It turns out that because of their greater heliocentric distances, the comets were too faint to show {\it after\/} sunrise on
both November~1 (upper panels) and November~25 (lower panels).  However, on both dates
three sungrazers were visible over a very limited period of time (less than 1~hour)
{\it before\/} sunrise very low (about 10$^\circ$) above the southeastern horizon.
The reader will notice that this behavior is exact opposite to the situation on
November~15, when no sungrazers could be seen {\it before\/} sunrise (Figure~5) but
became gradually visible {\it after\/} sunrise (Table~C-1).  However, one should
point out that the effect of a post-perihelion tail, whose presence could potentially
increase the brightness dramatically beyond sunrise, has not been taken into
consideration in the computations.

Table C-2 presents a picture of the swarm on November~13 and 14, similar to, but more
compact than, that on November~15 in Table~C-1.  On the 13th, the appearance at 7:12
resembled the appearance 48~hr later, except that the single visible sungrazer was
Pre-I (which at the time had 1~hr to go to perihelion) rather than IIa{\boldmath
$^\ast$}.  By noon, six comets were visible, Pe and I extremely bright and less
than 1$^\circ$ from the Sun.  By 15:36 the number of visible sungrazers declined by
only one.

A swarm of three comets was marginally visible already at 7:12 on the morning of
November~14; the Sun was then merely 3$^\circ$ above the horizon.  By noon a record
seven sungrazers were seen (cf.\ Figure~6), five of them --- Ia, I, IIa{\boldmath
$^\ast$}, III, and IIIa --- about equally bright; an eighth piece, Pe, was near the
visibility limit.  By 15:36 the number of visible sungrazers dropped to four; again,
they all had set before sunset. 

Table C-3 shows the evolution of the swarm on November~16 and 17.  On the 16th at
7:12 there was only one potentially visible sungrazer.  By noon Fragment~II was by far
the brightest, reaching perihelion about 1.5~hr later; also seen were three additional
comets and two more potentially.  By 15:36, Fragment~II was the only visible one.  On
the 17th, the situation further deteriorated, Fragment~II clearly fading though still
visible.  An noon, only two sungrazers, II and IIa, were clearly visible, surviving
through 15:36.

From this review of the five days, November~13--17, it is apparent that the number
of visible sungrazers correlates closely with the distribution of their perihelion
times.  Since seven of the ten comets passed perihelion in the first half of the
4.6~day long period, the number of the visible ones was higher on November~13--14
than on November 16--17.  As expected, the number is also seen to correlate with the
angular distance from the Sun; statistically, most visible sungrazers were very close
to the Sun, mostly within 4$^\circ$ or so.  Their largest number could be seen for
several hours approximately centered on local noon, when the swarm was high enough
above the horizon.

The next step was to examine the rising and setting of the swarm over a longer period
of time, up to two weeks, before the sungrazers' perihelion times.  It turns out that
on November~1 (Table~C-4 and Figure~7) the comets rose about two hours before the
Sun, but their apparent magnitude was only near 0.  At 5:24, when the Sun was
more than 15$^\circ$ below the horizon, they were just a few degrees above it and
the model shows that only one was about 1~mag above the limiting magnitude.  The
situation was not improving with time; at 6:00 the lesser effect of atmospheric
extinction was compensated by advancing twilight.  By 6:36, when twilight was
nearly over, no sungrazer was visible any longer.  By November~7 (Table~C-4), the
sunrise was lagging behind the sungrazers' rise by only an hour or so and they were
not more than a few degrees above the horizon when civil twilight began.  The
comets were too faint to be seen in broad daylight and they had set long before
the Sun.  On November 11 (Table~C-5), the noon appearance was nearly equivalent to
that on November~17 (cf.\ Figure~6 and Table~C-3), with two sungrazers then visible;
none of them was visible on November~11 in the morning at 7:12 and 9:00 and in the
afternoon at 15:00.  Table~C-5 also shows that at noon of November~19 only one of
the two sungrazers seen two days earlier, Fragment~II, was still visible, the
brightness of the other (IIa) being marginal. 

Examination of the arrangement of the swarm relative to the Sun on November~25
(Table~C-6 and Figure~7), some ten days after perihelion, shows that the overall
visibility of the sungrazers was comparable to that in the period of time between
November~1 and 7.  Fragments~I and III may have been seen over a fraction of an hour
at altitudes of several degrees above the horizon in the southeast around 6:00 (when
the Sun was 12$^\circ$ below the horizon) and Fragment~II at an altitude of less than
10$^\circ$ around 6:36 (when the Sun was 5$^\circ$ below the horizon), but either
was no longer seen at 7:12, after sunrise (cf.\ Figure~6).  However, as already
noted, prominent post-perihelion tails may have already been developed enough to
substantially improve the visibility during the day.  The sungrazers again set long
before sunset, the tails then pointing to the south.
%

%
\begin{figure*}
\vspace{-3.05cm}
\hspace{-0.25cm}
\centerline{
\scalebox{0.86}{
\includegraphics{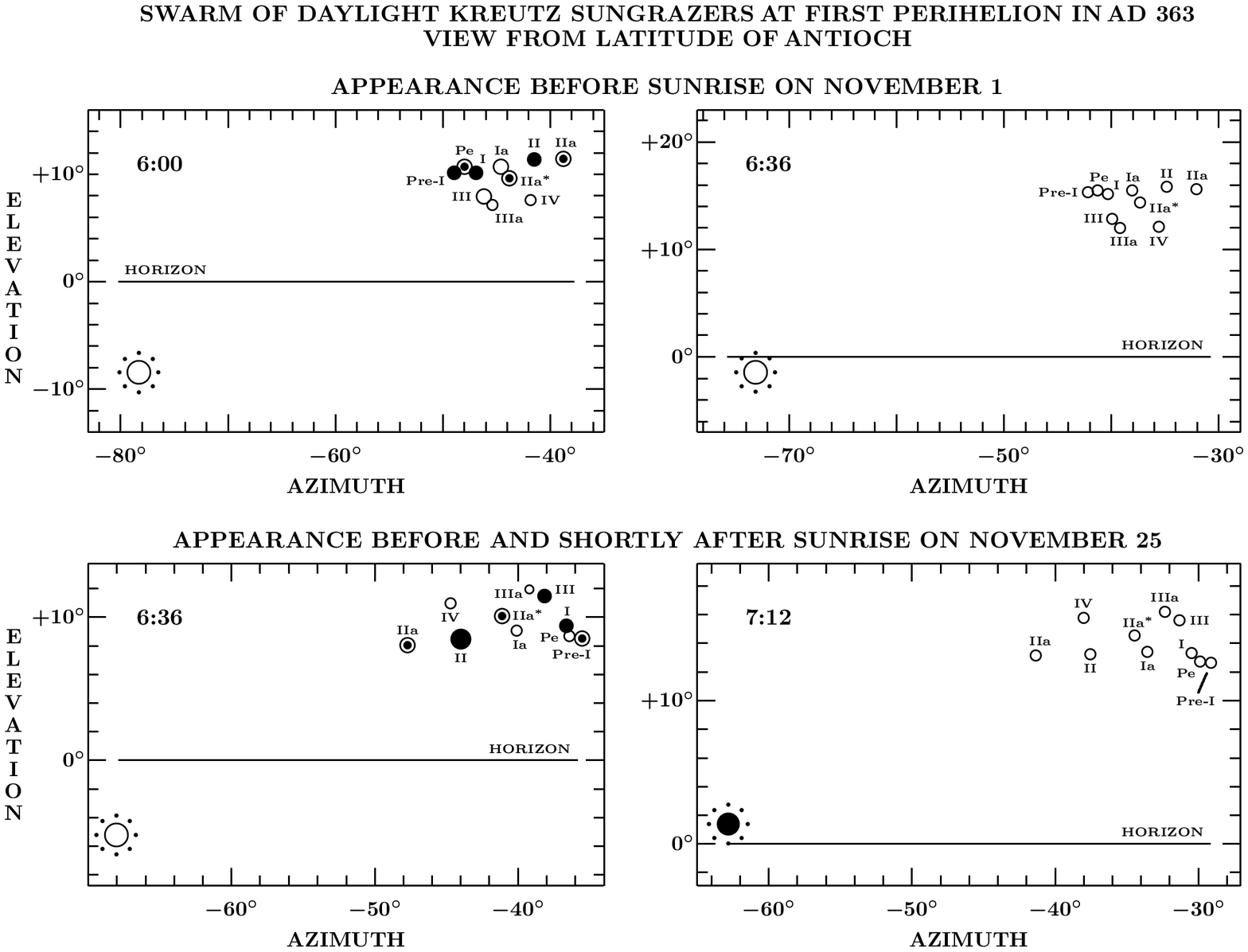}}}
\vspace{-8.75cm}
\caption{Simulation of the swarm of ten Kreutz sungrazers in projection
onto the plane of the sky, viewed from the latitude of Antioch
at, respectively, 6:00 and 6:36 local solar time on November 1 (at the
top) and at 6:36 and 7:12 local solar time on November~25.  See the
caption to Figure~5 for more explanation.{\vspace{0.6cm}}}
\end{figure*}

The results of this exercise are summarized in Table 8, which shows that at their
first arrival at perihelion, a swarm of Kreutz sungrazers, generated in the
framework of the proposed contact-binary model, is capable to produce a spectacle
of several brilliant daylight comets, which lasts over a few days, each day for
several hours about local noon (when high enough above the horizon).  The apparent
brightness of at least some of the comets is magnitude $-$10 or more, exceeding
the limiting magnitude for the naked eye by up to about 7~magnitudes.  The display
could certainly be described by Ammianus' words.  The number, distribution, and
identities of the sungrazers vary from day to day and even from the morning to
late afternoon on each day.  The swarm is most impressive on November~13 to 15.

Two weeks before perihelion, at the beginning of November, up to three comets
may briefly (for less than an hour) be visible before sunrise very low above the
southeastern horizon; they are too faint to remain visible toward the end of
twilight, when they are overwhelmed by approaching daylight.  In late November
the sungrazers may be visible because of the post-perihelion tail development,
several of them could continue to be daytime objects, but the prime of the show
would move to pre-noon hours, with the tails then pointing toward the horizon.
For example, at 10:12 on November 25, five sungrazers with tails more than 5~mag
brighter than the head --- II, I, III, IIa, and IIa{\boldmath $^\ast$} in the
order of decreasing brightness --- could be visible at elevations lower than
that of the Sun; by 14:30 they all would be below the horizon,{\nopagebreak}
their tails preceding them in the southerly direction.{\pagebreak}

\begin{table*}
\vspace{-4.2cm}
\hspace{0.6cm}
\centerline{
\scalebox{1}{
\includegraphics{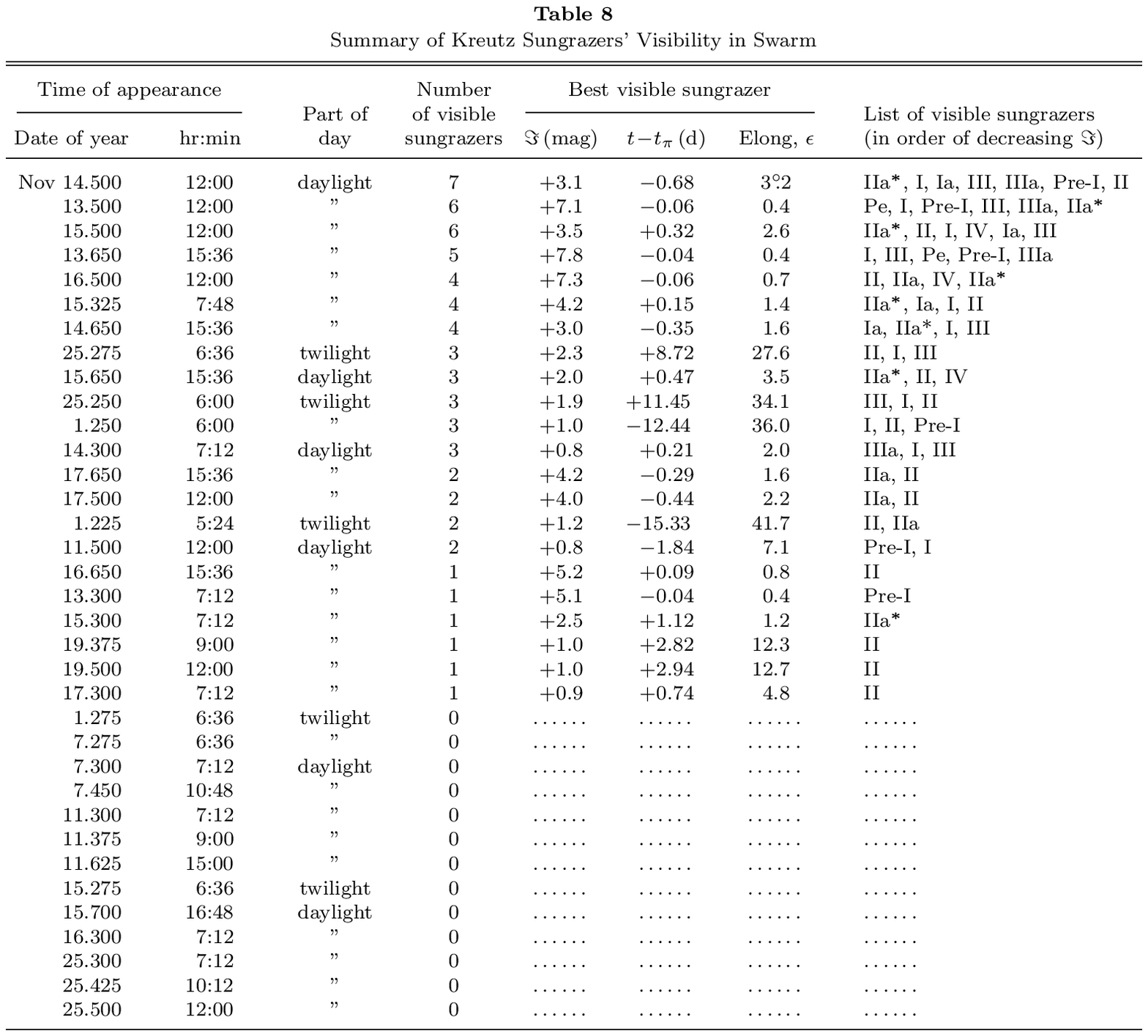}}}
\vspace{-10.3cm}
\end{table*}
\section{Conclusions} 
The results of this investigation offer the following conclusions:\\[0.1cm]
\indent
(1) Even though the developments in the past decade revealed weaknesses of the
two-superfragment model, two of its fundamental properties --- the progenitor's
fragmentation taking place at large heliocentric distance and the fragments
arriving nearly simultaneously at their first perihelion in the mid-4th century
AD --- are still deemed absolutely essential for reaching a self-consistent
solution to the problem of origin and evolution of the Kreutz system, the
division into Populations~I and II in particular; accordingly, they have been
incorporated into the new contact-binary model in Paper~1.

(2) The cascading nature of the fragmentation process, which the Kreutz system
is subjected to, is of utmost importance because the fact that a sungrazer can
break up at any point of the orbit --- rather than at perihelion only --- does
dramatically boost the fragmentation {\it rate\/}, which, in turn, severely
curtails the {\it lifespan\/} of sungrazers, including the giant ones.        

(3) The remark by the Roman historian Ammianus Marcellinus that ``in broad daylight
comets were seen'' in late AD~363 is highly relevant to the problem of the Kreutz
system in that (i)~the time closely agrees with the perihelion time anticipated by
the two-superfragment model; (ii)~the plural implies an unrivaled, essentially
simultaneous appearance of spectacular daylight comets, a scenario expected for the
progenitor's fragments; and (iii)~the timing, halfway between the arrivals of the
Aristotle's comet of 372~BC and the probable Kreutz member X/1106~C1 --- with a
further link to C/1843~D1 --- suggests a periodicity of about 740~years, connecting
four generations of sungrazers.

(4) I argue that Ammianus' remark must have been motivated by a lasting impression
that the event left on him and that there is no chance that the spectacle was
fabricated by him to serve as a portent.  A simultaneous or nearly simultaneous
appearance of two or more brilliant comets is historically unprecedented, in broad
daylight even more so.  It could never have crossed his mind that a celestial
show of the kind was at all possible, until he witnessed it.  Besides, if
fabricating it as a portent, Ammianus would certainly have dated the event before
the death of his beloved Emperor Julian, not months after it.

(5) Under the circumstances it is beneficial to test whether, or to what extent,
could the event noted by Ammianus be broadly consistent with the appearance
of a swarm of Kreutz sungrazers.  In the modeled scenario, the swarm is generated
by a process of cascading fragmentation of the contact-binary progenitor near
aphelion of its orbit, the individual fragments arriving at perihelion over a
period of a few days in late AD~363.

(6) Inspection of Hasegawa \& Nakano's list of potential historical members of
the Kreutz system shows a total of seven comets first sighted late in the year,
between mid-October and mid-December: 943, 1232, 1381, 1588, 1663, 1666, and 1695
(C/1695~U1).  The temporal distribution of their first sightings is strongly
skewed toward the early part of the period, late October and early November,
apparently an effect of the gradually increasing deep southerly declinations.
None of the seven comets appears to have been detected before perihelion.

(7) The apparent magnitude of Halley's comet at times of its preperihelion first
sightings recorded in returns before the comet's identity was recognized, starting
in 12~BC, is revised.  It is shown that the existing brightness limit, based on
the comet's inaccurate light curve from the 1910 apparition, is underestimated.
A revised calibration, employing the Green \& Morris light curve from the 1986
apparition, implies that at times of the first naked-eye sighting, Halley's comet
was on the average of an apparent magnitude 2.6, rather than the previously
proposed magnitude 3.5 to 4.

(8) Hasegawa \& Nakano's assumption that the potential historical Kreutz sungrazers
had an apparent magnitude~5 when last sighted with the naked eye after perihelion
is significantly off the mark.  The magnitude of the head is found instead to have
been near 7, because the brightest part of the comet was then its tail, consisting
primarily of massive, early post-perihelion emissions of microscopic dust.  A week
or so after perihelion the tail could be up to 5~magnitudes brighter than the head;
at the time of the last naked-eye sighting the difference could still amount to
about 2~magnitudes.  The implication is that the intrinsic brightness of the head
of potential historical Kreutz sungrazers in the Hasegawa \& Nakano paper is
significantly overestimated.

(9) Among the bright members of the Kreutz system, the best determined light
curve is available for comet Ikeya-Seki.  The comet was brightening and fading
symmetrically relative to perihelion, following an inverse fourth power of
heliocentric distance, with the absolute magnitude of 5.9.  Recently published
brightness estimates for the Great September Comet of 1882 from New Zealand,
combined with the previously available spotty data, suggests that this comet's
nuclear condensation brightened on the way to perihelion as an inverse fourth
power of heliocentric distance as well.  I assume in this paper that the $r^{-4}$
variation is the universal preperihelion light-curve law for all sungrazers
comparable with, or exceeding, Ikeya-Seki in intrinsic brightness, including the
potential sungrazers on the Hasegawa-Nakano list and the proposed Kreutz comets
in AD~363.     

(10) Comparison of the light curve of comet Ikeya-Seki with Schaefer's algorithm
for the limiting magnitude of a stellar object by the naked eye shows that the
comet brightened with decreasing solar elongation less steeply than the limiting
magnitude in advanced twilight but more steeply than the limiting magnitude in
broad daylight.  These findings imply that if Ikeya-Seki had arrived before the
era of telescopic comet hunters, the chance of its detection with the naked eye
was decreasing from early twilight (when it was modest at best) until the comet
was nearing the Sun's disk when, very briefly, the detection probability was up.

(11) Comet Ikeya-Seki would certainly have been discovered with the naked eye
days after perihelion when its brilliant tail was rapidly developing.  Suddenly
blazing into view, this massive dust feature also explains the detection times of
the Kreutz candidates on Hasegawa \& Nakano's (2001) list that passed perihelion
in October--November, even though they were, on the average, intrinsically fainter
than Ikeya-Seki.  None of them was discovered before perihelion and neither would
have been, quite possibly, Ikeya-Seki in the pre-telescope era.

(12) The first naked-eye sighting just days after perihelion appears to be a
general rule of a thumb for most historical Kreutz sungrazers, confirming the
importance of the post-perihelion tail.  The brightest among the bright objects
were detected in broad daylight, the rest in twilight.  The brightness supremacy
of the Great March Comet of 1843, the main surviving fragment of the progenitor's
Lobe~I (and the largest mass of Population~I), and the Great September Comet
of 1882, the main surviving fragment of Lobe~II (and the largest mass of
Population~II), is demonstrated by the fact that they are the only two Kreutz
sungrazers discovered --- in the telescope era --- with the naked eye {\it
weeks before perihelion\/}.\footnote{The apparent discovery of comet C/1843~D1
in early February makes it eligible for the designation C/1843~C1.}     

(13) Based on the times of their first sightings and expected compliance with
the $r^{-4}$ law, the two giant sungrazers are found to have been of equivalent
intrinsic brightness when approaching perihelion.  Calibrated with the Halley
data, the absolute magnitudes come nominally out to be \mbox{$H_0^- \!= 3.5$ mag}
for C/1843~D1 and \mbox{$H_0^- \!= 3.4$ mag} for C/1882~R1, with uncertainty
greater than the difference.

(14) The dramatically different behavior of the two giant sungrazers after
perihelion is attributed to the nuclear fragmentation of the 1882 comet in
close proximity of perihelion, while the nucleus of the 1843 comet appeared
intact throughout its apparition.  In broader terms, this distinction may
imply that, as a rule, members of Population~II are more susceptible to
perihelion fragmentation than members of Population~I, a tendency confirmed
by other sungrazers, such as Ikeya-Seki and Pereyra (C/1963~R1), as well.
The fragmentation of C/1882~R1 resulted in major effects on the comet's
post-perihelion light curve (making it distinctly flatter and the absolute
magnitude much brighter than before perihelion) as well as on the time of
the last sighting (with both the naked eye and telescopically).  Indeed,
the head of the 1882 sungrazer is known to have been fading after perihelion
according to an $r^{-3.3}$~law, its absolute magnitude having reached $-$0.2
(Sekanina 2002).  Its tail was seen with the naked eye for nearly six months
after perihelion!  The post-perihelion light curve of the nuclear condensation
of the 1843 sungrazer shows a rate of fading a little steeper than the
inverse fourth power of heliocentric distance.  I adopted for this comet an
$r^{-4.2}$~law, which combined with the condition for the last sighting with
the naked eye implied a post-perihelion absolute magnitude of 4.6, nominally
1.1~mag fainter than before perihelion.  The last naked-eye sighting of its
tail occurred 1$\frac{1}{2}$ months after perihelion.

(15) Based on this limited evidence, I adopted a rule that the post-perihelion
light curve of an intrinsically bright Kreutz sungrazer {\it not fragmenting
at perihelion\/} followed an $r^{-4.2}$~law and that the exponent of the
light-curve law got flatter by 0.2 for each additional major secondary nucleus
observed.  The preperihelion and post-perihelion absolute magnitudes were linked
by requiring a smooth light-curve transition at perihelion.  This rule was
applied to the Kreutz sungrazers in the proposed AD~363 scenario to define
their light curves.

(16) The enormous difference between C/1843~D1 and C/1882~R1 in the time of the
last sighting with the naked eye is believed to be hereditary, transferred from
the parent to its fragments, thereby offering a test of the contentious issue of
whether the comet X/1106~C1 --- granted it was a Kreutz sungrazer --- constituted
the previous appearance of the 1843 comet or the 1882 comet.  Given that the 1106
comet was seen for only about 50~days, the test unequivocally suggests a close
relationship between X/1106~C1 and C/1843~D1.  The post-perihelion and preperihelion
absolute magnitudes of the 1106 comet were determined under these circumstances to
have equaled 4.0 and 2.9, respectively.  

(17) The main computations in this study were to simulate a swarm of 10~daylight
comets in AD~363, equated with fragments of the Kreutz progenitor modeled as a contact
binary.  For each fragment I adopted its orbital elements from Paper~1 and assigned
it a set of light-curve parameters.  November~15.00 was adopted as a reference time,
defined by the perihelion passage of the fragmented progenitor's center of mass.  The
perihelion times of the comets, reckoned from this reference time, were determined
by the perturbations produced by the transverse component of the separation velocities
acquired in a sequence of fragmentation events that marked the birth of each of the
10~sungrazers.  The range of their perihelion times was 4.6~days. 

(18) Employing approximations to simulate the AD~363 scenario, in particular the use
of local solar time to eliminate the dependence on the observing site's geographic
longitude and local sidereal time, I determined, for each of the 10~Kreutz sungrazers
in the swarm:\ (i)~the horizontal coordinates for the latitude of Antioch on the
Orontes, Ammianus' residence at the time, and (ii)~the visibility index $\Im$ in
terms of Schaefer's limiting magnitude for a stellar object, as a function of the
sungrazer's solar elongation and elevation and the Sun's elevation.  The outcome
was the swarm of sungrazers projected onto the plane of the sky relative to the
Sun and the discrimination between the ones that are and are not expected to be
naked-eye detectable.  The computations were made for every day, starting
November~1 and ending November~25, for 25~instances every day, between 4:48 and
19:12 at 36~minute intervals.

(19)  The results of this exercise show that, in projection onto the plane of the
sky, the Kreutz sungrazers were approaching the Sun from the southwest, passing
between the Sun and Earth, and were receding to the west-southwest on the far side
of the Sun.  Up to seven comets were simultaneously visible with the naked eye
when near the Sun and high enough above the horizon.  Their number, identity,
and arrangement of display varied from day to day, their apparent brightness being
up to more than 7~mag above the limiting magnitude for the naked eye.  As expected,
in most cases the brightest and best seen comets were those within a day or so of
their perihelion passage; they were located mostly within a few degrees of the Sun.

(20) The impressive show continued over four or five consecutive days (essentially
equaling the range of the sungrazers' perihelion times), the number of visible
comets decreasing rather steeply from the peak of the seven on November~14 and
six on the 13th and 15th on both earlier and later days.  Except when within a
few hours of perihelion, the comets always set before sunset.  Around November~1
and November~25 a few sungrazers could be seen with the naked eye rather briefly
very low above the southeastern horizon in morning twilight, but they disappeared
by daybreak.

(21) The tested scenario certainly does not contradict Ammianus' narrative.  The
narrow range of the perihelion times is obviously the decisive factor in making
the swarm of daylight Kreutz sungrazers look compact.  To a degree this range
depends on the orbital locations of the fragmentation events; if they should occur
after aphelion, the degree of compactness of the swarm would still be increased;
if before aphelion, the perihelion times of fragments would be spread out over a
little longer interval and so would the swarm.

(22)  The range of the fragments' perihelion times would grow rapidly --- and the
swarm of comets would tend to vanish --- with increasing magnitude of the separation
velocity's radial component, which was in the tested scenario assumed to be nil.  A
radial velocity as low as 0.1~m~s$^{-1}$ would change a perihelion time by about
5~days, so that a systematic effect of this or greater magnitude in several consecutive
fragmentation events could easily extend the range of perihelion times over months,
in which case it is highly unlikely that Ammianus would use the plural in his remark.
The individual sungrazers would arrive at perihelion far apart from one another and
their close physical relationship would be harder to recognize. \\

This research was carried out at the Jet Propulsion Laboratory under contract with
the National Aeronautics and Space Administration.\\[-0.2cm]

%
\begin{center}
{\large \bf Appendix A} \\[0.3cm]
EFFECTS OF THE SEPARATION VELOCITY\\OF A DETACHING FRAGMENT ON\\ITS ORBITAL
ELEMENTS \\[0.3cm]
\end{center}
Let a Kreutz sungrazer move about the Sun unaffected by an outgassing-driven
nongravitational acceleration and/or the planetary perturbations, and let its
orbit be determined by the argument of perihelion $\omega$, the longitude
of the ascending node $\Omega$, the inclination $i$, the perihelion distance
$q$, and the orbital period $P$.  Let the comet pass through perihelion at
time $t_0$ and break up at time $t_{\rm frg}$, where \mbox{$t_0 \leq t_{\rm
frg} < t_\pi$} and $t_\pi$ is the time of next perihelion, \mbox{$t_\pi =
t_0 + P$}.  Let the comet's heliocentric distance and true anomaly at
$t_{\rm frg}$ be, respectively, \mbox{$r(t_{\rm frg}) = r_{\rm frg}$} and
\mbox{$u(t_{\rm frg}) = u_{\rm frg}$}.  The outcome of the breakup event is
the birth of a fragment that separates at a rate of $V_{\rm sep}$, usually on
the order of 1~m~s$^{-1}$, from the main mass whose orbital motion is deemed
unaffected by the fragmentation episode.

Next, I introduce an RTN right-handed orthogonal coordinate system, whose
origin is in the center of mass of the primary, the R axis pointing radially
away from the Sun, the T axis in the orbital plane, and the N axis normal to
this plane.  Let the components of the fragment's separation velocity vector
{\boldmath $V_{\bf sep}$} in the cardinal directions of the RTN system be
$V_{\rm R}$, $V_{\rm T}$, and $V_{\rm N}$.  This velocity vector prompts the
fragment to get into a new orbit, determined by the asterisk-labeled elements
\begin{eqnarray*}
\omega^{\displaystyle \ast}\! & = & \omega + \Delta \omega(\mbox{\boldmath
 $V_{\bf sep}$}), \nonumber \\
\Omega^{\displaystyle \ast}\! & = & \Omega + \Delta \Omega(\mbox{\boldmath
 $V_{\bf sep}$}), \nonumber \\
i^{\displaystyle \ast}\! & = & i + \Delta i(\mbox{\boldmath $V_{\bf sep}$}),
 \nonumber \\[-0.02cm]
q^{\displaystyle \ast}\! & = & q + \Delta q(\mbox{\boldmath $V_{\bf sep}$}),
 \nonumber \\[0.02cm]
P^{\displaystyle \ast}\! & = & P + \Delta P(\mbox{\boldmath $V_{\bf sep}$}),
 \nonumber \\[-0.02cm]
t_\pi^{\displaystyle \ast} & = & \, t_\pi + \Delta t_\pi(\mbox{\boldmath
 $V_{\bf sep}$}). \rlap{\hspace{1.85cm}{\rm (A-1)}}
\end{eqnarray*}
The aim of this exercise is to derive $\Delta \omega$, \ldots, $\Delta t_\pi$
as a function of the separation velocity vector {\boldmath $V_{\bf sep}$}.

To undertake this task, I begin with the determination of the primary comet's
position vector ({\boldmath $S_{\bf frg}$}) and orbital-velocity vector
({\boldmath $U_{\bf frg}$}) at the fragmentation time, $t_{\rm frg}$, in
the ecliptic coordinates:
\begin{displaymath}
{\hspace{-0.3cm}}\mbox{\boldmath $S_{\bf frg}$} = (x_{\rm frg}, y_{\rm frg},
 z_{\rm frg}), \;
 \mbox{\boldmath $U_{\bf frg}$} = (\dot{x}_{\rm frg}, \dot{y}_{\rm frg},
 \dot{z}_{\rm frg}). \rlap{\hspace{0.12cm}{\rm (A-2)}}
\end{displaymath}
They are given by the expressions:
\begin{displaymath}
\left( \!\!
\begin{array}{c}
x_{\rm frg} \\
y_{\rm frg} \\
z_{\rm frg}
\end{array}
\!\! \right) \!= r_{\rm frg} \!\left( \!\!
\begin{array}{cc}
P_x & Q_x \\
P_y & Q_y \\
P_z & Q_z
\end{array}
\!\! \right) \!\! \times \!\!\left( \!\!
\begin{array}{c}
\cos u_{\rm frg} \\
\sin u_{\rm frg}
\end{array}
\!\! \right), \rlap{\hspace{0.67cm}{\rm (A-3)}} \\[-0.05cm]
\end{displaymath}
and
\begin{displaymath}
\left( \!\!
\begin{array}{c}
\dot{x}_{\rm frg} \\
\dot{y}_{\rm frg} \\
\dot{z}_{\rm frg}
\end{array}
\!\! \right) \!=\! \frac{k_0}{\sqrt{p}} \!\left( \!\!
\begin{array}{cc}
P_x & Q_x \\
P_y & Q_y \\
P_z & Q_z
\end{array}
\!\! \right) \!\!\times \!\!\left( \!\!
\begin{array}{c}
-\sin u_{\rm frg} \\
e \!+\! \cos u_{\rm frg}
\end{array}
\!\! \right), \rlap{\hspace{0.5cm}{\rm (A-4)}}
\end{displaymath}
where the directional cosines $P_x$, \ldots, $Q_z$, are, together with $R_x$,
$R_y$, and $R_z$ (employed below), the components of the unit vectors {\boldmath
$P$}, {\boldmath $Q$}, and {\boldmath $R$} in the orthogonal coordinate system
tied to the orbital plane and the line of apsides:
\begin{eqnarray*}
\left( \!
\begin{array}{ccc}
P_x & \,P_y & \,P_z \\
Q_x & \,Q_y & \,Q_z \\
R_x & \,R_y & \,R_z
\end{array}
\! \right) & = & \left( \!\!
\begin{array}{ccc}
\;\;\:\cos \omega & \;\sin \omega & \;0 \\
-\!\sin \omega    & \;\cos \omega & \;0 \\
        0         &         0     & \;1
\end{array}
\right) \nonumber \\[0.15cm]
& \times & \! \left(
\begin{array}{ccc}
1 & \:    0        & \;   0   \\
0 & \;\;\;\:\cos i & \;\sin i \\
0 & \ -\!\sin i    & \;\cos i
\end{array}
\right) \!\!\times \!\! \left( \!\!
\begin{array}{ccc}
\;\;\:\cos \Omega & \;\sin \Omega & \;0 \\
   -\!\sin \Omega & \;\cos \Omega & \;0 \\
       0          & \;     0      & \;1
\end{array}
\right) \!. \nonumber \\[-0.1cm]
& & \rlap{\hspace{5.15cm}{\rm (A-5)}}
\end{eqnarray*}

At the time of separation, $t_{\rm frg}$, the position vector{\vspace{-0.04cm}}
of the fragment, {\boldmath $S$}$_{\bf frg}^{\displaystyle \ast}$, coincides
{\vspace{-0.08cm}}with the primary's position vector, while the fragment's
{\vspace{-0.04cm}}orbital-velocity vector, {\boldmath $U$}$_{\bf
frg}^{\displaystyle \ast}$, is the sum of the primary's orbital-velocity
vector and the fragment's separation velocity vector, whose ecliptic
components are denoted by $V_x$, $V_y$, and $V_z$,
\begin{displaymath}
\hspace{-0.22cm}\left( \!\!
\begin{array}{c}
x_{\rm frg}^{\displaystyle \ast} \\
y_{\rm frg}^{\displaystyle \ast} \\
z_{\rm frg}^{\displaystyle \ast}
\end{array}
\!\! \right) \!=\! \left( \!\!
\begin{array}{c}
x_{\rm frg} \\
y_{\rm frg} \\
z_{\rm frg}
\end{array}
\!\! \right) \!, \;\;
\left( \!\!
\begin{array}{c}
\dot{x}_{\rm frg}^{\displaystyle \ast} \\
\dot{y}_{\rm frg}^{\displaystyle \ast} \\
\dot{z}_{\rm frg}^{\displaystyle \ast}
\end{array}
\!\! \right) \!=\! \left( \!\!
\begin{array}{c}
\dot{x}_{\rm frg} \\
\dot{y}_{\rm frg} \\
\dot{z}_{\rm frg}
\end{array}
\!\! \right) \!+\! \left( \!\!
\begin{array}{c}
V_x \\
V_y \\
V_z
\end{array}
\!\! \right)\! .\rlap{\hspace{0.18cm}{\rm (A-6)}}
\end{displaymath}
The ecliptic components of the separation velocity vector are related to its
components in the RTN coordinate system by
\begin{displaymath}
\left( \!\!
\begin{array}{c}
V_x \\
V_y \\
V_z
\end{array}
\!\! \right) \!= \! \left( \!\!
\begin{array}{ccc}
P_x & Q_x & R_x \\
P_y & Q_y & R_y \\
P_z & Q_z & R_z
\end{array}
\!\! \right) \!\times \! \left( \!\!
\begin{array}{ccc}
\cos u_{\rm frg} &  \!-\!\sin u_{\rm frg} & 0 \\
\sin u_{\rm frg} & \;\:\,\cos u_{\rm frg} & 0 \\
      0          &         0          & 1
\end{array}
\! \right) \!\times \!\left( \!\!
\begin{array}{c}
V_{\rm R} \\
V_{\rm T} \\
V_{\rm N}
\end{array}
\!\! \right) \!.
\end{displaymath}

\vspace{-0.17cm}
\hspace{7.47cm}(A-7)\\[0.2cm]
One is now ready to initiate the determination of the fragment's orbital
elements by introducing the angular-momentum vector components:
\begin{eqnarray*}
\Im_{xy} & = & \left|\,
\begin{array}{cc}
x_{\rm frg} & \; y_{\rm frg} \\[0.1cm]
\dot{x}_{\rm frg}^{\displaystyle \ast} & \; \dot{y}_{\rm frg}^{\displaystyle
 \ast}
\end{array}
\, \right| \!, \nonumber \\[0.2cm]
\Im_{yz} & = & \left| \,
\begin{array}{cc}
y_{\rm frg} & \; z_{\rm frg} \\[0.1cm]
\dot{y}_{\rm frg}^{\displaystyle \ast} & \;\dot{z}_{\rm frg}^{\displaystyle
 \ast}
\end{array}
 \,\right| \!, \nonumber \\[0.2cm]
\Im_{zx} & = & \left|\,
\begin{array}{cc}
z_{\rm frg} & \; x_{\rm frg} \\[0.1cm]
\dot{z}_{\rm frg}^{\displaystyle \ast} & \;\dot{x}_{\rm frg}^{\displaystyle
 \ast}
\end{array}
 \,\right| \! . \rlap{\hspace{2.2cm}{\rm (A-8)}}
\end{eqnarray*}
%
%

The following computations allow to incorporate, if deemed desirable, a
nongravitational acceleration into the orbital motion of the fragment on the
assumptions that it points in the antisolar direction and varies inversely
as the square of heliocentric distance, the constraints employed by
Hamid \& Whipple (1953) in their investigation and likewise integrated
into the standard model for the split comets (Sekanina 1982).  Let $k_0$
be{\vspace{-0.05cm}} the Gaussian gravitational constant expressed in
AU$^{\frac{3}{2}}$\,day$^{-1}$ and $\gamma_0$ a dimensionless parameter that
describes the fragment's nongravitational acceleration in units of 10$^{-5}$\,the
Sun's gravitational acceleration.  The fragment's {\it effective\/} Gaussian
gravitational constant $k_0^{\displaystyle \ast}$, which replaces $k_0$
below, is equal to
\begin{displaymath}
k_0^{\displaystyle \ast} \!= \sqrt{k_0^2 - 10^{-5} \gamma_0 k_0^2} =
 k_0 \sqrt{1 \!-\! 10^{-5} \gamma_0}. \rlap{\hspace{0.4cm}{\rm (A-9)}}
\end{displaymath}

The fragment's longitude of the ascending node, $\Omega^{\displaystyle
\ast}\!$, inclination, $i^{\displaystyle \ast}\!$, and semilatus rectum,
$p^{\displaystyle \ast}\!$, related to the perihelion distance,
$q^{\displaystyle \ast}\!$, via the orbital eccentricity, $e^{\displaystyle
\ast}\!$, by \mbox{$p^{\displaystyle \ast} = q^{\displaystyle \ast}
(1 \!+\! e^{\displaystyle \ast}\:\!\!)$}, follow from the relations, 
\begin{displaymath}
\left( \!\!
\begin{array}{c}
\Im_{xy} \\
\Im_{yz} \\
\Im_{zx}
\end{array}
\!\! \right) = k_0^{\displaystyle \ast} \sqrt{p^{\displaystyle \ast}} \left( \!\!
\begin{array}{c}
\cos i^{\displaystyle \ast} \\
\sin \Omega^{\displaystyle \ast} \sin i^{\displaystyle \ast} \\
-\cos \Omega^{\displaystyle \ast} \sin i^{\displaystyle \ast}
\end{array}
\!\!\! \right) \!.\rlap{\hspace{0.55cm}{(A-10)}}
\end{displaymath}
For the longitude of the ascending node one finds
\begin{displaymath}
\hspace{-0.43cm}\tan \Omega^{\displaystyle \ast} \!= -\frac{\Im_{yz}}{\Im_{zx}}, \;\;\;
 {\rm sign}(\sin \Omega^{\displaystyle \ast}\:\!\!) = {\rm sign}(\Im_{yz});\rlap{\hspace{0.2cm}{(A-11)}}
\end{displaymath}
for the inclination
\begin{displaymath}
\tan i^{\displaystyle \ast} \!= \frac{\sqrt{\Im_{yz}^2 \!+\!
 \Im_{zx}^2}}{\Im_{xy}},\rlap{\hspace{1.65cm}{(A-12)}}
\end{displaymath}
where the sign of the denominator determines the quadrant of the inclination;
and for the semilatus rectum
\begin{displaymath}
p^{\displaystyle \ast} = \frac{\Im_{xy}^2 \!+\! \Im_{yz}^2 \!+\!
 \Im_{zx}^2}{(k_0^{\displaystyle \ast\:\!\!})^2} .\rlap{\hspace{1.63cm}{(A-13)}}
\end{displaymath}
Next, the fragment's orbital eccentricity is
\begin{displaymath}
e^{\displaystyle \ast} \! = \sqrt{1 + \! p^{\displaystyle \ast} \! \left\{ \:\!\!
 \frac{1}{(k_0^{\displaystyle \ast\:\!\!})^2} \!\left[ \! \left( \! \dot{x}_{\rm
 frg}^{\displaystyle \ast} \!\right)^{\!2} \!\!+\! \left(\! \dot{y}_{\rm
 frg}^{\displaystyle \ast} \!\right)^{\!2} \!\!+\! \left(\! \dot{z}_{\rm
 frg}^{\displaystyle \ast} \!\right)^{\!2} \right] \!-\!  \frac{2}{r_{\rm frg}}
 \!\right\}}\, ,
\end{displaymath}

\vspace{-0.2cm}
\hspace{7.29cm}(A-14)\\[0.15cm]
so that the perihelion distance comes out from\\[-0.15cm]
\begin{displaymath}
q^{\displaystyle \ast} \!= \frac{p^{\displaystyle \ast}}{1 \!+\!
 e^{\displaystyle \ast}} \rlap{\hspace{2.46cm}(A-15)}
\end{displaymath}
and the orbital period from
\begin{displaymath}
P^{\displaystyle \ast} \!= \frac{2\pi}{k_0^{\displaystyle \ast}}
 \! \left(\!\frac{q^{\displaystyle \ast}}{1 \!-\! e^{\displaystyle
 \ast}} \!\! \right)^{\!\!\frac{3}{2}} \!\!. \rlap{\hspace{1.92cm}(A-16)}
\end{displaymath}
The fragment's true anomaly at the time of separation from the primary is given by
\begin{displaymath}
\hspace{-0.69cm}\sin u_{\rm frg}^{\displaystyle \ast} = \frac{\sqrt{p^{\displaystyle
 \ast}}}{k_0^{\displaystyle \ast} e^{\displaystyle \ast}
 r_{\rm frg}} \! \left( x_{\rm frg} \dot{x}_{\rm frg}^{\displaystyle
 \ast} \!+\! y_{\rm frg} \dot{y}_{\rm frg}^{\displaystyle \ast}
 \!+\! z_{\rm frg} \dot{z}_{\rm frg}^{\displaystyle \ast} \right) \rlap{\hspace{0.11cm}(A-17)}
\end{displaymath}
with \mbox{${\rm sign}(\cos u_{\rm frg}^{\displaystyle \ast}) =
{\rm sign}(p^{\displaystyle \ast} \!\!-\! r_{\rm frg})$}.  The argument of
peri\-helion is involved with the true anomaly at~separation~by\\[-0.3cm]
\begin{displaymath}
\hspace{-0.06cm}\cos (\omega^{\displaystyle \ast} \!\!+\! u_{\rm frg}^{\displaystyle
 \ast}) = \frac{x_{\rm frg} \cos \Omega^{\displaystyle
 \ast} \!+\! y_{\rm frg} \sin \Omega^{\displaystyle \ast}}{r_{\rm frg}}
 \rlap{\hspace{0.38cm}(A-18)}
\end{displaymath}
with \mbox{${\rm sign}[\sin (\omega^{\displaystyle \ast} \!\!+\! u_{\rm
frg}^{\displaystyle \ast})] = {\rm sign}(z_{\rm frg})$}.  Equations~(A-17)
{\vspace{-0.04cm}}and (A-18) isolate the argument of perihelion.
Finally, to derive the fragment's next perihelion time, $t_\pi^{\displaystyle
\ast}$, {\vspace{-0.02cm}}one first gets the eccentric anomaly at
separation, $\epsilon_{\rm frg}^{\displaystyle \ast}$,
\begin{displaymath}
\epsilon_{\rm frg}^{\displaystyle \ast} = 2 \arctan \! \left( \! \sqrt{ \frac{1
 \!-\! e^{\displaystyle \ast}}{1 \!+\! e^{\displaystyle \ast}}}
 \tan {\textstyle \frac{1}{2}} u_{\rm frg}^{\displaystyle \ast} \! \right) \!,
 \rlap{\hspace{0.58cm}(A-19)} \\[-0.05cm]
\end{displaymath}
which provides the following relation for the time of the fragment's next
perihelion passage:
\begin{displaymath}
t_\pi^{\displaystyle \ast} = t_{\rm frg} - \frac{\epsilon_{\rm
 frg}^{\displaystyle \ast} \!-\! e^{\displaystyle \ast} \! \sin
 \epsilon_{\rm frg}^{\displaystyle \ast}}{k_0^{\displaystyle \ast}}
 \! \left( \! \frac{q^{\displaystyle \ast}}{1 \!-\! e^{\displaystyle
 \ast}} \:\!\!\! \right)^{\!\!\frac{3}{2}} \!\!. \rlap{\hspace{0.61cm}(A-20)}
\end{displaymath}
The eccentric anomaly $\epsilon_{\rm frg}^{\displaystyle \ast}$ in
{\vspace{-0.08cm}}this equation is in radians and for fragmentation
{\vspace{-0.05cm}}times $t_{\rm frg}$ between $t_0$ and $t_\pi^{\displaystyle
\ast}$ its range is \mbox{$-2\pi \leq \epsilon_{\rm frg}^{\displaystyle \ast}
\leq 0$}.  Inserting{\vspace{-0.04cm}} $\omega^{\displaystyle \ast}$, \ldots,
$t_\pi^{\displaystyle \ast}$ into (A-1) completes the derivation of the effects
of the fragment's separation velocity on its orbital elements. \\[-0.1cm]

%
\begin{figure}[b]
\vspace{0.8cm}
\hspace{-0.2cm}
\centerline{
\scalebox{0.59}{
\includegraphics{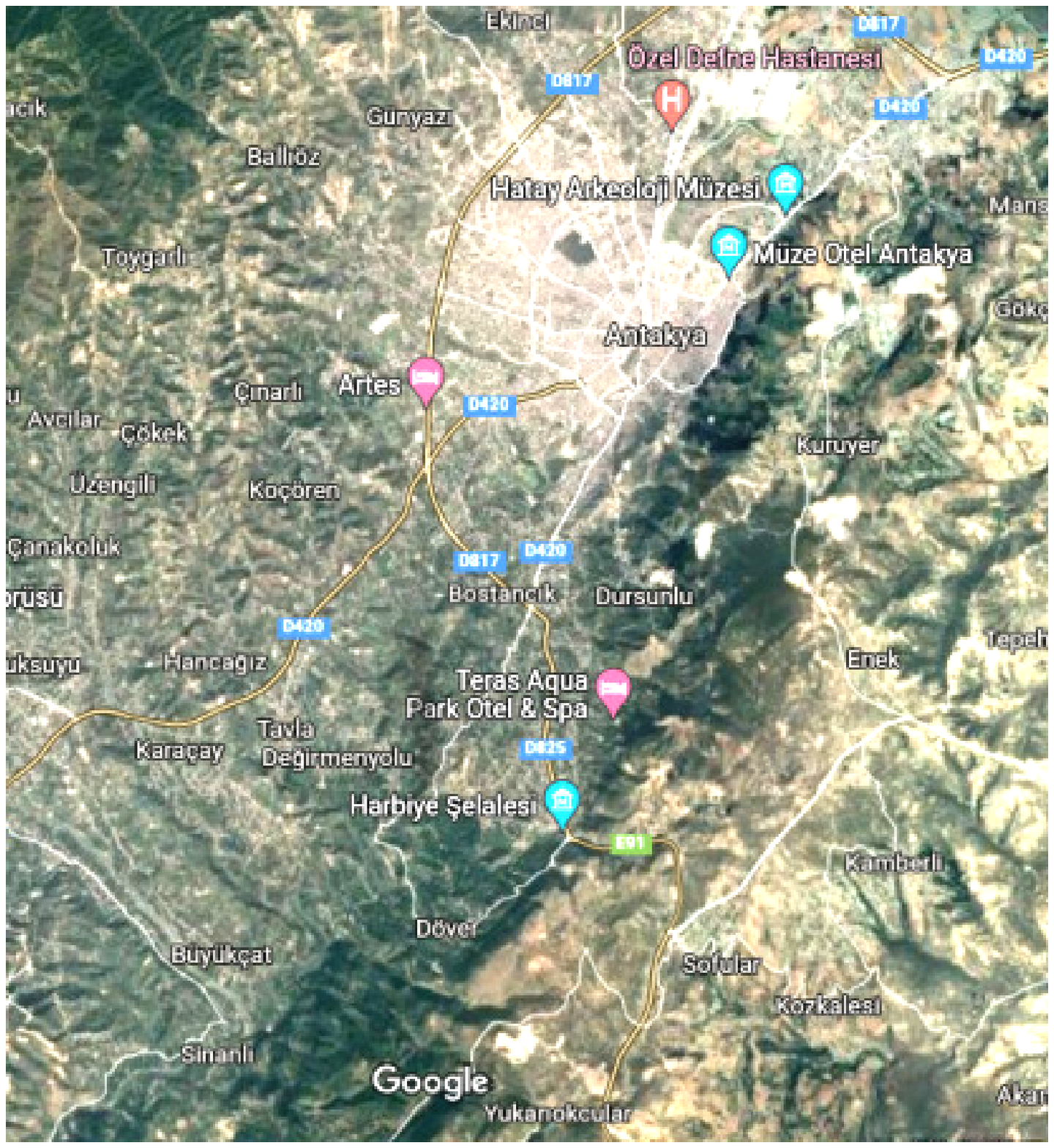}}}
\vspace{0cm}
\vspace{0.2cm}
\parbox{8.6cm}{{\footnotesize \bf Figure B-1}.  Satellite image of the city of
Antakya and its environs.  North is up, east to the left.  The diagonal measures
about 26 km.  The dark green elongated feature just to the city's east and
southeast, extending from the north-northeast to the south-southwest is Mount
Silpius.  The river Orontes, flowing through the city also from the north-northeast
to the south-southwest, is too narrow to show on this scale.\vspace{0cm}}
\end{figure}

\begin{center}
{\large \bf Appendix B}\\[0.3cm]
ANCIENT ANTIOCH:\ THE CITY, GEOGRAPHY, AND LOCAL HORIZON\\[0.3cm]
\end{center}
The ancient city of Antioch
('$\:\!\!\!${\it A\/}$\nu\tau\iota\acute{o}\chi\varepsilon\iota\alpha$) was located
on the site of the modern city of Antakya, the capital of the Hatay Province,
Turkey, at latitude 36$^\circ\!$.2\,N and longitude 36$^\circ\!$.2\,E.

\begin{figure*}[t]
\vspace{0.18cm}
\hspace{-0.17cm}
\centerline{
\scalebox{1.32}{
\includegraphics{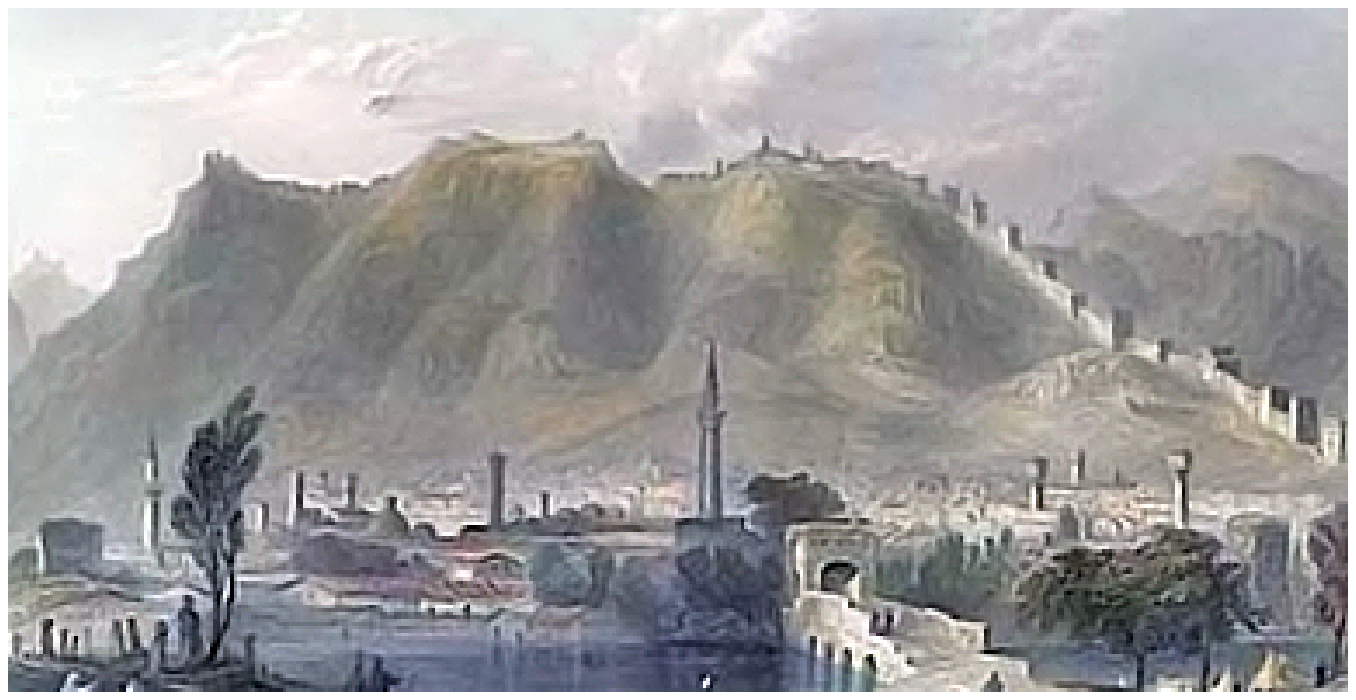}}}

%

\vspace{0.5cm} 
\hspace{-0.13cm}
\centerline{
\scalebox{0.663}{
\includegraphics{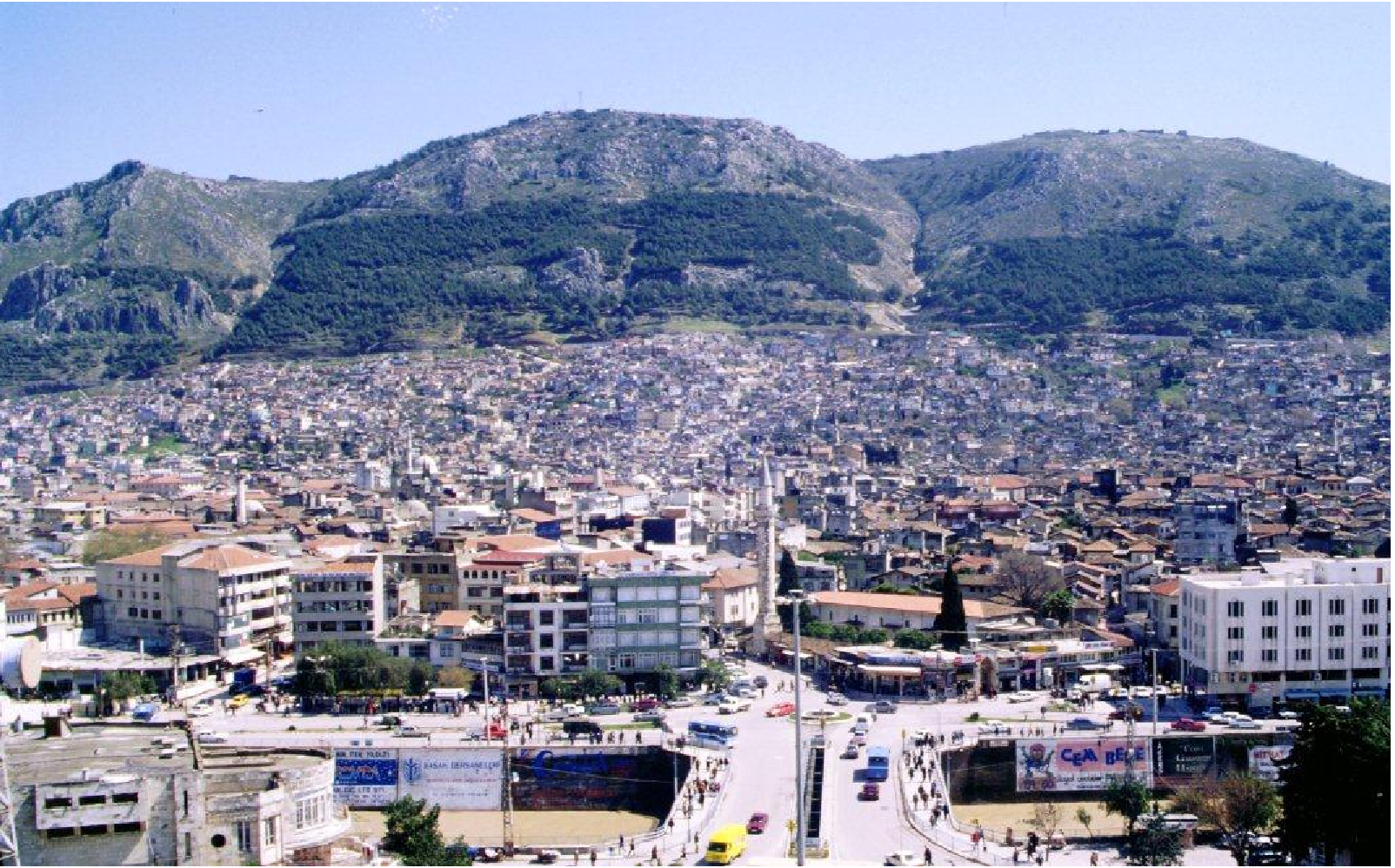}}}

\vspace{0.2cm}
\parbox{18.05cm}{{\bf Figure B-2 (top)}.  Antioch on the Orontes and Mount Silpius
in the early 19th century, seen in a close-up picture made at a location in the city's
southwestern suburb.  The artist's line of sight subtends a small angle with the
direction along which the hill extends, making its main dimension project greatly
foreshortened.  Note the southern section of the defensive wall on the right,
climbing up the slopes of the hill, and the citadel at the summit on the upper
left. (W. H. Bartlett 1836; steel engraved print).{\hspace{5cm}\vspace{0.15cm}}
{\bf Figure B-3 (bottom)}.  The modern city of Antakya, with Mount
Silpius (Habib-i Neccar) in the background, displayed in a photograph taken from a
more distant location to the northwest of downtown.  Unlike in Figure~B-2, the hill
now projects broadside.  Note the river Orontes (Asi Nehn) in the foreground.}
\end{figure*}

In a mountainous terrain and near the border with Syria to the east and south, the
location is a fertile valley on the river Orontes, which flows through the municipality
from the north-northeast to the south-southwest (albeit with numerous minor bends),
eventually feeding the Levantine Sea, the easternmost part of the Mediterranean.
Antioch's distance from the sea is 17~km.
  
The Antioch valley is surrounded, on the north side, by Mount Amanus, which is connected
to the Taurus range further north, and, on the south side, by the Mount Casius (also Kel
Da\v{g}i or Jebel al-Aqra) range (peaking at 1700~meters), which terminates, just to
the southeast and east of the city, by a 440-meter high triple-peaked hill, extending
parallel to the valley and known in the antiquity as Mount Silpius (now Habib-i Neccar).
This geography is clearly apparent from Figure~B-1, a satellite image of the area.
Although the valley between the mountains is about 16~km long and 8~km wide, the city
of Antioch is said to have been built on the eastern side of the river, in part at the
very foot of Mount Silpius, covering in the early period an area of about 4.5~km$^2$,
its extent apparently growing with time.

%
Completed in the 1st century AD was a nearly 20~km long defensive wall, encircling the
entire city and climbing, on the eastern side, the slopes of Mount Silpius.  Along the
entire length of the wall, separated from one another by about \mbox{100--150 meters},
were multi-storied bastions staffed by mercenaries and at the summit of Mount Silpius the
rampart was dominated by a fortified citadel that was overlooking the city.  A panorama
of Antioch and the surrounding area in the early 19th century --- with the commanding
appearance of Mount Silpius and clear evidence of the surviving wall with the bastions,
especially its southern sector, and the citadel --- is displayed in Figure~B-2
(Bartlett 1836), while a view of the modern city of Antakya is offered in Figure~B-3.

With the city at an altitude of 70~meters and Mount Silpius summit at 440~meters, the
eastern and southeastern horizon of Antioch was greatly elevated.  This applied to a wide
range of azimuths, as Figure~B-2 shows that the ridge connecting the three peaks was not
significantly depressed.  For example, the city's locations that were about 3~kilometers
distant from the foot of Mount Silpius, objects in the eastern and southeastern sky at
elevations of up to \mbox{arctan[(440\,--\,70)/3000] = 7$^\circ$} remained unobservable,
hidden behind the hill.  For locations closer to the hill, the horizon's elevation at
the critical azimuths rose rapidly; at 1~kilometer from the hill the horizon was at
20$^\circ$!  It happens that the southeastern azimuths are exactly the directions at
which the Kreutz sungrazers are computed to have briefly been visible in twilight, just
a few degrees above the ideal horizon, more than a week or so before or after the time of
the spectacle's anticipated peak performance.  It is obvious that Ammianus was prevented
by Mount Silpius from seeing the Kreutz sungrazers low in the southeastern sky even if
the observing conditions were otherwise favorable.  By contrast, the southwestern
horizon was essentially unobstructed.

By strange coincidences, Antioch appears to be related, one way or another, to
what I consider three generations of the Kreutz sungrazers.  Besides the spectacle
of AD~363, Antioch is obliquely linked to the comet of 372~BC, known as Aristotle's
comet, as the city was founded by Seleucus I.\ Nicator, a former general in the army
of Aristotle's former student, Alexander the Great.  There is also an indirect
connection between Antioch and the probable Kreutz sungrazer of the third generation,
the Great Comet of 1106, by virtue of the city's becoming a Crusader state in 1098.
Both the comet and Antioch, especially its conquest, were, side by side, commented
on in {\it Historia Hierosolymitana\/} (History of Jerusalem) and other Crusader
chronicles (e.g., by Fulcher of Chartres; see Ryan 1969). \\[-0.2cm]


\vspace{-0.6cm}
\begin{center}
{\large \bf Appendix C} \\[0.25cm]
TABULATION OF\\[0.01cm]
POSITIONAL AND VISIBILITY MODELS FOR\\[0.01cm]
SWARM OF KREUTZ SUNGRAZERS SIMULATING\\[0.01cm]
AMMIANUS' DAYLIGHT COMETS IN AD 363 \\[0.1cm]
\end{center}
Below is a series of six tables, C-1 through C-6, showing the positional and
brightness predictions for the ten Kreutz sungrazers in a swarm, used to simulate
Ammianus' {\it comets seen in broad daylight\/}.  The positions are compared
with that of the Sun.  Each of Tables~C-1 through C-5 consists of six sections,
Table~C-6 of five sections.  Each section provides computed data for a particular
time of appearance on a particular date, arranged into 12~columns, which offer: \\[0.1cm]
\indent (1) date of appearance in local solar time (LST); \\[0.03cm]
\indent (2) hour and minute LST; \\[0.03cm]
\indent (3) object identification --- the Sun or a sungrazer; \\[0.03cm]
\indent (4) difference --- time of appearance minus perihelion time (days); \\[0.03cm]
\indent (5) azimuth, reckoned positive from the south to the west (deg); \\[0.03cm]
\indent (6) elevation, reckoned positive from the flat local horizon toward the zenith
 (deg); \\[0.03cm]
\indent (7) solar elongation (deg); \\[0.03cm]
\indent (8) zenithal position angle, reckoned positive clockwise from the direction to
 zenith (deg); \\[0.03cm]
\indent (9) phase angle, Sun-comet-Earth (deg); \\[0.03cm]
\indent (10) apparent magnitude, computed from the light-curve parameters in
 Table~7;\\[0.03cm]
\indent (11)~Schaefer's (1998) limiting magnitude, derived as described in the
 text; \\[0.03cm]
\indent (12) visibility index, defined by Eq.\,(17) as a difference of columns~11 and
10 (mag).\\[-0.22cm]

For the Sun, only columns 1--3 and 5--6 are relevant.  For the sungrazers, columns
5, 6, 8, and 11 refer to the latitude of Antioch, but to an arbitrary longitude.  The
sidereal time is defined by Eq.\,(20), an approximation.\\[-0.22cm]

\begin{table*}[h]
\vspace{-4.2cm}
\hspace{0.6cm}
\centerline{
\scalebox{1}{
\includegraphics{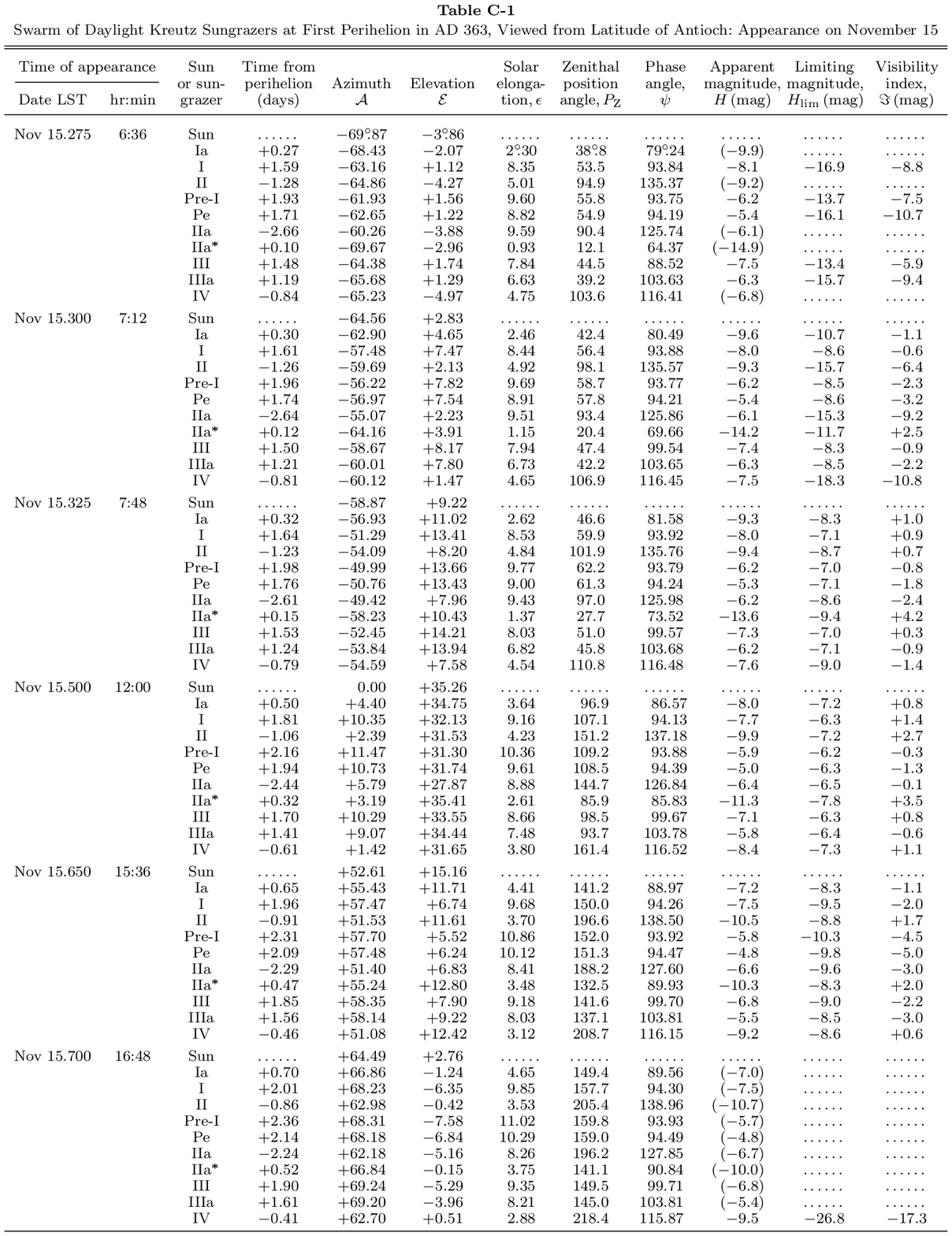}}}
\end{table*}
\begin{table*}
\vspace{-4.2cm}
\hspace{0.6cm}
\centerline{
\scalebox{1}{
\includegraphics{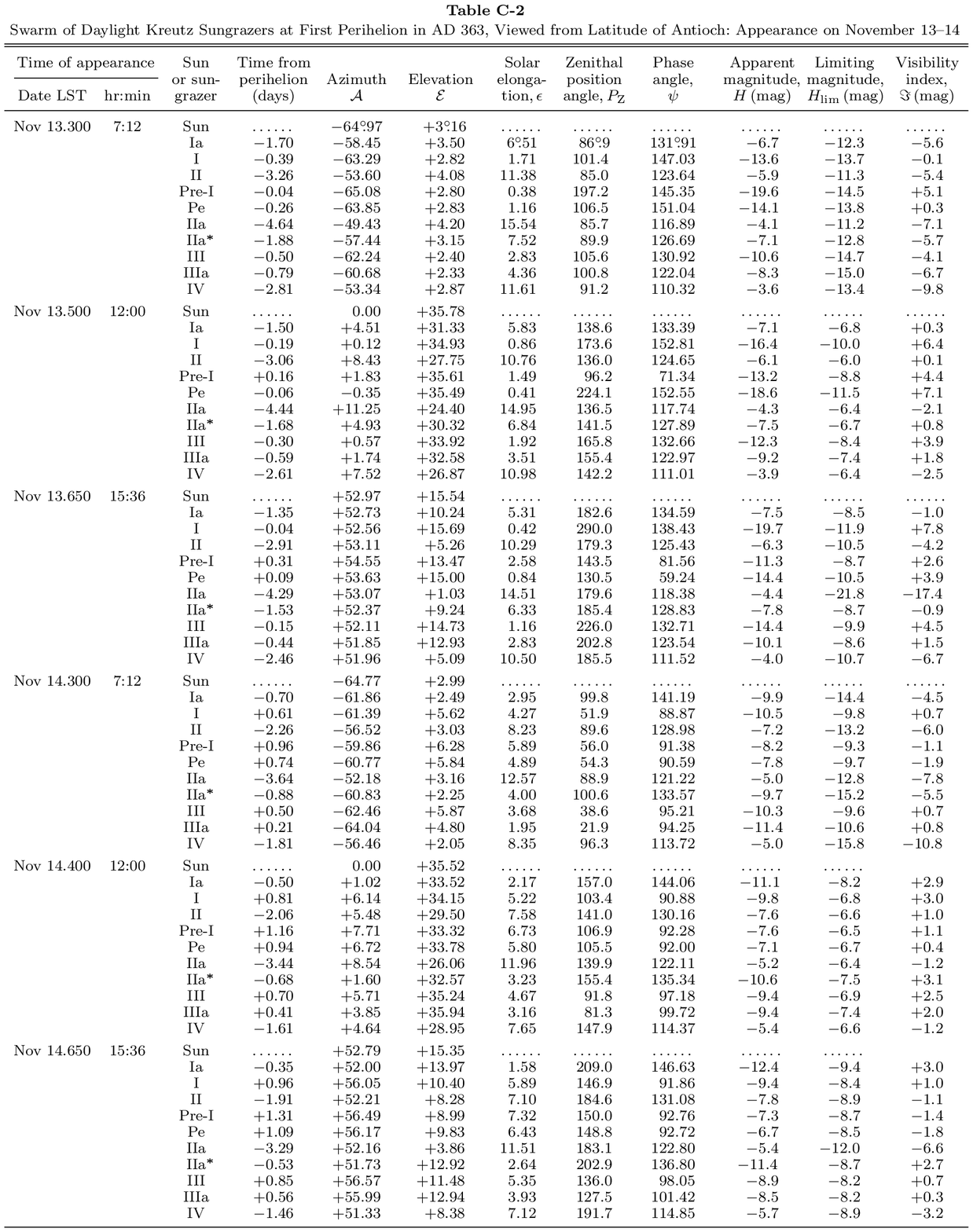}}}
\end{table*}
\begin{table*}
\vspace{-4.2cm}
\hspace{0.6cm}
\centerline{
\scalebox{1}{
\includegraphics{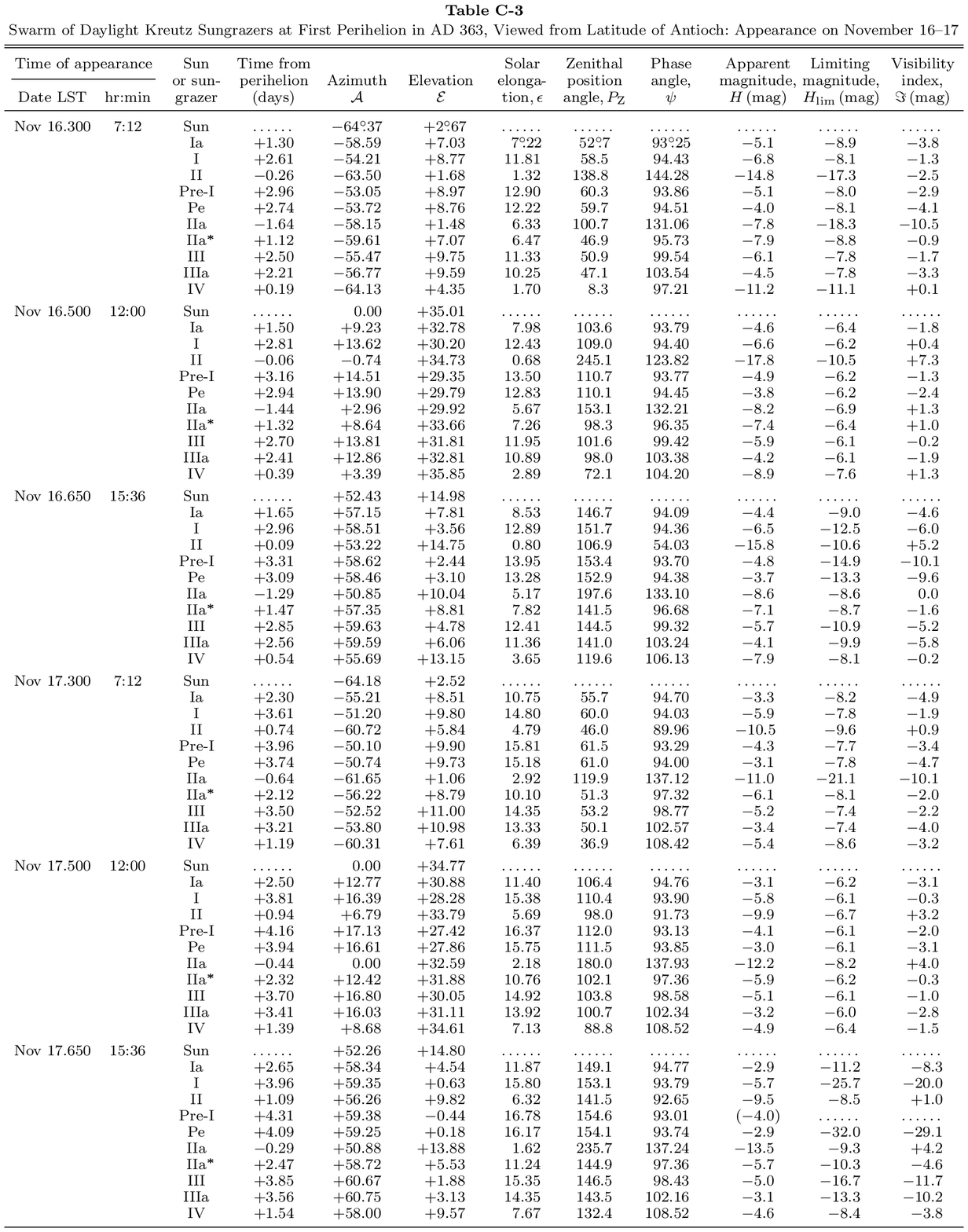}}}
\end{table*}
\begin{table*}
\vspace{-4.2cm}
\hspace{0.6cm}
\centerline{
\scalebox{1}{
\includegraphics{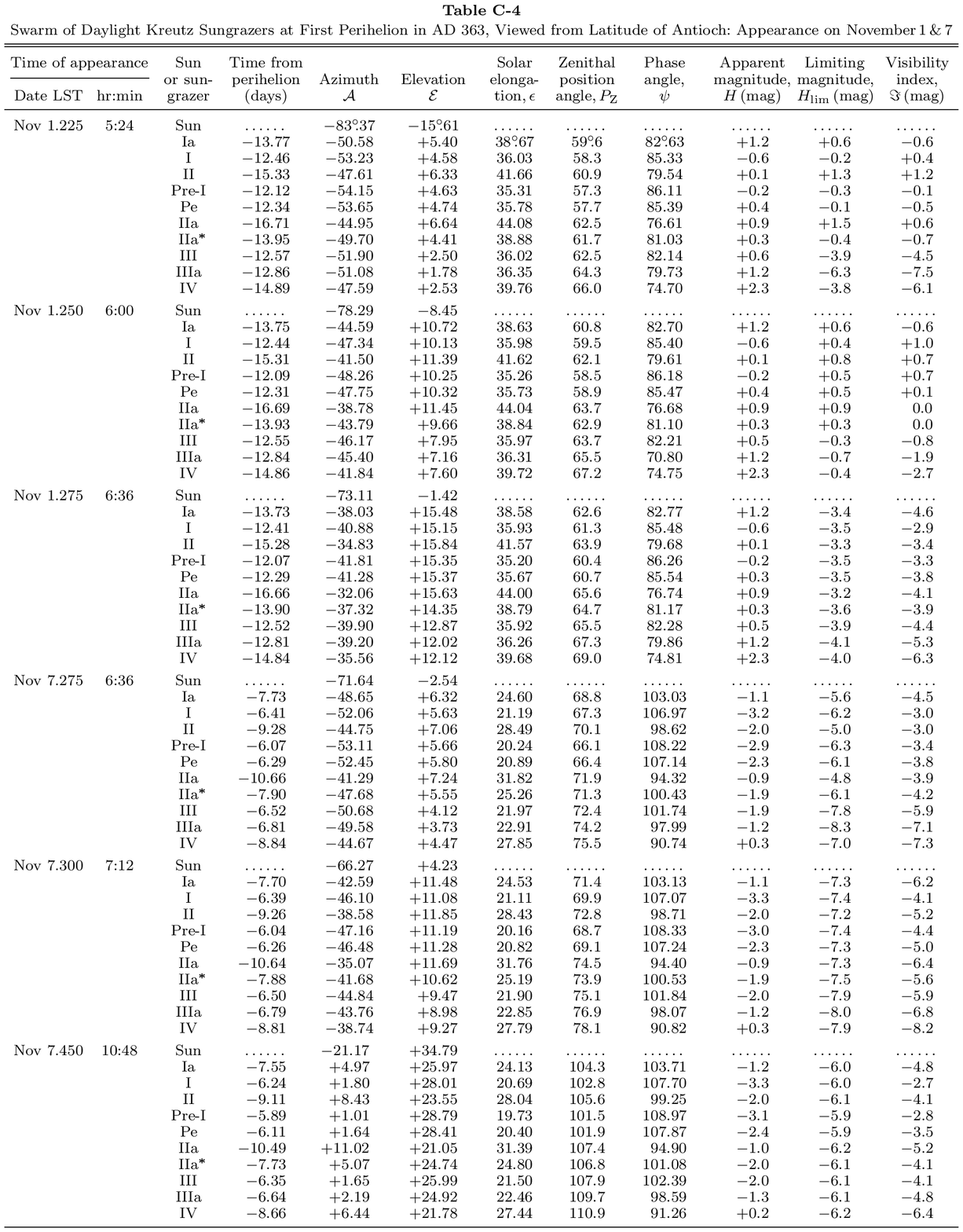}}}
\end{table*}
\begin{table*}
\vspace{-4.2cm}
\hspace{0.6cm}
\centerline{
\scalebox{1}{
\includegraphics{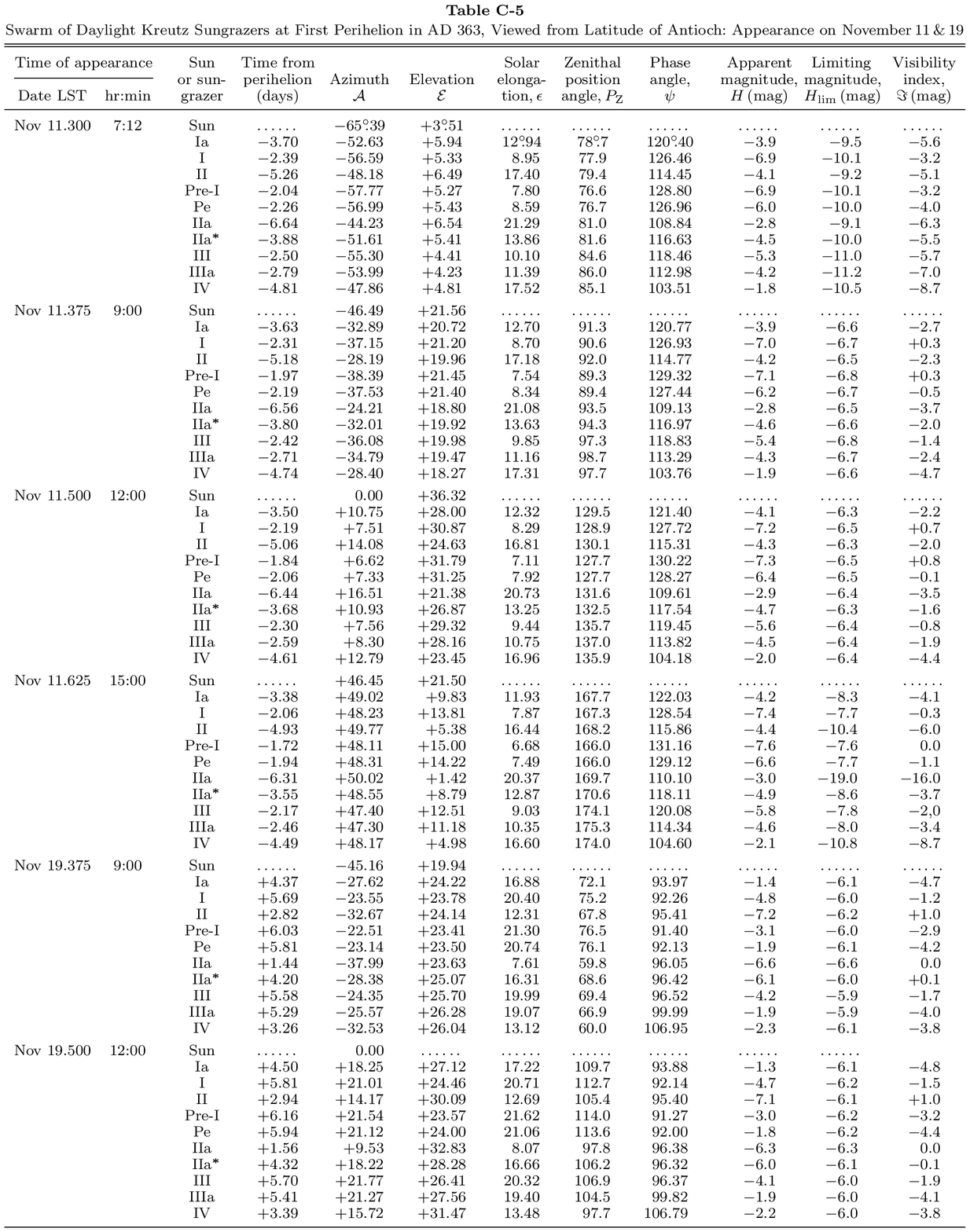}}}
\vspace{-0.2cm}
\end{table*}
\begin{table*}[t]
\vspace{-4.7cm} 
\hspace{0.6cm}
\centerline{
\scalebox{1}{
\includegraphics{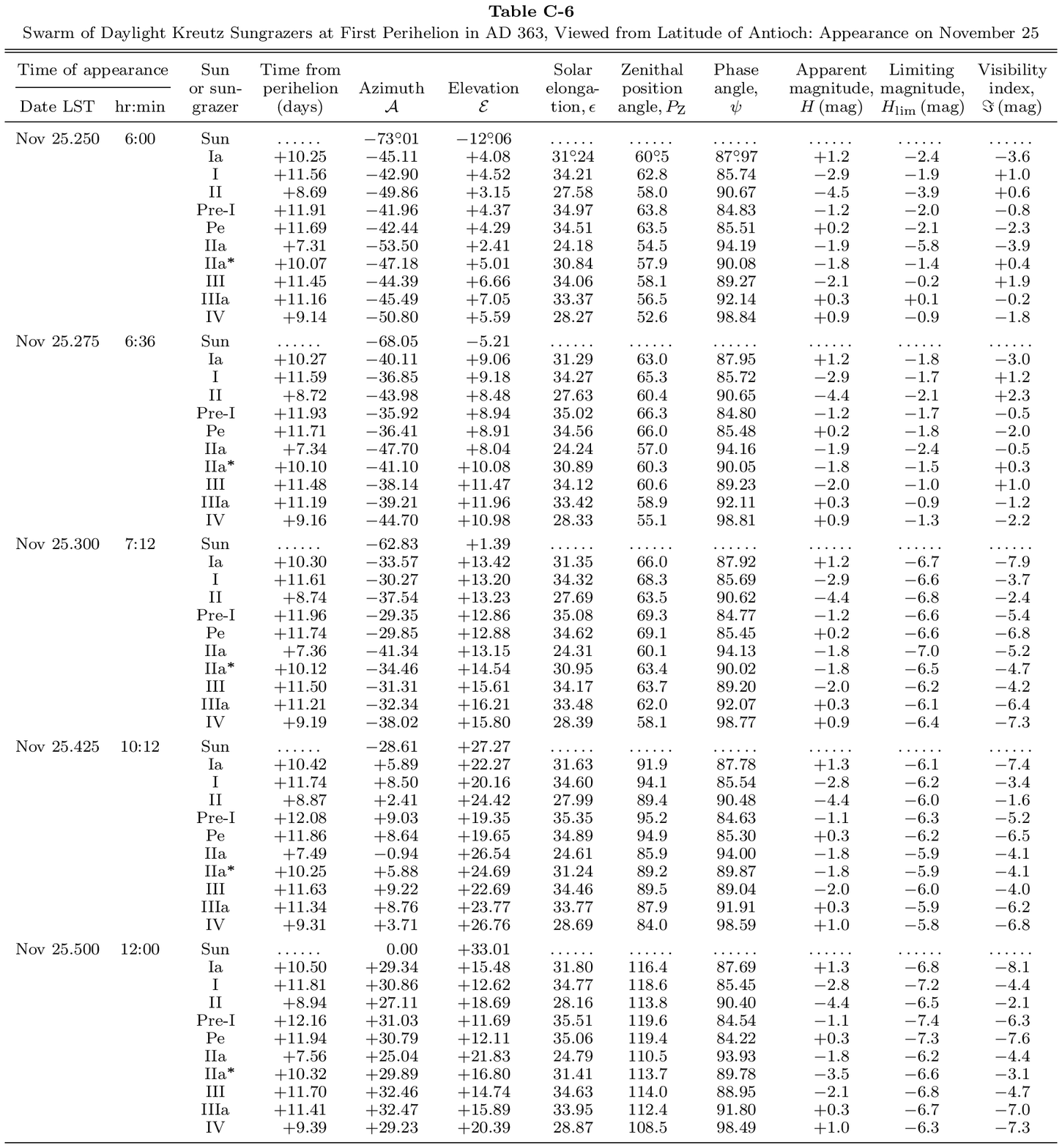}}}
\vspace{-5.15cm} 
\end{table*}

{\small \bf Table C-1} provides the data on the swarm appearance for six times on
November~15.  The first time is {\sl 6:36\/} (civil twilight), when five sungrazers
were above the horizon (none visible to the naked eye) and five below it.  The average
solar elongation, {\elng}, of the five comets above the horizon is 8$^\circ\!$.3,
their average visibility index, {\vis}, is $-$8.5~mag.  The second time is {\sl
7:12\/} (daylight from now on), when all ten sungrazers were above the horizon, one
visible with the naked eye (\mbox{$\Im > +0.5$ mag}); \mbox{{\elng} = 6$^\circ\!$.4}
and \mbox{{\vis} = $-$3.4 mag}.  By the next time, {\sl 7:48\/}, four sungrazers
were visible with the naked eye, \mbox{{\elng} = 6$^\circ\!$.5},
\mbox{{\vis} = 0.0 mag}.  At noon, the fourth time, six sungrazers could be seen,
\mbox{{\elng} = 6$^\circ\!$.8} and \mbox{{\vis} = +0.8 mag}.  By {\sl 15:36} only
two sungrazers remained visible, \mbox{{\elng} = 7$^\circ\!$.1} and \mbox{{\vis} =
$-$1.6 mag}.  At the last tabulated time, {\sl 16:48\/}, minutes before sunset,
only a single sungrazer was just barely above horizon. \\[-0.25cm]

{\small \bf Table C-2} lists the data for three particular times on November~13 and 14
each.  The first time on both dates is {\sl 7:12\/}.  On the 13th all comets as well
as the Sun were at this time at elevations between 2$^\circ\!$.4 and 4$^\circ\!$.2
and only one sungrazer was visible with the naked eye.  Average solar elongation and
visibility index were 6$^\circ\!$.3 and $-$3.9~mag.  Six sungrazers were visible at
noon, when \mbox{{\elng} = 5$^\circ\!$.8} and \mbox{{\vis} = +2.0~mag}.  At {\sl
15:36\/} the numbers were five comets, 5$^\circ\!$.5, and $-$1.0~mag.

The next day, November 14, at {\sl 7:12\/} the number of visible sungrazers equaled
three, while the two averages were \mbox{{\elng} = 5$^\circ\!$.7} and \mbox{{\vis} =
$-$3.5 mag}.  At {\sl 12:00\/} the number of visible comets peaked at seven, with
\mbox{{\elng} = 5$^\circ\!$.8} and \mbox{{\vis} = +1.4 mag}.  By {\sl 15:36\/} the
number of visible sungrazers dropped to four, the average solar elongation continued to
climb to 5$^\circ\!$.9 and the average visibility index was down to $-$0.6~mag.\\[-0.25cm]

{\small \bf Table C-3} covers the dates November 16 and 17.  The number of displayed
sungrazers was clearly on a downturn.  On the 16th, no comet was visible at {\sl
7:12\/}, four at {\sl 12:00\/}, and one at {\sl 15:36\/}.  The average solar
elongations at the three times were, respectively, 8$^\circ\!$.2, 8$^\circ\!$.6, and
9$^\circ\!$.0, indicating that most sungrazers were clearly receding from the Sun;
the visibility index equaled, respectively, $-$3.1~mag, +0.4~mag, and $-$3.8~mag.

The trend continued on November 17.  Although one comet was visible at {\sl 7:12\/},
the average visibility index declined by 0.5~mag to $-$3.6~mag and \mbox{{\elng} =
10$^\circ\!$.8}.  At noon and {\sl 15:36\/} one could see the same two sungrazers (II
and IIa, which were close to their perihelia).  At {\sl 12:00\/} the averages {\elng}
and {\vis} were, respectively, 11$^\circ\!$.4} and $-$0.7~mag.  Even though the Sun
was still about 15$^\circ$ above the horizon at {\sl 15:36\/}, one sungrazer already
set and three others were about to set.  This strongly affected the visibility index,
whose average for the nine comets was $-$9.2~mag, while \mbox{{\elng} =
11$^\circ\!$.2}. \\[-0.25cm]

{\small \bf Table C-4} provides data for a pair of dates long before the spectacle reached
its peak.  The comets were too faint to be visible in broad daylight, but they had
risen before the Sun over the southeastern horizon.  For November~1 I list the swarm
appearance at {\sl 5:24\/} in astronomical twilight, at {\sl 6:00\/} in nautical
twilight, and at {\sl 6:36\/} in civil twilight.  Two sungrazers were visible to the
naked eye in astronomical twilight, three in nautical twilight, but none in civil
twilight, the changes being due to the interplay between atmospheric extinction
(peaking in the first case) and increasing interference from sunlight (peaking in
the last case).  The average solar elongations varied insignificantly, equaling
38$^\circ\!$.3, 38$^\circ\!$.2, and 38$^\circ\!$.1.  On the other hand, the average
visibility index showed a peak in nautical twilight, the numbers at the three times
being $-$1.8, $-$0.3, and $-$4.2~mag, respectively.

By November 7, much of the edge relative to the Sun in elevation, enjoyed by the
sungrazers six days earlier, was lost.  This is seen clearly in their positions on the
two dates at {\sl 6:36\/}, both in civil twilight.  The outcome of this difference
is that, even though closer to the time of the best display, no sungrazer was visible
on November~7.  The average visibility index was then $-$4.6~mag at {\sl 6:36\/},
$-$5.8~mag at {\sl 7:12\/}, and $-$4.3~mag at {\sl 10:48\/}.  The average solar
elongations at the three times were, respectively, 24$^\circ\!$.5, 24$^\circ\!$.4,
and 24$^\circ\!$.0. \\[-0.25cm]
   
{\small \bf Table C-5} describes the appearance of the swarm at selected times on
November~11 and 19.  On the first date two sungrazers were visible at noon, but
none at {\sl 7:12\/}, {\sl 9:00}, and again at {\sl 15:00}.  The average solar
elongations and visibility indices were 12$^\circ\!$.9 and $-$5.4~mag at {\sl 7:12\/};
12$^\circ\!$.7 and $-$1.9~mag at {\sl 9:00\/}; 12$^\circ\!$.4 and $-$1.5~mag at
{\sl 12:00\/}; and 12$^\circ\!$.0 and $-$4.5~mag at {\sl 15:00}.{\pagebreak}

{\vspace*{0.2cm}}
The conditions on November 19 are compared with those on November~11 employing the
display data at {\sl 9:00\/} and {\sl 12:00\/}.  On the 19th one sungrazer,
Fragment~II, was visible at either time.  However, while \mbox{{\elng} = 16$^\circ\!$.8}
and \mbox{{\vis} = $-$2.1 mag} at {\sl 9:00\/}, the respective numbers were 17$^\circ\!$.1
and $-$2.3~mag at {\sl 12:00\/}, so that the conditions at noon were, surprisingly,
slightly inferior to those three hours earlier. \\[-0.25cm]

{\vspace*{0.1cm}}
{\small \bf Table C-6} presents the data for the final date, November~25.  The
visibility conditions appear to have been more favorable than expected that long
after the peak display.  They may in fact be compared with those on November~1
examined in Table~C-4.  The data are shown for the end of astronomical twilight
at {\sl 6:00\/}, early civil twilight at {\sl 6:36\/}, just after sunrise at
{\sl 7:12\/}, mid-morning at {\sl 10:12\/}, and noon. During this period of time
the visibility conditions are seen to have suddenly deteriorated at sunrise.  Three
comets were visible at the two twilight times, but none later on.  As expected, the
average solar elongation was very slowly increasing from 31$^\circ\!$.3 at {\sl
6:00\/} to 31$^\circ\!$.9 at {\sl 12:00}, but the visibility index dropped abruptly
at sunrise, its values equaling $-$0.9 and $-$0.5~mag at the two twilight times,
but $-$5.4, $-$4.9, and $-$5.5~mag for the three daylight times.  As noted in the
text, the conditions could at these times be more favorable for any sungrazer, whose
brightness was substantially enhanced by a contribution from the post-perihelion
dust tail. \\[-0.1cm]

\begin{center}
{\footnotesize        
REFERENCES} 
\end{center}
\begin{description}
{\footnotesize
\item[\hspace{-0.15cm}]
Bartlett, W. H. 1836, Syria, the Holy Land, Asia Minor, etc.\,Illus-{\linebreak}
 {\hspace*{-0.6cm}}trations.  London:\ Fisher, Son, \& Co.
\\[-0.57cm]
\item[\hspace{-0.3cm}]
Broughton, R. P. 1979, J. Roy. Astron. Soc. Canada, 73, 24
\\[-0.57cm]
\item[\hspace{-0.3cm}]
Close, M. 1843, Mon. Not. Roy. Astron. Soc., 5, 293
\\[-0.57cm]
\item[\hspace{-0.3cm}]
Drijvers, J. W. 2011, J. Late Antiq., 4, 280
\\[-0.57cm]
\item[\hspace{-0.3cm}]
Drijvers, J. W. 2022, The Forgotten Reign of the Emperor Jovian{\linebreak}
 {\hspace*{-0.6cm}}(363--364):\ History and Fiction.  New York:\ Oxford
 University{\linebreak}
 {\hspace*{-0.6cm}}Press, 264\,pp
\\[-0.57cm]
\item[\hspace{-0.3cm}]
Eddie, L. A. 1883, Mon. Not. Roy. Astron. Soc., 43, 289
\\[-0.57cm]
\item[\hspace{-0.3cm}]
Elkin, W. L. 1882, Mon. Not. Roy. Astron. Soc., 43, 22
\\[-0.57cm]
\item[\hspace{-0.3cm}]
Encke, J. F. 1843, Astron. Nachr., 20, 349 
\\[-0.57cm]
%
%
%
\item[\hspace{-0.3cm}]
Ernst, M. 1911, Astron. Nachr., 187, 303 
\\[-0.57cm]
%
%
\item[\hspace{-0.3cm}]
Ferrin, I., \& Gil, C. 1988, A\&A, 194, 288 
\\[-0.57cm]
\item[\hspace{-0.3cm}]
Finlay, W. H. 1883, Mon. Not. Roy. Astron. Soc., 43, 319
\\[-0.57cm]
%
%
\item[\hspace{-0.3cm}]
Galle, J. G. 1894, Verzeichniss der Elemente der bisher be-{\linebreak}
 {\hspace*{-0.6cm}}rechneten Cometenbahnen nebst Anmerkungen und Literatur-{\linebreak}
 {\hspace*{-0.6cm}}Nachweisen (neu bearbeitet, erg\"{a}nzt und fortgesetzt bis zum{\linebreak}
 {\hspace*{-0.6cm}}Jahre 1894).  Leipzig: W. Engelmann, 315pp
\\[-0.57cm]
\item[\hspace{-0.3cm}]
Gill, D. 1882, Mon. Not. Roy. Astron. Soc., 43, 19
\\[-0.57cm]
\item[\hspace{-0.3cm}]
Gould, B. A. 1883a, Astron. Nachr., 104, 129
\\[-0.57cm]
\item[\hspace{-0.3cm}]
Gould, B. A. 1883b, Astron. Nachr., 106, 273
\\[-0.57cm]
\item[\hspace{-0.3cm}]
Green, D. W. E. 2012, CBET 2967
\\[-0.57cm]
%
%
\item[\hspace{-0.3cm}]
Green, D. W. E., \& Morris, C. S. 1987, A\&A, 187, 560 
\\[-0.57cm]
%
%
\item[\hspace{-0.3cm}]
Hamid, S. E., \& Whipple, F. L. 1953, AJ, 58, 100
\\[-0.57cm]
\item[\hspace{-0.3cm}]
Hasegawa, I., \& Nakano, S. 2001, PASJ, 53, 931
\\[-0.57cm]
\item[\hspace{-0.3cm}]
Herrick, E. C. 1843a, Amer. J. Sci. Arts, 43, 412
\\[-0.57cm]
\item[\hspace{-0.3cm}]
Herrick, E. C. 1843b, Amer. J. Sci. Arts, 45, 188 
\\[-0.57cm]
\item[\hspace{-0.3cm}]
Ho, P.-Y. 1962, Vistas Astron., 5, 127
\\[-0.57cm]
\item[\hspace{-0.3cm}]
Holetschek, J. 1896, Denk. Akad. Wiss. Wien (Math.-Naturwiss.{\linebreak}
 {\hspace*{-0.6cm}}Kl.), 63, 317
\\[-0.57cm]
%
%
\item[\hspace{-0.3cm}]
Kendall, E. O. 1843, Astron. Nachr., 20, 387
\\[-0.57cm]
\item[\hspace{-0.3cm}]
Kiang, T. 1972, Mem. Roy. Astron. Soc., 76, 27
\\[-0.57cm]
\item[\hspace{-0.3cm}]
Kreutz, H. 1888, Publ. Sternw. Kiel, 3
\\[-0.62cm]
\item[\hspace{-0.3cm}]
Kreutz, H. 1891, Publ. Sternw. Kiel, 6}
%
%
\vspace{0.15cm}
\end{description}
{\pagebreak}
\begin{description}
{\footnotesize
\item[\hspace{-0.3cm}]
Kreutz, H. 1901, Astron. Abhandl., 1, 1
\\[-0.57cm]
\item[\hspace{-0.3cm}]
Kronk, G. W. 1999, Cometography:\ Volume 1 (Ancient--1799).{\linebreak}
 {\hspace*{-0.6cm}}Cambridge, UK:\ Cambridge University Press, 580\,pp
\\[-0.57cm]
\item[\hspace{-0.3cm}]
Kronk, G. W. 2004, Cometography:\ Volume 2 (1800--1899).  Cam-{\linebreak}
 {\hspace*{-0.6cm}}bridge, UK:\ Cambridge University Press, 852\,pp
\\[-0.57cm]
%
%
\item[\hspace{-0.3cm}]
Lynn, W. T. 1903, Observatory, 26, 326
\\[-0.57cm]
\item[\hspace{-0.3cm}]
Maclear, T. 1851, Mem. Roy. Astron. Soc., 20, 62
\\[-0.57cm]
\item[\hspace{-0.3cm}]
Marcus, J. N. 2007, Internat. Comet Quart., 29, 39 
\\[-0.57cm]
\item[\hspace{-0.3cm}]
Marsden, B. G. 1967, AJ, 72, 1170
\\[-0.57cm]
%
%
\item[\hspace{-0.3cm}]
Marsden, B. G. 1989, AJ, 98, 2306
\\[-0.57cm]
\item[\hspace{-0.3cm}]
Milon, D. 1966, Stroll. Astron., 19, 206 
\\[-0.57cm]
%
%
\item[\hspace{-0.3cm}]
Milon,\,D.,\,Solberg,\,G.,\,\&\,Minton,\,R.\,B.\,1967, Stroll.\,Astron.,\,20,\,165
\\[-0.57cm]
\item[\hspace{-0.3cm}]
Morris, C. S., \& Green, D. W. E. 1982, AJ, 87, 918 
\\[-0.57cm]
\item[\hspace{-0.3cm}]
\"{O}pik, E. 1966, Ir. AJ, 7, 141
\\[-0.57cm]
\item[\hspace{-0.3cm}]
Orchiston, W., Drummond, J., \& Kronk, G. 2020, J. Astron. Hist.{\linebreak}
 {\hspace*{-0.6cm}}Herit., 23, 628
\\[-0.57cm]
\item[\hspace{-0.3cm}]
Peirce, B. 1844, American Almanac.  Boston:\ J.\,Munroe, 94
\\[-0.57cm]
\item[\hspace{-0.3cm}]
Plummer, W. E. 1889, Observatory, 12, 140
\\[-0.57cm]
%
%
\item[\hspace{-0.3cm}]
Roemer, E. 1966, PASP, 78, 83 
\\[-0.57cm]
\item[\hspace{-0.3cm}]
Rolfe, J. C. 1940, The Roman History of Ammianus Marcellinus,{\linebreak}
 {\hspace*{-0.6cm}}{\it Res Gestae\/}, Book~25, Section 10.2 (Online Translation
 from La-{\linebreak}
 {\hspace*{-0.6cm}}tin),
 {\tt https://penelope.uchicago.edu/Thayer/E/Roman/Texts/{\linebreak}
 {\hspace*{-0.6cm}}Ammian/25$^\ast\!$.html}
\\[-0.57cm]
\item[\hspace{-0.3cm}]
Ryan, F. R. 1969, A History of the Expedition to Jerusalem,{\linebreak}
 {\hspace*{-0.6cm}}1095--1127; by Fulcher of Chartres (Translation from Latin),
 ed.{\linebreak}
{\hspace*{-0.6cm}}H.\,S.\,Fink.  Knoxville,\,TN:\ University of Tennessee
 Press,\,348\,pp.
\\[-0.57cm]
\item[\hspace{-0.3cm}]
Schaefer, B. E. 1993, Vistas Astron., 36, 311 
\\[-0.57cm]
\item[\hspace{-0.3cm}]
Schaefer, B. E. 1998, Sky Tel., 95, 57; code version by L. Bogan{\linebreak}
{\hspace*{-0.6cm}}at {\tt https://www.bogan.ca/astro/optics/vislimit.html}
\\[-0.57cm]
\item[\hspace{-0.3cm}]
Seargent, D. 2009, The Greatest Comets in History:\ Broom Stars{\linebreak}
 {\hspace*{-0.6cm}}and Celestial Scimitars.  New York:\ Springer Science+Business{\linebreak}
 {\hspace*{-0.6cm}}Media, LLC, 260pp
\\[-0.57cm]
{\nopagebreak}
\item[\hspace{-0.3cm}]
Sekanina, Z. 1982, in Comets, ed. L. L. Wilkening (Tucson, AZ:{\linebreak}
 {\hspace*{-0.6cm}}University of Arizona), 251
\\[-0.57cm]
\item[\hspace{-0.3cm}]
Sekanina, Z. 2000, ApJ, 542, L147
\\[-0.57cm]
\item[\hspace{-0.3cm}]
Sekanina, Z. 2002, ApJ, 566, 577
\\[-0.57cm]
\item[\hspace{-0.3cm}]
Sekanina, Z. 2021, eprint arXiv:2109.01297
\\[-0.57cm]
\item[\hspace{-0.3cm}]
Sekanina, Z., \& Chodas, P. W. 2004, ApJ, 607, 620
\\[-0.57cm]
\item[\hspace{-0.3cm}]
Sekanina, Z., \& Kracht, R. 2015, ApJ, 801, 135
\\[-0.57cm]
\item[\hspace{-0.3cm}]
Simms, W. 1845, Mon. Not. Roy. Astron. Soc., 6, 22
\\[-0.57cm]
\item[\hspace{-0.3cm}]
Stephenson, F.\,G., Yau, K.\,K., \& Hunger, H. 1985, Nature,\,314,\,587
\\[-0.57cm]
\item[\hspace{-0.3cm}]
Strom, R. 2002, A\&A, 387, L17
\\[-0.57cm]
\item[\hspace{-0.3cm}]
Tebbutt, J. 1882, Observatory, 5, 367 
\\[-0.57cm]
%
%
%
\item[\hspace{-0.3cm}]
Warner, B. 1980, Mon.\,Not.\,Roy.\,Astron.\,Soc.\,South Africa, 39, 69
\\[-0.57cm]
%
%
\item[\hspace{-0.3cm}]
Weinberg, J. L., \& Beeson, D. E. 1976, in Study of Comets,{\linebreak}
 {\hspace*{-0.6cm}}NASA\,SP-393,\,ed.\,B.\,Donn,\,M.\,Mumma,\,W.\,Jackson,\,M.\,A'Hearn,{\linebreak}
 {\hspace*{-0.6cm}}\& R. Harrington.  Washington, DC:\ U.S. GPO, 92
\\[-0.57cm]
\item[\hspace{-0.3cm}] 
Yau,\,K., Yeomans,\,D., \& Weissman,\,P.\,1994, Mon.\,Not.\,Roy.\,Astron.{\linebreak}
 {\hspace*{-0.6cm}}Soc., 266, 305
\\[-0.57cm]
\item[\hspace{-0.3cm}]
Ye, Q.-Z., Hui, M.-T., Kracht, R., \& Wiegert, P. A. 2014, ApJ,{\linebreak}
 {\hspace*{-0.6cm}}796, 83 (8\,pp)
\\[-0.64cm]
\item[\hspace{-0.3cm}]
Yeomans, D. K., \& Kiang, T. 1981, {\vspace{-0.07cm}}Mon.\ Not.\ Roy.\ Astron.\ Soc.,{\linebreak}
 {\hspace*{-0.6cm}}197, 633}
\vspace{-0.2cm}
\end{description}   
\end{document}